\documentclass[preprint,aps,nofootinbib,tightenlines]{revtex4}

\usepackage{graphicx}
\usepackage{amsmath}
\usepackage[hypertex]{hyperref}

\newcommand{\eq}[1]{Eq.~\eqref{#1}}
\newcommand{\eqs}[2]{Eqs.~\eqref{#1} and \eqref{#2}}
\renewcommand{\sec}[1]{Sec.~\ref{#1}}

\newcommand{\fig}[1]{Fig.~\ref{#1}}

\newcommand{\ie}{\emph{i.e.}}
\newcommand{\eg}{\emph{e.g.}}

\newcommand{\geneva}{\texttt{GenEvA}}
\newcommand{\GenEvA}{\texttt{GenEvA}}
\newcommand{\madevent}{\texttt{MadEvent}}
\newcommand{\madgraph}{\texttt{MadGraph}}

\newcommand{\alpgen}{\texttt{ALPGEN}}
\newcommand{\fortran}{Fortran}
\newcommand{\aalpha}{\texttt{ALPHA}}

\newcommand{\nn}{\nonumber}

\newcommand{\lab}[1]{\mathrm{#1}}
\newcommand{\CM}{\mathrm{CM}}
\renewcommand{\min}{\mathrm{min}}
\renewcommand{\max}{\mathrm{max}}
\newcommand{\start}{\mathrm{start}}
\newcommand{\cut}{\mathrm{cut}}
\newcommand{\match}{\mathrm{match}}

\newcommand{\df}{\mathrm{d}}
\newcommand{\ha}{\hat{\alpha}}
\newcommand{\cP}{\mathcal{P}}
\newcommand{\vS}{{\Sigma}}
\newcommand{\sud}{\Delta}
\newcommand{\GeV}{\:\mathrm{GeV}}

\begin{document}


\title{GenEvA (II): A phase space generator \\ from a reweighted parton shower}

\author{Christian W.~Bauer\footnote{Electronic address: cwbauer@lbl.gov}}

\author{Frank J.~Tackmann\footnote{Electronic address: ftackmann@lbl.gov}}

\author{Jesse Thaler\footnote{Electronic address: jthaler@jthaler.net}}

\affiliation{Ernest Orlando Lawrence Berkeley National Laboratory,
University of California, Berkeley, CA 94720
\vspace{2ex}}

\begin{abstract}
We introduce a new efficient algorithm for phase space generation. A parton shower is used to distribute events across all of multiplicity, flavor, and phase space, and these events can then be reweighted to any desired analytic distribution. To verify this method, we reproduce the $e^+ e^- \to n \text{ jets}$ tree-level result of traditional matrix element tools. We also show how to improve tree-level matrix elements automatically with leading-logarithmic resummation. This algorithm is particularly useful in the context of a new framework for event generation called \GenEvA. In a companion paper~\cite{genevaphysics}, we show how the \GenEvA\ framework can address contemporary issues in event generation.
\end{abstract}

\maketitle

\vspace*{-9ex}%
\tableofcontents

\newpage
\section{Introduction}

\subsection{Motivation}

Precision theoretical calculations are vital for any comparison between experimental measurements and underlying theoretical models. While in many cases we are interested in the effects of physics Beyond the Standard Model, a precise understanding of Standard Model backgrounds is always required for such a comparison to be useful. To be able to study the effect of experimental cuts and detector effects, the most useful form for theoretical predictions is as fully exclusive events with a high multiplicity of hadrons in the final state.

Ignoring the issue of hadronization and other nonperturbative effects, the challenge for constructing exclusive event generators at the perturbative level is that full quantum field theoretic results for high-multiplicity final states become prohibitively difficult both to calculate and to implement. At tree level, the number of required Feynman diagrams grows factorially with the number of partons in the final state, so that a direct calculation is not feasible. Even if an expression for the partonic calculation can be obtained, the large volume of high-multiplicity phase space makes it difficult to efficiently distribute events according to those distributions. Both of these constraints imply that in practice, it is impossible to use full quantum information to calculate typical partonic final states arising in collider environments.

It has been known for some time that for emissions at small transverse momentum, much of the quantum nature of QCD becomes subdominant, such that emissions in QCD can be approximated by classical splitting functions~\cite{Gribov:1972ri, Altarelli:1977zs, Dokshitzer:1977sg}. These splitting functions form the basis for so-called parton showers, which build up arbitrarily complicated final states by starting with low-multiplicity final states and recursively splitting one final state parton into two. The recursive nature of parton showers allows them to be cast into a Markov chain algorithm. Parton showers form the basis for multi-purpose event generators like \texttt{Pythia}~\cite{Sjostrand:2000wi, Sjostrand:2006za}, \texttt{Herwig}~\cite{Corcella:2000bw, Gieseke:2003hm, Gieseke:2006ga}, \texttt{Ariadne}~\cite{Lonnblad:1992tz}, \texttt{Isajet}~\cite{Paige:2003mg}, and \texttt{Sherpa}~\cite{Gleisberg:2003xi}, which build up QCD-like events from simple hard-scattering matrix elements.

While parton shower algorithms easily generate high-multiplicity events, the resulting distributions are not based on full QCD, and for example do not accurately describe emissions at large transverse momentum. For this reason event generators based on full matrix element information have been constructed. Examples for generators based on tree-level calculations are \texttt{Alpgen}~\cite{Mangano:2002ea}, \texttt{MadEvent}~\cite{Maltoni:2002qb}, \texttt{CompHEP}~\cite{Boos:2004kh}, \texttt{AMEGIC}~\cite{Krauss:2001iv}, \texttt{Whizard}~\cite{Kilian:2007gr}, and \texttt{Helac-Phegas}~\cite{Kanaki:2000ey, Papadopoulos:2000tt, Cafarella:2007pc}, while programs using next-to-leading order (NLO) calculations also exist, examples being \texttt{MCFM}~\cite{MCFM}, \texttt{NLOJet}~\cite{NLOJET}, \texttt{PHOX}~\cite{PHOX}, and \texttt{VBFNLO}~\cite{VBFNLO}. In the last few years, significant progress has been made towards combining full QCD calculations, distributed with traditional phase space generators, with parton shower algorithms~\cite{Catani:2001cc, Lonnblad:2001iq, Krauss:2002up, MLM, Mrenna:2003if, Schalicke:2005nv, Lavesson:2005xu, Hoche:2006ph, Alwall:2007fs, Giele:2007di, Lavesson:2007uu, Nagy:2007ty, Nagy:2008ns}. Encouraged by these successes, it is worth considering whether additional progress could be made by considering a fundamentally new approach to event generation.

In this paper, we introduce a novel use of the parton shower as a phase space generator for generic partonic calculations. The idea is to generate an event with a given final state multiplicity using a simple parton shower algorithm, and then to reweight this event to any desired partonic distribution. This algorithm is a key component of the \GenEvA\ framework for event generation that we discuss in a companion paper~\cite{genevaphysics}, so we will refer to the phase space generator presented here as the \GenEvA\ algorithm---for \textbf{Gen}erate \textbf{Ev}ents \textbf{A}nalytically.

The \GenEvA\ framework asserts that the choice of partonic calculations should not be determined by algorithmic needs but rather by the underlying full theory information we want to describe. For this to be possible, the \GenEvA\ framework requires a ``projectable'' phase space generator with a built-in notion of a matching scale, and the \GenEvA\ algorithm is an example of such a generator. In the companion paper~\cite{genevaphysics}, we implement the following partonic results for $e^+ e^- \to n \text{ jets}$:
\begin{enumerate}
\item \textbf{LO/LL}: Tree-level (LO) matrix elements with up to 6 final state partons, improved by leading logarithms (LL),
\item \textbf{NLO/LL}: Next-to-leading order (NLO) matrix elements including the virtual 2-parton diagram, improved by leading logarithms,
\item \textbf{NLO/LO/LL}: A combination of the above two results, where the NLO calculations are supplemented by higher-multiplicity tree-level calculation, all with leading-logarithmic resummation.
\end{enumerate}
While both LO/LL~\cite{Catani:2001cc, Lonnblad:2001iq, Krauss:2002up, MLM, Mrenna:2003if, Schalicke:2005nv, Lavesson:2005xu, Hoche:2006ph, Alwall:2007fs, Giele:2007di, Lavesson:2007uu, Nagy:2007ty, Nagy:2008ns} and NLO/LL~\cite{Collins:2000gd,Collins:2000qd,Potter:2001ej, Dobbs:2001dq, Frixione:2002ik, Frixione:2003ei, Nason:2004rx, Nason:2006hfa, LatundeDada:2006gx, Frixione:2007vw, Kramer:2003jk, Soper:2003ya, Nagy:2005aa, Kramer:2005hw} results have been implemented elsewhere, the third result has never been implemented before. Furthermore, these distributions are obtained without the need for negative-weight events.

From  a technical point of view, reweighting a parton shower has become possible with the introduction of the analytic parton shower in Ref.~\cite{Bauer:2007ad}. There, the probabilities to obtain individual branches are independent, so that the total probability of generating a given shower history is just the product of multiple splitting probabilities. This allows one to analytically calculate the exact probability for generating a given event. Because the parton shower covers Lorentz-invariant phase space multiple times, we also introduce a method to eliminate this double-counting problem while making maximal use of the full event sample given by the parton shower.

Some of the techniques in this paper are readily applicable to existing Monte Carlo programs, and could be implemented via a simple change in bookkeeping instead of a complete code rewrite. Indeed, the idea of reweighting parton showers to matrix elements appeared in the literature over 20 years ago \cite{Seymour:1994df,Seymour:1994we,Bengtsson:1986hr,Bengtsson:1986et,Miu:1998ut,Miu:1998ju,Norrbin:2000uu,Lonnblad:2001iq} and is implemented in Monte Carlo programs like \texttt{Pythia} \cite{Sjostrand:2000wi, Sjostrand:2006za}. The \GenEvA\ algorithm offers a generalization of this well-known method. As we will see, some algorithmic and conceptual ideas are taken directly from \alpgen\ \cite{Mangano:2002ea} and \madevent~\cite{Maltoni:2002qb}, so the code optimizations present in more mature Monte Carlo programs could aid in further increasing the efficiency of the \GenEvA\ algorithm.

As  in the companion paper \cite{genevaphysics}, we will focus on the process $e^+ e^- \to n \text{ jets}$, though in the conclusions in \sec{sec:conclusions} we discuss obvious generalizations to hadronic collisions with heavy resonances. In \sec{sec:overview}, we introduce the main ingredients of the \GenEvA\ algorithm: the parton shower probability, the Jacobian transformation to Lorentz-invariant phase space, the overcounting factor, and the desired matrix element. Secs.~\ref{sec:ps}, \ref{sec:jac}, \ref{sec:oc}, and \ref{sec:me} discuss these ingredients in turn, and \sec{sec:me} also contains a discussion how the structure of the \GenEvA\ algorithm allows for an automatic leading-logarithmic improvement of tree-level matrix elements. In \sec{sec:results} we verify that the \GenEvA\ algorithm works as a proper phase space generator, showing as an example that it can reproduce the results of traditional tree-level event generators, and discuss the efficiency of the algorithm. The reader not interested in the technical details may skip Secs.~\ref{sec:ps} through \ref{sec:me} and directly go from \sec{sec:overview} to \sec{sec:results}.

Before going into the details of the \GenEvA\ algorithm, we first want to discuss why the parton shower is an appropriate and beneficial starting point for event generation.

\subsection{Phase Space from a Parton Shower}

What does it mean to generate phase space from a parton shower? Ignoring the fact that the parton shower is derived from the soft-collinear limit of QCD, there are three main distinguishing features. First, it is well known that $n$-body phase space can be built recursively from $(n-1)$-body phase space, and a parton shower is an explicit implementation of this recursion. Second, a single parton shower run covers all of multiplicity, flavor, and phase space (hereafter referred to just as ``phase space'') in one shot, which is an example of ``single-channel'' integration. Third, the parton shower preserves probability, meaning that during an event run, the fraction (or ``grid'') of events of a given multiplicity, flavor, and kinematic structure will remain constant. Thus, as a phase space generator, the parton shower is a recursive, single-channel, fixed-grid algorithm.

The \GenEvA\ algorithm is then diametrically opposed to modern phase space integrators like \madevent~\cite{Maltoni:2002qb}, which use non-recursive, multi-channel, adaptive-grid algorithms, based on \texttt{Vegas}~\cite{Lepage:1977sw} and the \texttt{Bases/Spring} package~\cite{Kawabata:1995th}. While event generators like \texttt{CalcHEP}/\texttt{CompHEP}~\cite{Pukhov:1999gg,Boos:2004kh} do use recursive methods (though not a parton shower) to generate phase space, algorithms such as \texttt{RAMBO}~\cite{Kleiss:1985gy} and \texttt{SARGE}~\cite{Draggiotis:2000gm} were developed in order to \emph{avoid} using a recursive definition of phase space because of problems of efficiency. Almost every phase space generator uses some notion of adaptation, in order to avoid the problem of large over- or under-sampling of phase space regions. And advanced phase space algorithms use multi-channel integration to create a single event sample culled from separate runs.

To see why \GenEvA\ is a step forward rather than a step backward, it is important to note one crucial commonality between \texttt{MadEvent} and \GenEvA: Both know about the singularity structure of QCD from $1/Q^2$ poles. In practice, we will see that for $e^+ e^- \to n \text{ jets}$, \GenEvA\ performs at or above \texttt{MadEvent} efficiencies. This suggests that the use of advanced non-recursive, multi-channel, adaptive-grid algorithms by other phase space generators is compensating for something that \GenEvA\ has already built-in, namely, an excellent approximation to QCD-like events through the parton shower. In other words, despite the fact that the recursive, single-channel, fixed-grid machinery of \GenEvA\ is so primitive, the parton shower is sufficiently close to the right answer that more advanced methods are simply not necessary. Whether this fact will continue to be true in hadronic collisions is an open question.

While the machinery of \GenEvA\ is simple-minded, the technical details needed to turn a parton shower into a phase space generator are not. The fact that the parton shower preserves probability means that one needs a thorough understanding of exactly what that probability distribution is, and the bulk of this paper is devoted to explaining in detail how the analytic parton shower in Ref.~\cite{Bauer:2007ad} can become an efficient generator of Lorentz-invariant phase space.

For example, a crucial technical detail is that since the parton shower generates the entire phase space out of successive $1\to 2$ splittings, it will usually distribute events with more partons in the final state than we have full QCD calculations available for. The \GenEvA\ algorithm allows such events to be ``truncated" to lower-multiplicity final states in a way that still allows for an analytic calculation of the resulting truncation probability. This is a concrete realization of the ``phase space projection'' requirement of the \GenEvA\ framework \cite{genevaphysics}, though the form implemented here is a ``reversible projection'' that has additional analytic properties beyond what is strictly speaking necessary in that context.

\subsection{Benefits of GenEvA}
\label{subsec:benefits}

There are numerous benefits to using the parton shower as a phase space generator, and many of these benefits actually derive from the fact that \GenEvA\ is a recursive, single-channel, fixed-grid algorithm.

First, the parton shower is based on the symmetry structures of QCD. All possible flavor and color structures are generated with approximately the right probabilities, meaning that processes that are naturally linked, such as $e^+ e^- \to q\bar{q} + n \text{ gluons}$ and $e^+ e^- \to q\bar{q}q'\bar{q}' + m \text{ gluons}$, need not have separate phase space integrations. The single-channel nature of the \GenEvA\ algorithm makes the program trivial to parallelize, because a single \GenEvA\ run contains a complete sampling of all desired physics processes.

Second, the parton shower is based on the singularity structures of QCD. \GenEvA\ reproduces both the singular pole and leading-logarithmic behavior of full QCD distributions, yielding events with moderate weights. More importantly, the program distributes events with appropriate Sudakov factors~\cite{Sudakov:1954sw} already, allowing for an automatic leading-log improvement of any tree-level matrix element. Of course, these generated Sudakov factors can also be calculated analytically and removed or exchanged if desired.

Third, because the phase space generation is recursive, a single event can have a nested set of consistent $n$-body phase space variables. This allows considerable flexibility in the types of parton-level calculations that can be handled, and allows a simple implementation of the NLO/LO/LL merged sample from the companion paper \cite{genevaphysics}.

Fourth, a single event can have numerous different weights corresponding to different theoretical assumptions and tunable parameters. Using the feature of truncation, \GenEvA\ allows the same event to be interpreted as coming from an $n$-body or $m$-body matrix element with $n$ not necessarily the same as $m$.\footnote{In a traditional phase space generator, it is possible to reweight events distributed according to one $n$-body matrix element to a different $n$-body matrix element, but the value of $n$ must be fixed.} This makes it possible to have a single set of events that is capable of describing multiple theoretical distributions, greatly simplifying the process of parton-level systematic error estimation.

Moreover, by running this single event set through the required detector simulation, one obtains the equivalent of multiple detector-improved distributions, each with roughly the statistical significance of the full number of events. Given the large computational cost of full detector simulation, it is highly desirable to reuse Monte Carlo in this way.\footnote{Because detector simulation is a linear operation, it is even possible to try different theoretical distributions \emph{after} detector simulation.}
Because the \GenEvA\ algorithm includes the proper singularities and Sudakov factors out-of-the-box, this reweighting strategy will potentially be feasible even on large Monte Carlo sets.

Finally, \GenEvA\ is well-suited to situations where it is difficult or tedious to distribute events according to a certain kind of theoretical calculation. For example, the application of soft-collinear effective theory (SCET)~\cite{Bauer:2000ew,Bauer:2000yr,Bauer:2001ct,Bauer:2001yt} to event generation~\cite{Bauer:2006qp,Bauer:2006mk,Schwartz:2007ib} is most useful once the final state kinematics and matching scales $\mu$ are already known. If we are interested in calculating subleading logarithms using SCET, it is hard to analytically determine the proper SCET amplitudes over all of phase space. However, as a template for QCD-like events, \GenEvA\ determines the kinematics and scales without reference to the amplitudes, allowing in principle numeric methods to be used to calculate SCET operator mixing.

Because of the above benefits, it is tempting to identify the \GenEvA\ framework introduced in the companion paper \cite{genevaphysics} with the \GenEvA\ algorithm that will be presented here. This is particularly true because we will see that in the \GenEvA\ algorithm, it is actually easier and more efficient to include leading-logarithmic resummation in matrix elements than not. However, we emphasize that the \GenEvA\ algorithm can be used as phase-pace generator independently of any particular strategy for event generation, and similarly the \GenEvA\ framework could work with an alternative phase space algorithm.

\section{Overview of the GenEvA Algorithm}
\label{sec:overview}

The idea of the \GenEvA\ algorithm is straightforward: Generate events using a parton shower and reweight these events to any desired distribution in a way that avoids double-counting. In this section, we develop the master formula for the event weight and then summarize the various ingredients used in this master formula. The details of the algorithm are then given in Secs.~\ref{sec:ps}, \ref{sec:jac}, \ref{sec:oc}, and \ref{sec:me}. Readers not interested in the technical details should get a basic understanding of the algorithmic issues involved from this section, and can then safely skip to \sec{sec:results} for a comparison to other event generators and the efficiency of the algorithm.

\subsection{The Master Formula}

Let $\Phi$ denote a point in Lorentz-invariant phase space generalized to include all additional particle information, like flavor, helicity, and color, that fully characterize an event. We want to distribute events according to a general distribution $\sigma(\Phi)$. That is, we want the contribution of an event with value $\Phi$ to be $\df\sigma = \sigma(\Phi) \,\df\Phi$. Even though the value of $\sigma(\Phi)$ is calculable for fixed values of $\Phi$, it is usually very complicated to directly distribute events according to $\sigma(\Phi)$.

\begin{figure}
\includegraphics[scale=0.7]{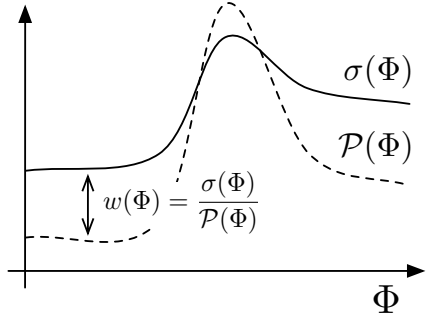}
\caption{The simplest version of reweighting. Events are distributed according to a normalized probability distribution $\cP(\Phi)$ and given weights $w(\Phi) = \sigma(\Phi)/\cP(\Phi)$ such that the resulting weighted events are effectively distributed according to $\sigma(\Phi)$. In \GenEvA, $\Phi$ corresponds to Lorentz-invariant phase space variables and $\sigma(\Phi)$ is a differential cross section.}
\label{fig:reweighta}
\end{figure}

If we have a known normalized probability distribution $\cP(\Phi)$ for which it is easy to distribute according to, we can choose values of $\Phi$ according to $\cP(\Phi)$ and assign the weight
\begin{equation}
w(\Phi) = \frac{\sigma(\Phi)}{\cP(\Phi)}
\end{equation}
to each event with value $\Phi$, as in \fig{fig:reweighta}. Since the probability of choosing a value $\Phi$ is $\df\cP = \cP(\Phi)\,\df\Phi$, the weighted set of events with values $\Phi$ will be distributed according to
\begin{equation}
\df\sigma = w(\Phi) \,\df\cP = \frac{\sigma(\Phi)}{\cP(\Phi)} \cP(\Phi) \,\df\Phi = \sigma(\Phi) \,\df\Phi
\,,\end{equation}
as desired.

\begin{figure}
\includegraphics[scale=0.7]{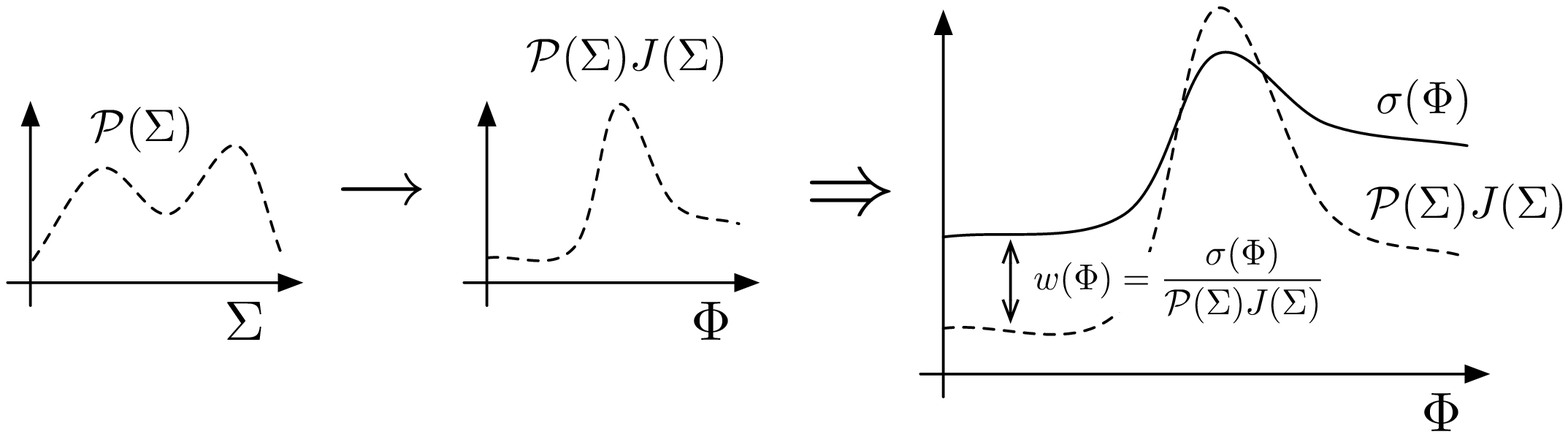}
\caption{Reweighting using a different distribution variable $\vS$, where there is a one-to-one and onto map $\vS \to \Phi(\vS)$. Events are distributed according to a normalized probability distribution $\cP(\vS)$, which together with the Jacobian $J(\vS)$ from the variable transformation $\vS\to \Phi(\vS)$ defines an effective probability distribution $\cP(\Phi) \equiv \cP(\vS) J(\vS)$. The weights $w(\Phi)$ can now be defined analogously to \fig{fig:reweighta}.}
\label{fig:reweightb}
\end{figure}

Now imagine we have an easy way to distribute events in terms of a different variable $\vS$ according to a known normalized distribution $\cP(\vS)$, and there is a mapping $\Phi \equiv \Phi(\vS)$ which is one-to-one and onto, as in \fig{fig:reweightb}. Using $\cP(\vS)$ as sampling function, the probability for choosing a value of $\Phi$ is now
\begin{equation}
\df\cP = \cP(\vS) \,\df\vS = \cP(\vS) J(\vS) \,\df\Phi \equiv \cP(\Phi) \,\df\Phi
\,,\end{equation}
where $\vS \equiv \vS(\Phi)$ is the inverse of $\Phi(\vS)$ and
\begin{equation}
\label{eq:jacobian}
\frac{1}{J(\vS)} = \biggl\lvert\frac{\partial(\Phi)}{\partial(\vS)} \biggr\rvert(\vS)
\end{equation}
is the Jacobian in the transformation from $\vS$ to $\Phi$. Thus, we obtain the event weight
\begin{equation}
w(\Phi) = \frac{\sigma(\Phi)}{\cP[\vS(\Phi)] J[\vS(\Phi)]}
\,.\end{equation}

\begin{figure}
\includegraphics[scale=0.7]{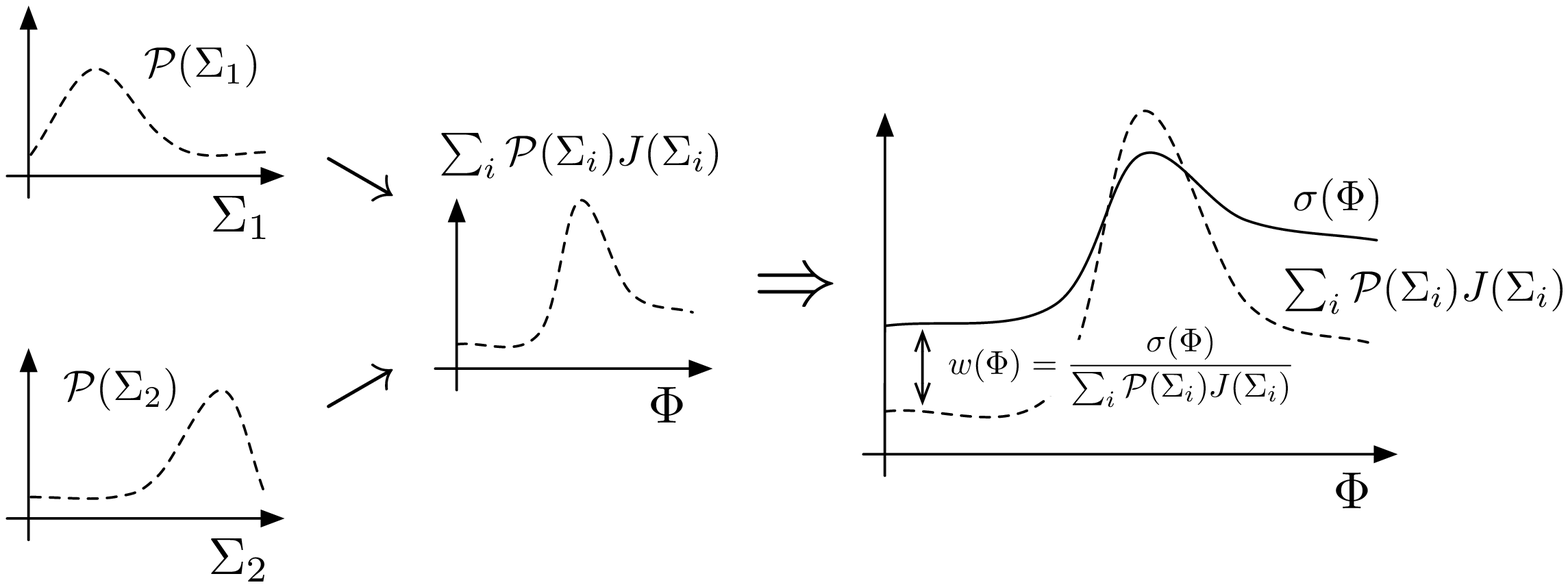}
\caption{Reweighting when there are multiple values $\vS_i$ that map to the same point $\Phi(\vS_i)$. If all $\vS_i$ that map to a given $\Phi$ are known, then there is again an effective probability distribution $\cP(\Phi) \equiv \sum_i \cP(\vS_i) J(\vS_i)$ that can be used to define an appropriate weight $w(\Phi)$. In \GenEvA, the parton shower provides a $\cP(\vS)$, and it is well-known that multiple parton shower histories $\vS_i$ can correspond to the same phase space point $\Phi(\vS_i)$. For this reweighting strategy to be computationally feasible, there must be an efficient way to calculate the sum over all parton shower histories $\sum_i \cP(\vS_i) J(\vS_i)$.}
\label{fig:reweightc}
\end{figure}

It is important that the mapping $\Phi(\vS)$ is onto, otherwise there would be values of $\Phi$ which are never covered by $\cP(\vS)$. On the other hand, $\Phi(\vS)$ need not be one-to-one. Instead, for a given value of $\Phi$ there could be a \emph{discrete} number $n(\Phi)$ of $\vS$ values which all map to that same $\Phi$ as in \fig{fig:reweightc}. In this case, the probability of choosing a value $\Phi$ is
\begin{equation}
\df\cP = \sum_{i = 1}^{n(\Phi)} \cP(\vS_i) J(\vS_i)\, \df\Phi
\,,\end{equation}
where here and in the following $\vS_i \equiv \vS_i(\Phi)$ denotes the $i$-th point $\vS$ for which $\Phi(\vS_i) = \Phi$ and $i$ runs from $1$ to $n(\Phi)$. The weight as a function of $\Phi$ is thus
\begin{equation}
\label{eq:optimalweight}
w(\Phi) = \frac{\sigma(\Phi)}{\sum_i \cP[\vS_i(\Phi)] J[\vS_i(\Phi)]}
\,.\end{equation}

\begin{figure}
\includegraphics[scale=0.7]{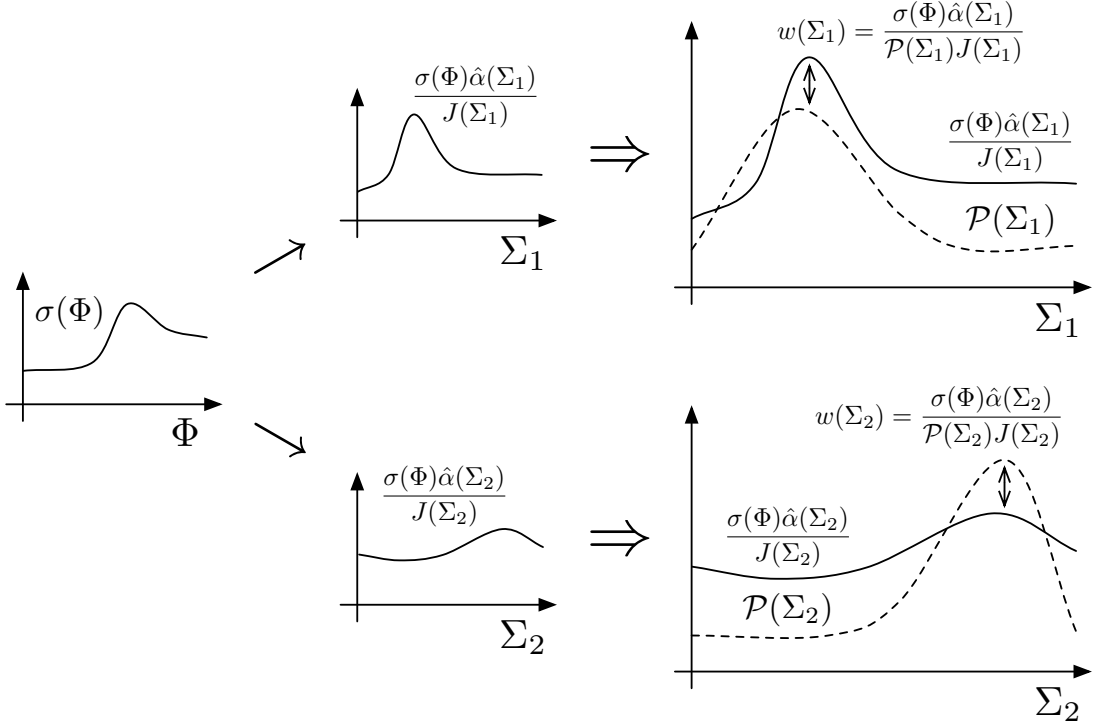}
\caption{The reweighting approach used by \GenEvA\ when the map $\Phi(\vS)$ is not one-to-one. Instead of taking the event weight to depend on $\Phi$ only, the event weight $w(\vS)$ is effectively a function of $\vS$. Using the overcounting factor $\hat{\alpha}(\vS_i)$, the desired distribution $\sigma(\Phi)$ is split up among the various $\vS_i$ that map to the same $\Phi$, including the Jacobian factor $J(\vS_i)$. As long as $\sum_i \hat{\alpha}(\vS_i) = 1$, the resulting events will be distributed according to $\sigma(\Phi)$ without the need to calculate $\cP(\vS_i) J(\vS_i)$ for every $\vS_i$. In practice, there is a small computational cost to make sure that $\hat{\alpha}(\vS_i)$ is properly normalized.}
\label{fig:reweightd}
\end{figure}

In  \GenEvA, the parton shower provides the sampling function $\cP(\vS)$, where $\vS$ describes a complete parton shower history, and $\Phi(\vS)$ is given by the shower's final state. It is well-known that multiple parton shower histories $\vS_i$ can produce the same phase space point $\Phi(\vS_i)$. In practice, it may be computationally expensive to calculate the sum in the denominator of \eq{eq:optimalweight}, as it requires reconstructing all possible parton shower histories $\vS_i$ for a given point $\Phi$. However, \eq{eq:optimalweight} is dictated by our choice of characterizing events by $\Phi$, which forces us to treat the weight of an event as a function of $\Phi$ as well. In particular, $w(\Phi)$ in \eq{eq:optimalweight} cannot depend on the actual $\vS_i$ which was used to generate the event.

To relax this condition we can simply characterize events by $\vS$ and let $w \equiv w(\vS)$ be a function of $\vS$ as in \fig{fig:reweightd}. In this case, the total contribution to $\df\sigma$ for a given $\Phi$ is
\begin{equation}
\label{eq:dsigmaw}
\df\sigma
= \sum_i w(\vS_i) \,\df\cP_i
= \sum_i \frac{\sigma(\vS_i)}{\cP(\vS_i)} \cP(\vS_i) \,\df\vS
= \sum_i \sigma(\vS_i) \,\df\vS
\,.\end{equation}
At this point, we need to define what we mean by $\sigma(\vS)$. From \eq{eq:dsigmaw} we have to require
\begin{equation}
\label{eq:dsigmaw2}
\df\sigma
= \sum_i \sigma(\vS_i) J(\vS_i) \,\df\Phi
= \sigma(\Phi) \,\df\Phi
\,.\end{equation}
Note that simply taking $\sigma(\vS) = \sigma[\Phi(\vS)]/J(\vS)$ does not work, because it would lead to an overcounting by a factor of $n(\Phi)$ in \eqs{eq:dsigmaw}{eq:dsigmaw2}. Instead, we need
\begin{equation}
\label{eq:sigmavS}
\sigma(\vS)
= \frac{\sigma[\Phi(\vS)]\, \ha(\vS)}{J(\vS)}
\qquad\text{with}\qquad
\sum_i \ha(\vS_i) = 1
\,,\end{equation}
such that
\begin{equation}
\df\sigma
= \sum_i \sigma(\vS_i) J(\vS_i) \,\df\Phi
= \sigma(\Phi) \sum_i \ha(\vS_i) \,\df\Phi
= \sigma(\Phi) \,\df\Phi
\,,\end{equation}
as required.

Using \eq{eq:sigmavS}, the final weight as a function of $\vS$ is
\begin{equation}
\label{eq:finalweight}
w(\vS)
= \frac{\sigma(\vS)}{\cP(\vS)}
= \frac{\sigma[\Phi(\vS)]\, \ha(\vS)}{\cP(\vS) J(\vS)}
\,.\end{equation}
\eq{eq:finalweight} is the master formula underlying the \GenEvA\ algorithm.
 Alternatively, if we wish to specify $\sigma(\vS)$ directly, we use the notation $\sigma_i(\Phi)$, where $i$ labels the different shower histories mapping to $\Phi$, and the event weight is
\begin{equation}
\label{eq:finalweight_i}
w(\vS) = \frac{\sigma_i[\Phi(\vS)]}{\cP(\vS) J(\vS)}
\,.\end{equation}
It is clear that all the information in $\sigma_i(\Phi)$ could be absorbed into an ordinary $\sigma(\Phi)$ times an overcounting factor $\hat \alpha(\vS_i)$. However, note that the Jacobian $J(\vS)$ still appears in \eq{eq:finalweight_i}, meaning that $\sigma_i(\Phi)$ is still a differential function of $\Phi$ and not $\vS$.

\subsection{Calculating the Event Weight}

The master formula in \eq{eq:finalweight} contains several ingredients. Here, we will give a brief overview of each of the components, while the details are discussed in the remainder of the paper.

\subsubsection{The Parton Shower Probability $\cP(\vS)$}
\label{subsec:showerprob}

In the \GenEvA\ algorithm, the parton shower plays the role of the sampling function $\cP(\vS)$ for reweighting to any desired matrix element $\sigma(\Phi)$. The crucial point is that the parton shower automatically contains the right singularity structure of QCD matrix elements, which should in principle allow a reweighting with reasonable efficiency. However, for a parton shower to be used for the purposes of reweighting, it has to satisfy several additional requirements.

First, we have to know the parton shower's probability distribution $\cP(\vS)$, which means we must be able to compute the exact probability with which the parton shower generates a given shower history $\vS$. The analytic parton shower algorithm constructed in Ref.~\cite{Bauer:2007ad} has this property and is the algorithm used here and detailed in \sec{sec:ps}. The key feature of this shower is that energy-momentum is conserved by construction without any \emph{ad hoc} momentum shuffling.

Second,  the parton shower has to cover all of phase space. Of course, almost any parton shower already covers all possible final state configurations, in that it will generate 2-body final states, 3-body final states, and so on, with the flavors of the final state particles representing all possible allowed configurations. What is important is that the shower also has no kinematic dead zones, to ensure that the map $\Phi(\vS)$ is indeed onto and all of phase space is covered as necessary.

Finally, as already mentioned, the fact that the parton shower generates many final state particles seems at first glance like a liability, because matrix elements for $n$-body configurations with arbitrary $n$ are not readily available. In order to reweight the generated events, we need to be able to ``backtrack'' (not necessarily uniquely) a high-multiplicity final state to a lower-multiplicity intermediate state, while still being able calculate analytically the probability with which this intermediate state was generated. We call this feature ``truncation'', and it is possible because of the analytic properties of the algorithm of Ref.~\cite{Bauer:2007ad} in conjunction with the parton shower's built-in notion of probability preserving ordering. This truncation is a concrete implementation of the phase space projection required by a phase space generator suitable for the \geneva\ framework as explained in more detail in Ref.~\cite{genevaphysics}.

\subsubsection{The Jacobian $J(\vS)$}

An event produced by \geneva's parton shower is characterized by a parton shower history $\vS$ that consists of the set of $1 \to 2$ splittings that build up the event. In contrast, Lorentz-invariant phase space $\Phi$ characterizes an event by the set of three-momenta (symmetrized appropriately with respect to identical particles) of the on-shell final state particles. As needed for reweighting, there is a map from parton shower histories $\vS$ to Lorentz-invariant phase space $\Phi$, such that for every parton shower history $\vS$ there is a unique point $\Phi(\vS)$. A nontrivial Jacobian factor $J(\vS)$ is required in changing variables from the natural variables used to describe the $1\to 2$ splittings in the shower history to those that describe Lorentz-invariant phase space. This Jacobian will be given in \sec{sec:jac}.

\subsubsection{The Overcounting Factor $\hat \alpha(\vS)$}

It is well known that the parton shower does not map out phase space in a one-to-one way, because in general, multiple parton shower histories can lead to the same final state four-momenta. The overcounting factor $\ha(\vS)$ takes care of this overcounting, and essentially determines how $\sigma(\Phi)$ is split up among the different $\vS_i(\Phi)$ that map to the same point $\Phi$. Since we are eventually only interested in $\sigma(\Phi)$, we can choose $\ha(\vS)$ freely only subject to the constraint in \eq{eq:sigmavS}. The simplest possible choice would be $\ha(\vS) = 1/n(\Phi)$, where $n(\Phi)$ counts the number of $\vS_i$ mapping to the same $\Phi$. This choice will in general lead to a poor statistical efficiency, since the weights $w(\vS)$ will vary widely for the different $\vS_i$ contributing to the same point $\Phi$. To ensure well-behaved statistics, we need to ensure that for a given phase space point $\Phi$, all event weights are roughly the same, thus requiring that $\ha(\vS)$ scales like $\cP(\vS) J(\vS)$. We will discuss the choice of $\ha(\vS)$ in the context of \GenEvA\ in detail in \sec{sec:oc}.

\subsubsection{The Matrix Element $\sigma[\Phi(\vS)]$}

 \GenEvA\ can distribute events according to nearly any distribution $\sigma(\Phi)$, and this quantity will depend on the underlying physics one is trying to describe. Crucially, physics considerations alone can determine the best choice for $\sigma(\Phi)$, and the algorithmic details of \GenEvA\ need not influence this choice.

While $\sigma(\Phi)$ in \eq{eq:finalweight} only depends on a set of on-shell four-momenta, in many cases it can be advantageous to use \eq{eq:finalweight_i} and directly work in terms of $\sigma_i(\Phi)$. That is, we may want to assign different weights to events having different parton shower histories $\vS_i$ even though they contribute to the same point $\Phi(\Sigma_i)$. Two reasons to use $\sigma_i(\Phi)$ are to improve the efficiency of \GenEvA\ and to resolve logarithmic ambiguities~\cite{genevaphysics}. In \sec{sec:me}, we will discuss tree-level fixed-order QCD matrix elements using $\sigma(\Phi)$,
as well as calculations which properly sum the leading-logarithmic dependence, using $\sigma_i(\Phi)$.

\subsection{Reweighting Efficiency}
\label{subsec:eff}

To assess the practical usefulness of the \GenEvA\ algorithm, it is important to define what we mean by efficiency. In this paper, we will use statistical efficiency as the figure of merit in contrast to unweighting efficiency. Since unweighting efficiency depends on the value of the maximum weight generated, it is sensitive to how many events were generated in a given run. Here we argue that statistical efficiency is a much better measure of the strength of an algorithm.

If $N$ is the total number of events, then in the limit of large $N$ the weights satisfy%
\footnote{This definition of weights differs slightly from the one commonly used in Monte Carlo programs. Here, the weight of an event is independent of the number of events generated, whereas most Monte Carlo programs divide individual event weights by $N$, the total number of events generated in a run. We will use the run-independent definition of weights throughout this paper.}
\begin{equation}
\label{eq:largeN}
\sum_n w(\Phi) \to N \int \! \df\Phi \, \sigma(\Phi)
\,.\end{equation}
To see this, consider the normalized distribution $\hat\sigma(\Phi) = \sigma(\Phi) / \int \! \df\Phi \, \sigma(\Phi)$. Since both $\hat\sigma(\Phi)$ and $\cP(\Phi)$ are normalized to unity, the normalized weights $\hat{w}(\Phi) = \hat\sigma(\Phi)/\cP(\Phi)$ satisfy $\sum_n \hat{w}(\Phi) \to N$, from which \eq{eq:largeN} follows.

Dividing \eq{eq:largeN} by $N$ we have
\begin{equation}
\frac{1}{N} \sum_n w(\Phi) \equiv \langle w(\Phi) \rangle \to \int \! \df\Phi \, \sigma(\Phi)
\,,\end{equation}
where $\langle \cdot \rangle$ denotes the sample average. From the standard error on averages, the statistical uncertainty on the Monte Carlo estimate of $\int \! \df\Phi \, \sigma(\Phi)$ is thus given by
\begin{equation}
\frac{\sqrt{ \langle w(\Phi)^2 \rangle - \langle w(\Phi) \rangle^2}}{\sqrt{N}}
\,.\end{equation}
The numerator describes the width of the distribution of weights, which vanishes in the limit of all weights being equal. This is equivalent to the fact that for unweighted events, the uncertainty on the total number of events is always zero. Thus, the ratio
\begin{equation}
\label{eq:neff}
\eta_\lab{eff}
= \frac{\langle w(\Phi) \rangle^2}{\langle w(\Phi)^2 \rangle}
= \frac{1}{N} \,\frac{\bigl[\sum_n w(\Phi) \bigr]^2 }{\sum_n w(\Phi)^2}
\equiv \frac{N_\lab{eff}}{N} \leq 1
\,,\end{equation}
provides a statistical measure of the efficiency of the reweighting. We can interpret $N_\lab{eff}$ as the effective number of events after reweighting in the sense that the sample of reweighted events is statistically equivalent to an unweighted sample of $N_\lab{eff}$ events. In the limit of all weights being equal $\eta_\lab{eff}\to 1$ and $N_\lab{eff} \to N$. Hence, this way of distributing events by reweighting $\cP(\Phi)$ to $\sigma(\Phi)$ is efficient as long as $\cP(\Phi)$ is reasonably close to $\hat\sigma(\Phi)$. For example, if we wish to distribute events according to a function $\sigma(\Phi)$ which has large peaks, we have to choose a sampling function $\cP(\Phi)$ that has a similar peak structure.

The  definition of $N_\lab{eff}$ differs from that of $N_\lab{unw}$, which is the number of events that would remain after unweighting the sample,
\begin{equation}
N_\lab{eff} = \frac{\bigl[\sum_n w(\Phi) \bigr]^2 }{\sum_n w(\Phi)^2}
\qquad\text{vs.}\qquad
N_\lab{unw} = \frac{\sum_n w(\Phi)}{\max[w(\Phi)]}
\,.\end{equation}
The unweighting efficiency $\eta_\lab{unw} = N_\lab{unw}/N$ is highly sensitive to the tail of the weight distribution, such that even an exponentially small number of high-weight events can cause a dramatic decrease in $\eta_\lab{unw}$. The only way to deal with this is to throw away events with too large weights, which of course introduces a systematic error which is very hard to quantify.
In contrast, the statistical efficiency $N_\lab{eff}/N$ correctly reflects the full statistical power of the event sample, and being based on sample averages, is unaffected by individual weights.

The main disadvantage of considering the statistical efficiency is that since the statistical uncertainties are determined by $N_\lab{eff}$ and not by $N$, it is necessary to store and process a larger total number of events than in a statistically equivalent unweighted sample. However, if this becomes a serious issue one has always the option to partially unweight the sample by unweighting just those events that have small weights. That is, we can define a threshold weight $w_0$, and only unweight those events with weights smaller than $w_0$. The expected statistical efficiency of the partially unweighted sample is given by
\begin{equation}
\label{eq:etaeff_w0}
\eta_\lab{eff}(w_0)
= \frac{1}{N}\,\frac{\bigr[\sum_n w(\Phi)\bigr]^2}{\sum_{w > w_0} w(\Phi)^2 + w_0 \sum_{w < w_0} w(\Phi)}
\,.\end{equation}
Note that $\eta_\lab{eff}(0) = \eta_\lab{eff}$ and $\eta_\lab{eff}(w_\max) = \eta_{\rm unw}$. In \sec{subsec:efficiency}, we will report the efficiency of \GenEvA\ not only in terms of $\eta_\lab{eff}$, but also in terms of the time needed to achieve a partially unweighted sample with $\eta_\lab{eff}(w_0) = 0.9$.

\section{The Analytic Parton Shower}
\label{sec:ps}

As  discussed in \sec{subsec:showerprob}, a parton shower to be used as phase space generator has to fulfill three requirements: 1) It has to have an analytically calculable probability $\cP(\vS)$; 2) it has to cover all of phase space; and 3) we have to be able to analytically truncate the shower, which means we must be able to truncate any high-multiplicity final state produced by the shower to an $n$-body final state for any given $n$, while retaining analytic control over the truncation probability. In the following we will describe in detail how these requirements are met by \geneva's parton shower.

\subsection{Overview}

The  parton shower used by the \geneva\ algorithm to generate phase space is based on the analytic parton shower algorithm constructed in Ref.~\cite{Bauer:2007ad}. As we will only be discussing $e^+ e^- \to n \text{ jets}$, we will only discuss the final state parton shower. We comment on the extension to hadronic collisions and initial state showers in \sec{subsec:towardsLHC}.

Since  the main purpose of \GenEvA's parton shower is to generate $n$-body phase space efficiently, the current implementation is only accurate to leading-logarithmic accuracy, and neglects several important physics effects a realistic parton shower should take into account, such as the running of $\alpha_s$ and color coherence~\cite{Bengtsson:1986hr,Bengtsson:1986et,Mueller:1981ex}. In addition, the shower we consider here treats all final state particles as massless. While it is straightforward to construct an algorithm that does incorporate all of these physics issues and still maintains analytic control (see the discussion in Ref.~\cite{Bauer:2007ad}), for the purposes of this paper the simple analytic parton shower discussed here will suffice.

\begin{figure}
\includegraphics[scale=0.7]{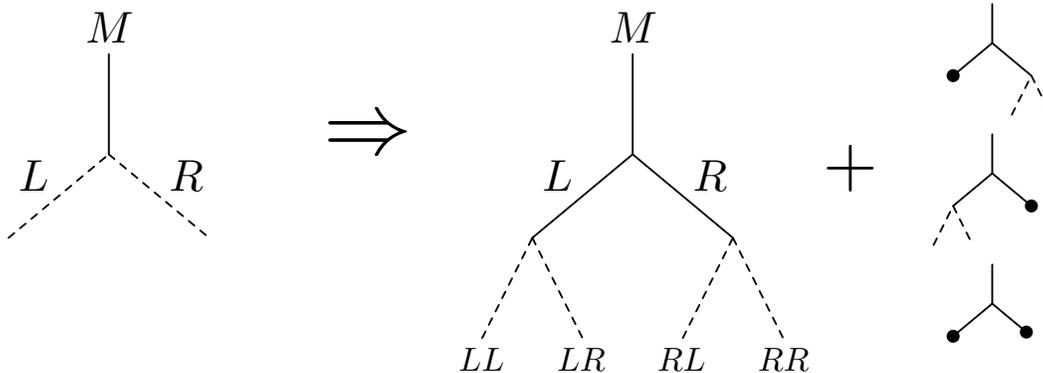}
\caption{The structure of the parton shower. A solid line indicates an already processed particle, while a dashed line indicates a particle that has not yet been processed. A dot at the end of a solid line indicates a particle that did not branch above the shower cutoff, \ie\ a final state particle. Starting from an existing branch $M \to LR$, where a mother particle $M$ has branched into daughter particles $L$ and $R$, \GenEvA's parton shower generates the double branch $M \to LR \to (LL/LR)(RL/RR)$, where it is understood that there is also a finite probability for $L$ or $R$ to remain unbranched. Understanding how the $L$ and $R$ branches are coupled is essential for being able to exactly compute the double-branch probability and to truncate the parton shower.}
\label{fig:showerlabel}
\end{figure}

An  event produced by \geneva's parton shower is characterized by the kinematics of the initial hard-scattering process (typically a $2\to2$ interaction), followed by a set of $1 \to 2$ particle splittings. We mostly follow the notation of Ref.~\cite{Bauer:2007ad}. We call a given $1 \to 2$ splitting a branch, with the original particle called the mother particle, and the two final particles called the daughter particles. We label the mother particle by $M$, and the daughter particles by $L$ and $R$. The daughters of these daughters (the granddaughters) are labeled by $LL$, $LR$, $RL$, and $RR$.

Starting  from an existing branch $M \to LR$, the shower generates a double branch $M \to LR \to (LL/LR)(RL/RR)$. This is illustrated in \fig{fig:showerlabel}, where here and in the following figures a solid line indicates an already processed particle, while a dashed line indicates a particle that has not yet been processed. A dot at the end of a solid line indicates a particle that did not branch above the shower cutoff, \ie\ a final state particle. The double-branch probability, \ie\ the probability to obtain a specific $1\to 2\to 4$ double branch from a given $1\to 2$ single branch, is the fundamental object in \GenEvA's parton shower, and allows it to have a well-defined notion of global evolution.

\subsection{Choice of Kinematic Variables}
\label{subsec:ChoiceofKinematicVariables}

\begin{figure}
\includegraphics[scale=0.7]{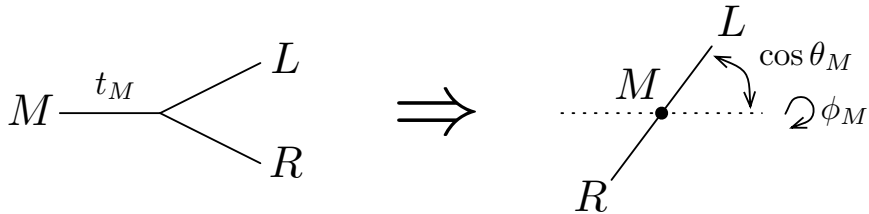}
\caption{The kinematic variables describing a single splitting in the parton shower. The splitting functions can be chosen to be functions of the mother's virtuality $t_M$ and the angles $\cos \theta_M$ and $\phi_M$, where $\cos \theta_M$ is the angle between the $L$ daughter and the boost axis to the mother's rest frame, and $\phi_M$ is the azimuth around the boost axis. Note that the $L$ and $R$ four-momenta cannot be determined until $t_L$ and $t_R$ are known, that is, only after the daughters themselves have been processed.}
\label{fig:showerkinematics}
\end{figure}

In general, the splitting functions describing the $1 \to 2$ parton splittings depend on three kinematic variables, whose precise definitions depend on the details of the parton shower. One of these variables is the evolution variable of the parton shower, which is always decreasing as the shower progresses, while the other two describe the remaining kinematics of the splitting. There are several choices possible for the evolution variables, such as virtuality~\cite{Bengtsson:1986et,Norrbin:2000uu}, the energy-weighted emission angle~\cite{Marchesini:1983bm,Corcella:2000bw}, transverse momentum~\cite{Gustafson:1986db,Gustafson:1987rq,Lonnblad:1992tz}, or more general variables~\cite{Giele:2007di}.

In  \geneva\, we use the analytic parton shower algorithm defined in Ref.~\cite{Bauer:2007ad}, in which the evolution variable is the virtuality $t_M$ of the mother, and the remaining two variables are chosen to be the angle $\cos \theta_M$ in the mother's rest frame between the $L$ daughter and the boost axis, and the azimuthal angle $\phi_M$ around this boost axis, as shown in \fig{fig:showerkinematics}. Thus, we can write the differential kinematics of the parton shower in terms of
\begin{equation}
\label{eq:showerspace}
\df\vS_2(M \to LR) = \df t_M \, \df\!\cos\theta_M \, \df \phi_M
\,,\end{equation}
with the phase space limits
\begin{equation}
\label{eq:showerspace_limits}
-1 \leq \cos\theta_M \leq 1
\,, \qquad
0 \leq \phi_M \leq 2 \pi
\,, \qquad
0 \leq \sqrt{t_L} + \sqrt{t_R} \leq \sqrt{t_M}
\,.\end{equation}
The primary reason to use the angles as the basic variables is that their phase space limits are trivial and completely independent of any other kinematic quantity, which also makes it trivial to satisfy energy-momentum conservation for each branch.

A complete shower history with $n$ final state partons is described by
\begin{equation}
\df \vS_n = \df\Phi^{(0)}_2 \prod_{i=1}^{n-2} \df\vS_2(M_i\to L_i R_i)
\,,\end{equation}
where $\df\Phi^{(0)}_2$ is the two-body phase space of the initial hard scattering and $\df\vS_2(M_i\to L_i R_i)$ are the subsequent branches from the shower. We will discuss the precise phase space measure and its relation to ordinary $n$-body phase space in more detail in \sec{sec:jac}.

Instead of the angle $\cos \theta_M$, many parton showers use the energy splitting $z_M = E_L/E_M$, describing how the energy of the mother is shared between the two daughters. Indeed, the splitting functions themselves are usually written in terms of $z$. The relation between $z$ and $\cos \theta$ is given by
\begin{align}
\label{eq:costheta_z}
\cos\theta_M
&= \frac{1}{\beta_M}\, \frac{2z_M - (1 + t_L/t_M - t_R/t_M)}{\lambda(t_M; t_L, t_R)}
\,,\nn\\
z_M &=
\frac{1}{2}\biggl[1 + \frac{t_L}{t_M} - \frac{t_R}{t_M} + \beta_M \cos\theta_M\,
 \lambda(t_M; t_L, t_R) \biggr]
\,,
\end{align}
where $t_M$ is the virtuality of the mother, $t_{L,R}$ are the virtualities of the daughters, and we have defined
\begin{equation}
\label{eq:betalambda}
\beta_M = \sqrt{1-\frac{t_M}{E_M^2}}
\,,\qquad
\lambda(t_M; t_L, t_R) = \frac{1}{t_M} \sqrt{(t_M - t_L - t_R)^2 - 4t_L t_R}
\,,\end{equation}
with $E_M$ being the energy of the mother.

In \geneva's shower, $\cos\theta$ is the fundamental splitting variable, while $z$ is a derived quantity. However, since the usual splitting functions are functions of $z$, we need to translate $\cos\theta$ into $z$. At the time the mother is branched, both daughters are still massless, so \eq{eq:costheta_z} gives the relation
\begin{equation}
\label{eq:zmother}
z(\cos\theta_M, E_M) = \frac{1}{2} + \frac{1}{2} \beta_M(E_M) \cos\theta_M
\,,\end{equation}
where we made the dependence on $E_M$ explicit. The problem is that the final value for the mother's energy $E_M$ cannot be determined until after the mother itself and its sister have been processed and their final invariant masses are known. This is equivalent to the fact that in ordinary parton showers based on $1\to2$ splittings, the value of $z_M$ has to be ``re-shuffled'' at a later stage in the algorithm in order to satisfy momentum conservation. For a more detailed discussion see Ref.~\cite{Bauer:2007ad}. In contrast,  once a value for $\cos\theta_M$ is determined, it never has to be changed afterwards, which crucially relies on the fact that its phase space limits in \eq{eq:showerspace_limits} are independent from the rest of the shower. Nevertheless, we still need to pick a value of $E_M$ in \eq{eq:zmother}, which we denote as $E_M^\start$. From \eq{eq:zmother} we see that the choice of $E_M^\start$ effectively determines the range of $z$ allowed in the splitting functions. Therefore, to avoid artificially large subleading-logarithmic effects, one should pick an $E_M^\start$ of the order of $E_M$. In our case, the precise choice of $E_M^\start$ is dictated by the requirement to have a simple analytic truncation, and will be discussed in \sec{subsec:truncation}.

We are now ready to obtain the required splitting functions. Taking into account the appropriate Jacobian factors, we have
\begin{equation}
f_{M\to LR}(t_M,\cos\theta_M,\phi_M;\, E_M^\start)
= \frac{\alpha_s}{2\pi}\, \frac{\beta_M}{2}\, \frac{1}{2\pi}\, \frac{1}{t_M}\,
P_{M\to LR}[z(\cos\theta_M, E_M^\start)]
\,,\end{equation}
where $P_{M\to LR}(z)$ are the well-known Altarelli-Parisi~\cite{Altarelli:1977zs} splitting functions
\begin{align}
P_{q \to qg}(z) &= C_F\,\frac{1+z^2}{1-z}
\,,\nn\\
P_{g \to gg}(z) &= C_A\biggl[\frac{1-z}{z} + \frac{z}{1-z} + z(1-z)\biggr]
\,,\nn\\
P_{g \to q\bar{q}}(z) &= T_R\,\bigl[z^2+(1-z)^2\bigr]
\,.\end{align}
In the following, to simplify the notation, we will mostly suppress the dependence of the splitting functions on $E_M^\start$, but one should keep in mind that wherever the splitting functions appear there is also an implicit dependence on $E_M^\start$.

\subsection{The Single-Branch Probability}

\begin{figure}
\includegraphics[scale=0.7]{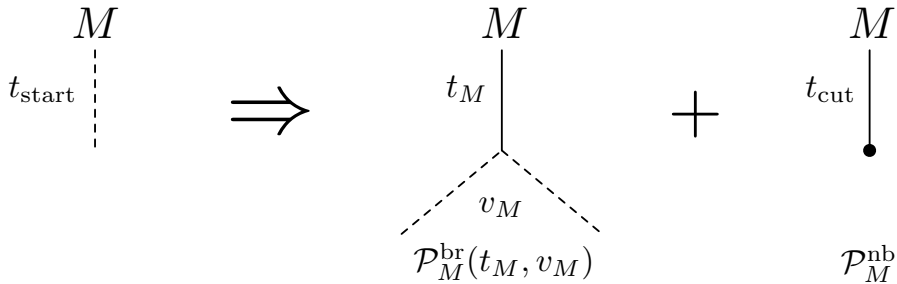}
\caption{The single-branch probability. Starting at the scale $t_\start$, the mother $M$ can either branch at the scale $t_M > t_\cut$ or remain unbranched down to the scale $t_\cut$,
which is denoted by a dot at the end of a line. The branching probability $\cP_M^{\rm br}(t_M, v_M)$ and the no-branching probability $\cP_M^{\rm nb}$ also depend on $t_\start$, $t_\cut$, and $E_M^\start$. The variable $v_M$ is a shorthand for the splitting angles and the daughters' flavors.}
\label{fig:showersingle}
\end{figure}

The basic ingredient in the parton shower is the single-branch probability, which determines if and how a given mother particle $M$ splits into two daughters $L$ and $R$, as illustrated in \fig{fig:showersingle}. The probability for a single branch to occur with given values of the splitting variables is a combination of the splitting function and a Sudakov factor (for a review of parton showers see for example Ref.~\cite{Sjostrand:2006za}):
\begin{align}
\label{eq:singlebranchprob}
\cP^{\rm br}_{M \to LR}(t_M,\cos\theta_M,\phi_M;\,t_\start, t_\cut)
&= f_{M \to LR}(t_M,\cos\theta_M,\phi_M) \, \sud_M(t_\start,t_M)
\nn\\
& \quad \times
\theta(t_\start > t_M)\, \theta(t_M > t_\cut)
\,.\end{align}
The Sudakov factor is defined as
\begin{equation}
\label{eq:sudakov}
\sud_M(t_2,t_1) = \exp\biggl[- \int_{t_1}^{t_2} \!\df t \int_{-1}^1 \!\df \!\cos\theta \int_0^{2\pi} \!\df \phi \,\sum_X f_{M \to X}(t,\cos\theta,\phi) \biggr]
\,,\end{equation}
where $X$ denotes all possible flavor structures allowed for the splitting of the mother $M$.
The Sudakov factor can be interpreted as a no-branching probability, since the probability of obtaining a value $t_M \leq t_\start$ is given by the probability to not branch anywhere between $t_M$ and $t_\start$ times the probability to branch at $t_M$, which is given by the splitting function evaluated at $t_M$. Thus the probability of having no branching anywhere above $t_\cut$, where $t_\cut$ is the cutoff scale of the parton shower, is determined by the no-branching probability directly
\begin{equation}
\label{eq:nobranch}
\cP^{\rm nb}_M(t_\start, t_\cut) = \sud_M(t_\start,t_\cut)
\,.\end{equation}

For simplicity, we will frequently use the shorthand notation
\begin{equation}
\cP_M^{\rm br}(t_M,v_M;\,t_\start,t_\cut) \equiv \cP^{\rm br}_{M \to LR}(t_M,\cos\theta_M,\phi_M;\,t_\start,t_\cut)
\,,\end{equation}
where $v_M$ denotes all variables other than the virtuality required to describe a single splitting. Thus, $v_M$ contains the angular variables $\cos\theta_M$ and $\phi_M$, as well as the flavor information $M \to LR$. We will also use the notation
\begin{equation}
\int \!\df v_M \equiv \int_{-1}^1 \!\df\!\cos\theta \int_0^{2\pi} \!\df \phi\, \sum_X
\,.\end{equation}

Of course, the probability for a particle to do anything (branch at any $t_M > t_\cut$ or not branch above $t_\cut$) is equal to unity
\begin{equation}
\label{eq:probabilityconserved}
\int_{t_\cut}^{t_\start} \!\df t_M \int \!\df v_M \,\cP_M^{\rm br}(t_M,v_M;\,t_\start,t_\cut)
+ \cP^{\rm nb}_M(t_\start, t_\cut) = 1
\,.\end{equation}
To see this explicitly, note that the Sudakov factor satisfies
\begin{align}
\label{eq:sudakovderivative}
\frac{\df}{\df t_1} \sud_M(t_2,t_1) &=
\sud_M(t_2,t_1) \int_{-1}^1 \! \df \!\cos\theta \int_0^{2\pi} \! \df\phi\, \sum_X f_{M \to X}(t_1,\cos\theta,\phi)
\nn\\
&= \int \! \df v_M \, \cP^{\rm br}_M(t_1,v_M;\,t_2,t_\cut)
\,,\end{align}
where we are assuming that $t_1 > t_\lab{cut}$. This implies
\begin{equation}
\label{eq:singlebranchintegral}
\int_{t_1}^{t_2} \!\df t_M \int \! \df v_M \, \cP^{\rm br}_M(t_M,v_M;\,t_\start,t_\cut)
= \sud_M(t_\start,t_2) - \sud_M(t_\start,t_1)
\,,\end{equation}
from which \eq{eq:probabilityconserved} follows.

\subsection{The Double-Branch Probability}
\label{subsec:doublebranchprob}

\begin{figure}
\includegraphics[scale=0.58]{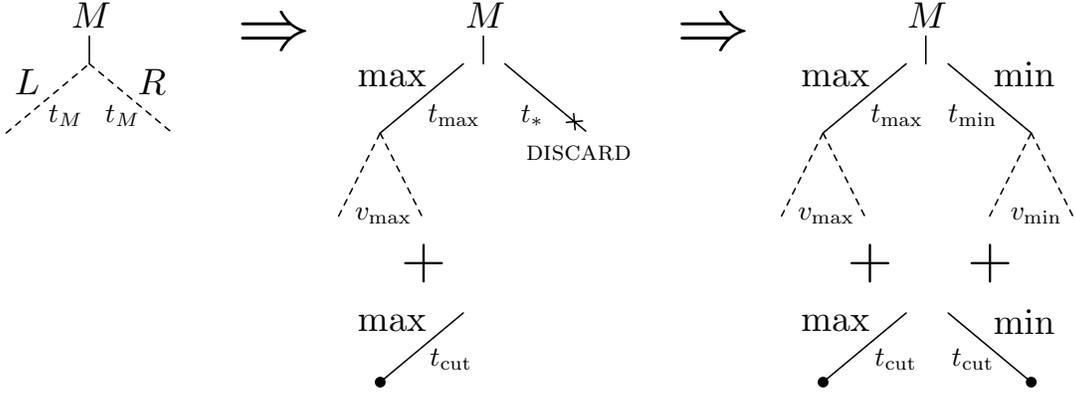}
\caption{The \GenEvA\ shower algorithm. Starting at the scale $t_M$, trial $L$ and $R$ branches are generated, but only the one with the larger of $t_L$ and $t_R$ is kept. Then, starting from the new scale $t_* = \min\bigl[t_\max, (\sqrt{t_M} - \sqrt{t_\max})^2\bigr]$, the other branch is generated. This coupling between the $L$ and $R$ branches is essential to obtain an analytic truncation, and means that the \GenEvA\ shower has a notion of global evolution.}
\label{fig:showerdoublea}
\end{figure}

\GenEvA's  parton shower algorithm always acts on an existing $M \to LR$ branch. It takes both the $L$ and $R$ daughters of the branch and considers their branching into two sets of granddaughters $LL$/$LR$ and $RL$/$RR$. The shower algorithm is illustrated in \fig{fig:showerdoublea}, and works as follows:
\begin{enumerate}
\item Pick a random branch with two unprocessed daughters $L$ and $R$ and a mother particle $M$ with invariant mass $t_M$.
\item Choose $E_{L,R}^\start$, set $t_\start = t_M$, and determine values of $\{t_L, v_L\}$ and $\{t_R, v_L\}$ independently according to the single-branch probability $\cP_i^{\rm br}(t_i,v_i;\,t_M,t_\cut)$, where $i = L,R$. If $t_i < t_\cut$, leave that particle unbranched.
\item Keep the branch of the daughter with the larger $t$, call this the $\max$-daughter, and define $t_\max = \max(t_L, t_R)$. Discard the branch of the $\min$-daughter.
\item Determine new values $\{t_\min, v_\min\}$ for the $\min$-daughter according to $\cP_\min^{\rm br}(t_\min,v_\min;\,t_*,t_\cut)$, with $t_\start$ set to
\begin{equation}
t_* = \min\bigl[t_\max, (\sqrt{t_M} - \sqrt{t_\max})^2 \bigr]
\,.\end{equation}
If $t_\min < t_\cut$, leave that particle unbranched.
\item Continue until all branches are processed.
\end{enumerate}

Since  \eq{eq:singlebranchprob} contains no restrictions on the angles $\cos\theta$ and $\phi$, their phase space is trivially covered. The range of $t_{L,R}$ allowed by the algorithm is
\begin{equation}
t_\cut \leq t_{L,R} \leq t_M
\,,\qquad
\sqrt{t_L} + \sqrt{t_R} \leq \sqrt{t_M}
\,,\end{equation}
where the second condition is enforced by $t_*$ in step 4 of the algorithm, as required by \eq{eq:showerspace_limits}. Hence, apart from the infrared shower cutoff $t_\lab{cut}$, the algorithm covers all of phase space without leaving any dead zones. In particular, a parton shower history with a cutoff value $t_\cut$ and $n$ particles in the final state covers the entire $n$-body phase space with the additional restriction that the minimum virtuality of any two particles is greater than $t_\cut$.\footnote{Two partons from different branches can also accidentally get closer than $t_\cut$, so there are additional regions of $n$-body phase space that are covered in practice. We will discuss this issue more in \sec{subsec:lo}.} As there are QCD singularities associated with two partons getting close in virtuality, the $t_\cut$ restriction is actually helpful to regulate infrared divergences in tree-level diagrams.

The above algorithm results in an analytically calculable parton shower. It is clear that the algorithm couples the splitting of the $L$ and the $R$ daughters together, resulting in a double-branch probability
\begin{equation}
\cP_M (t_L,v_L;\,t_R,v_R;\,t_M,t_\cut)
\equiv \cP_{M \to LR \to (LL/LR)/(RL/RR)}(t_L, \cos \theta_L, \phi_L;\,t_R,\cos \theta_R, \phi_R;\,t_M,t_\cut)
\,,\end{equation}
where $t_{L,R}$ are the final virtualities selected for the daughters, and $v_{L,R}$ denote all other variables describing the daughters' splittings, including the flavor and color structure for the $LL$, $LR$, $RL$, and $RR$ granddaughters.

\begin{figure}
\includegraphics[scale=0.58]{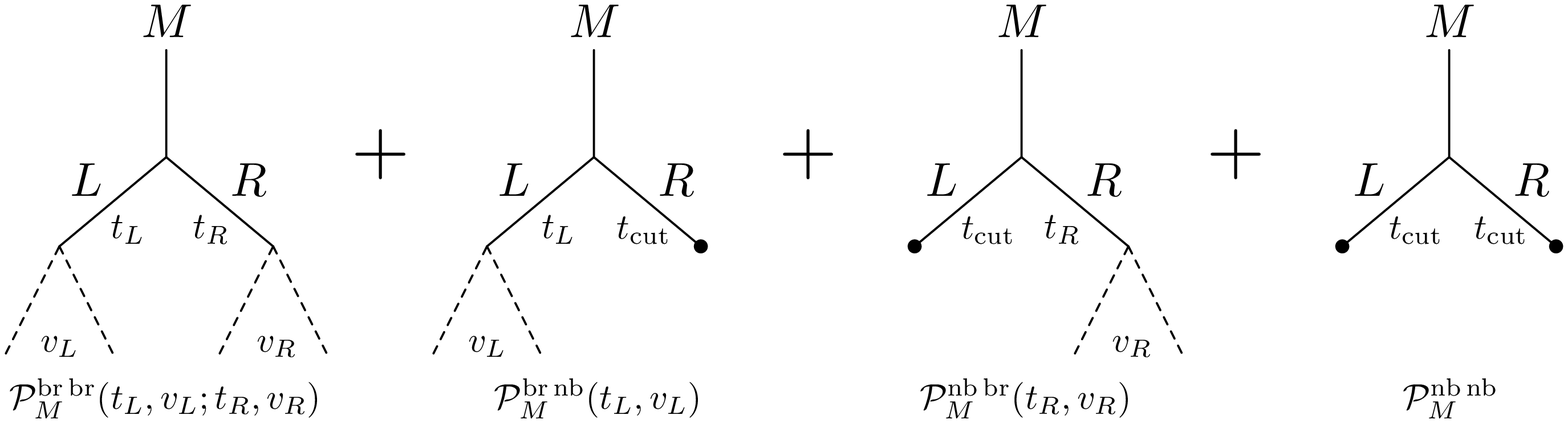}
\caption{The double-branch probability. The \GenEvA\ algorithm in \fig{fig:showerdoublea} yields four different branching/no-branching combinations. They are built out of products of single-branch probabilities, with nontrivial interdependence between $t_L$ and $t_R$.}
\label{fig:showerdoubleb}
\end{figure}

There are four possible cases for the double-branch probability as shown in \fig{fig:showerdoubleb}: 1) Both the left and right daughters branch; 2) the left daughter branches and the right one does not; 3) the right daughter branches and the left one does not; and 4) neither the left nor the right daughter branches. Following the shower algorithm, the double-branch probability for case 1) is given by
\begin{align}
\label{eq:doublebranch_brbr}
\cP^{\rm br\,br}_M(t_L,v_L;\,\,&t_R, v_R;\, t_M,t_\cut)
\nn\\
&= \theta(t_L > t_R) \Bigl[\cP_L^{\rm br}(t_L, v_L;\,t_M,t_\cut) \times \sud_R(t_M,t_L)\,
\cP_R^{\rm br}(t_R, v_R;\, t_*,t_\cut) \Bigr]
\nn\\ &\quad
+ \theta(t_R>t_L) \Bigl[ \cP_R^{\rm br}(t_R, v_R;\,t_M,t_\cut) \times \sud_L(t_M,t_R)\,
\cP_L^{\rm br}(t_L, v_L;\, t_*,t_\cut) \Bigr]
\,,\end{align}
where the extra factors of $\sud_{L,R}$ arise from discarding the first attempt to determine the smaller $t$ value. For the other three cases we have
\begin{align}
\label{eq:doublebranch_rest}
\cP^{\rm br\,nb}_M(t_L,v_L;\, t_M,t_\cut) &=
\cP_L^{\rm br}(t_L, v_L;\,t_M,t_\cut) \times \sud_R(t_M,t_L)\, \sud_R(t_*,t_\cut)
\,,\nn\\
\cP^{\rm nb\,br}_M( t_R, v_R;\,t_M,t_\cut) &=
\cP_R^{\rm br}(t_R, v_R;\,t_M,t_\cut) \times \sud_L(t_M,t_R)\, \sud_L(t_*,t_\cut)
\,,\nn\\
\cP^{\rm nb\,nb}_M(t_M,t_\cut) &= \sud_L(t_M,t_\cut) \times \sud_R(t_M,t_\cut)
\,,\end{align}
where it is understood that if $t_* < t_\cut$, then $\cP_i^{\rm br}(t_i, v_i;\, t_*,t_\cut) = 0$ and $\sud_i(t_*,t_\cut)= 1$.

We can formally combine all four cases by writing
\begin{align}
\label{eq:formaldoublebranch}
\cP_M(t_L,v_L;\,\, &t_R, v_R;\,t_M,t_\cut)
\nn\\
&= \cP^{\rm br\,br}_M(t_L,v_L;\, t_R, v_R;\,t_M,t_\cut)\, \theta(t_L>t_\cut)\, \theta(t_R>t_\cut)
\nn\\ &\quad
+ \cP^{\rm br\,nb}_M(t_L,v_L;\, t_M,t_\cut)\, \theta(t_L>t_\cut)\, \delta(t_R)\, \delta(v_R)
\nn\\ &\quad
+ \cP^{\rm nb\,br}_M(t_R, v_R;\,t_M,t_\cut)\, \theta(t_R>t_\cut)\, \delta(t_L)\, \delta(v_L)
\nn\\ &\quad
+ \cP^{\rm nb\,nb}_M(t_M,t_\cut)\, \delta(t_L)\, \delta(v_L)\, \delta(t_R)\, \delta(v_R)
\,.\end{align}
The $\delta$ functions are just formal crutches to be able to define values for $t_{L,R}$ and $v_{L,R}$ when no branching actually occurred. When a daughter does not branch, it is put on shell through the $\delta(t_{L,R})$ functions (for massless particles as we are considering here). The $\theta$ functions are not strictly necessary, as they are already contained in the single-branch probabilities, but to be explicit we write them out again.

As a check of the algorithm, it is straightforward to verify that the probability for anything to happen is equal to unity:
\begin{equation}
\label{eq:doubleunity}
\int_{0}^{t_M} \! \df t_L \int\! \df v_L \int_{0}^{t_M} \! \df t_R \int\! \df v_R \, \cP_M(t_L,v_L;\,t_R,v_R;\,t_M,t_\cut) = 1
\,.\end{equation}
To see this, integrate over $\min(t_L, t_R)$ using \eq{eq:singlebranchintegral}, and then examine the quantity
\begin{equation}
\frac{\df}{\df t} \bigr[\Delta_L(t_M,t)\, \Delta_R(t_M,t) \bigr]
\,,\end{equation}
to obtain a simple expression for the integral over $\max(t_L, t_R)$.

\begin{figure}[t]
\includegraphics[scale=0.7]{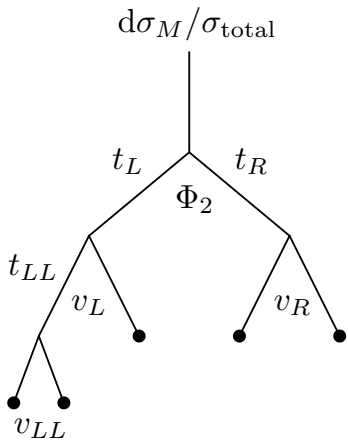}
\caption{An example parton shower history with 5 particles in the final state. The quantity $\df\sigma_{M\to LR}/\sigma_\lab{total}$ is the probability to obtain the initial $M\to LR$ branch from a $2 \to 2$ hard scattering as a function of $\Phi_2$. The total probability for obtaining the full 5-particle final state is given in \eq{eq:examplehistory}.}
\label{fig:showerhistory}
\end{figure}

It  is now straightforward to calculate the probability for generating an event with a given hard $2\to 2$ scattering and a history of branches with $n$ particles in the final state. It is simply given by the probability to obtain the hard scattering multiplied by the double-branch probabilities for each of the following branches. As an example, consider the parton shower history shown in \fig{fig:showerhistory}. The differential probability to obtain this event is given by
\begin{align}
\label{eq:examplehistory}
\df \cP(\vS_{\rm Fig.\,\ref{fig:showerhistory}})
= \frac{\df\sigma_{M \to LR}}{\df \sigma_\lab{total}}
&\times \cP^{\rm br\,br}_M(t_L,v_L;\,t_R,v_R;\,t_M,t_\cut)\, \df t_L \, \df v_L\, \df t_R\, \df v_R
\nn \\
&\times \cP^{\rm br\,nb}_L(t_{LL},v_{LL};\,t_L,t_\cut)\, \df t_{LL} \, \df v_{LL}
\times \cP^{\rm nb\,nb}_R(t_R,t_\cut)
\nn\\
&\times \cP^{\rm nb\,nb}_{LL}(t_{LL},t_\cut)
\,,\end{align}
where $\df\sigma_{M\to LR}/\sigma_\lab{total}$ is the probability to obtain the initial $M\to LR$ branch from a 2-body matrix element with $L$ and $R$ in the final state. Hence, \GenEvA's parton shower algorithm has indeed an analytic formula for the probability $\cP(\vS)$ to obtain a given parton shower history.

\subsection{Truncation}
\label{subsec:truncation}

The parton shower described above will produce final states with in principle arbitrarily many final state particles. The goal of \GenEvA\ is to reweight such an obtained final state to a known distribution. In practice, the exact distributions will only be available for a limited number of final state particles. Furthermore, for the purposes of the \GenEvA\ framework \cite{genevaphysics}, one might want to choose the scale that separates partonic calculations from phenomenological models to be different from the value $t_\cut$ used by the parton shower.

Consider the case where we only have partonic calculations available with up to $n_\max$ partons in the final state, and we want the minimum virtuality described by the partonic calculation to be $t_\match \geq t_\lab{cut}$. The parton shower algorithm will in general produce events with more than $n_\max$ partons in the final state and produce splittings with $t < t_\match$. By assumption, these events cannot be reweighted to the partonic calculation. One way to deal with such events is to simply reject them, and only reweight events that satisfy the criteria required by the partonic calculation. However, this would reject a large number of events and would obviously result in large inefficiencies. Furthermore, because the parton shower preserves probability, unless the user calculated the rejection probability, vetoing events would make it impossible to obtain total cross section information from the phase space generator.

Instead of rejecting events with too many final state partons or too low a splitting scale, we can truncate these events to a final state with at most $n_\max$ particles, in such a way that each branch has virtuality above $t_\match$. This is achieved by a simple truncation algorithm that truncates events to a given $\{t_\match, n_\max\}$:
\begin{enumerate}
\item Order the branches by the value of the evolution variable $t$.
\item Remove any branches with $t < t_\match$. That is, remove the daughter particles from any mother whose virtuality is $t < t_\match$, leaving the mother unbranched.
\item If the final state contains more than $n_\max$ particles remove additional branches, starting with the smallest value of $t$, until there are $n_\max$ final state particles.
\item Recalculate the kinematics of all remaining final state particles.
\item For events with $n_\max$ final state particles, redefine $t_\match$ to be the smallest value of the virtualities $t$ of the remaining branches.
\end{enumerate}
This algorithm clearly results in an event sample for which all branches have $t > t_\match$ and there are no more than $n_\max$ final states. The last step guarantees that if the shower restarts from the event-defined $t_\match$ scale, then the resulting distribution of events will be identical to those if truncation were not applied. Note that this algorithm preserves probability because no events are ever rejected.

To be able to reweight the resulting events after truncation, we of course need the probability with which the truncated event was generated. As expected from the notion of global evolution, the probability for finding a truncated event should be exactly equal to the probability of generating that event with the shower cutoff $t_\cut$ set equal to the event-defined $t_\match$, \ie\
\begin{equation}
\label{eq:truncgoal}
\cP^\lab{trunc}_M(t_L,v_L;\,t_R,v_R;\,t_M,t_\cut;\, t_\match) = \cP_M(t_L,v_L;\,t_R,v_R;\,t_M, t_\match)
\,.\end{equation}
To see this, we write out $\cP_M^\lab{trunc}$ in analogy with \eq{eq:formaldoublebranch} as
\begin{align}
\cP^\lab{trunc}_M(t_L,v_L;\,\,& t_R, v_R;\,t_M,t_\cut;\, t_\match)
\nn\\
&= \cP^{\rm trunc \, br\,br}_M(t_L,v_L;\, t_R, v_R;\,t_M,t_\cut;\, t_\match)\, \theta(t_L>t_\match)\, \theta(t_R>t_\match)
\nn\\ &\quad
+ \cP^{\rm trunc \, br\,nb}_M(t_L,v_L;\, t_M,t_\cut;\, t_\match)\, \theta(t_L>t_\match)\, \delta(t_R)\, \delta(v_R)
\nn\\ &\quad
+ \cP^{\rm trunc \, nb\,br}_M(t_R, v_R;\,t_M,t_\cut;\, t_\match)\,
\theta(t_R>t_\match)\, \delta(t_L)\, \delta(v_L)
\nn\\ &\quad
+ \cP^{\rm trunc \, nb\,nb}_M(t_M,t_\cut;\, t_\match)\,
\delta(t_L)\, \delta(v_L)\, \delta(t_R)\, \delta(v_R)
\,,\end{align}
and show that
\begin{equation}
\cP^{\rm trunc \, x\,y}_M(\cdots;\,t_M,t_\cut;\, t_\match)
= \cP^{\rm \, x\,y}_M(\cdots;\,t_M, t_\match)
\,.\end{equation}

\begin{figure}
\includegraphics[scale=0.58]{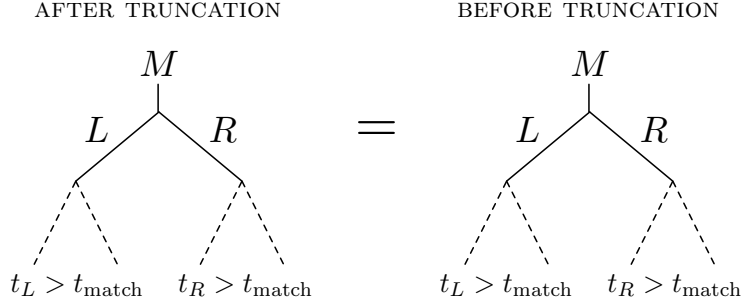}
\caption{To get a truncated event where both daughters are branched, the original event has to have $t_{L,R} > t_\match$.}
\label{fig:truncationexamplea}
\end{figure}

There are three cases to consider. First, take a double branch for which truncation keeps both the left and the right branches. This occurs if $t_L$ and $t_R$ are initially greater than $t_\match$, as in \fig{fig:truncationexamplea}. Thus, the probability of getting given $t_L$ and $t_R$ values after truncation is
\begin{align}
\cP^{\rm trunc \, br\,br}_M(t_L,v_L;\,t_R,v_R;\,t_M,t_\cut;\, t_\match)
&= \cP^{\rm br\,br}_M(t_L,v_L;\,t_R,v_R;\,t_M,t_\cut)
\nn\\
&\equiv \cP^{\rm br\,br}_M(t_L,v_L;\,t_R,v_R;\,t_M,t_\match)
\,,\end{align}
 where we used that the single-branch probability in \eq{eq:singlebranchprob} only depends on the shower cutoff through an overall $\theta$ function. Thus, the probability to get this double branch after truncation is equal to the probability to get this double branch from running the shower with $t_\cut = t_\match$.

\begin{figure}
\includegraphics[scale=0.58]{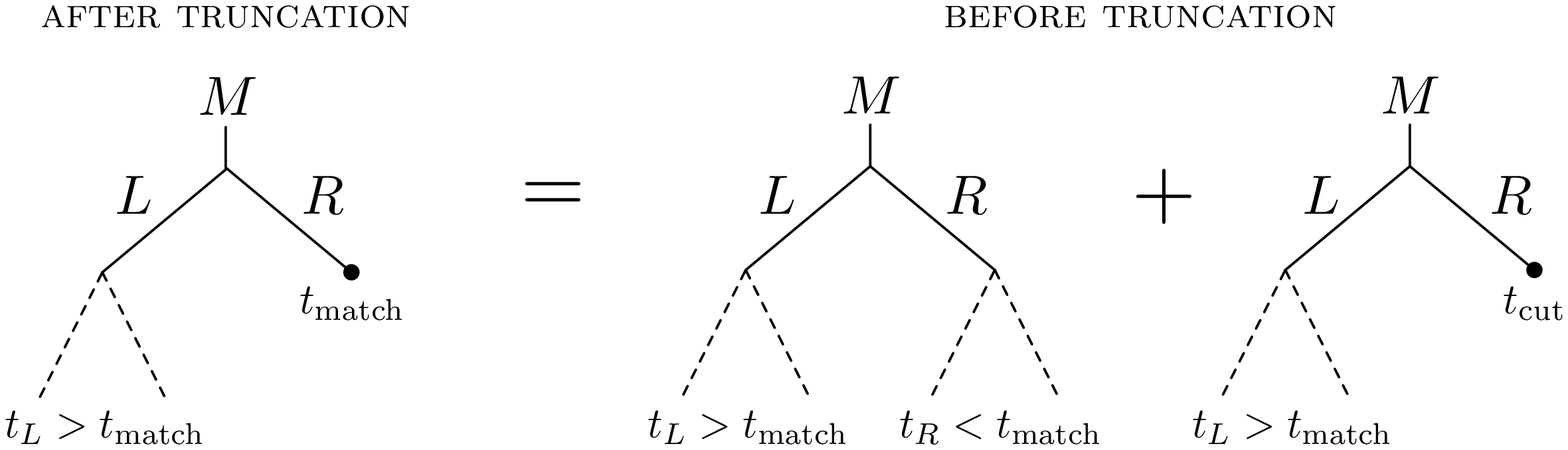}
\caption{To get a truncated event where only the left daughter is branched, the original event has to have $t_L > t_\match$ and $t_R < t_\match$. Since, formally, an unbranched particle has $t < t_\cut$, there are two different ways this can happen. For the truncated event, the shower cutoff is effectively raised from $t_\cut$ to $t_\match$. The same holds if the roles of $L$ and $R$ are interchanged.}
\label{fig:truncationexampleb}
\end{figure}

Next, take the case where after truncation one daughter is branched and one is unbranched. This occurs for example when $t_R < t_\match < t_L$, which happens if the left daughter branched above $t_\match$, while the right daughter did not branch at all or branched below $t_\match$, as in \fig{fig:truncationexampleb}. In this case, the truncation probability is
\begin{align}
\label{eq:truncation_integration}
&\cP^{\rm trunc \, br\,nb}_M\,(t_L,v_L;\, t_M,t_\cut;\, t_\match)
\nn\\
&= \int_{t_\cut}^{t_\match} \! \df t_R \int \!\df v_R\,\cP^{\rm br\,br}_M(t_L,v_L;\,t_R,v_R;\,t_M,t_\cut)
+ \cP^{\rm br\,nb}_M(t_L,v_L;\, t_M,t_\cut)
\nn \\
&= \cP_L^{\rm br}(t_L, v_L;\,t_M,t_\cut)
\times \sud_R(t_M,t_L)\biggl[
\int_{t_\cut}^{t_\match} \! \df t_R \int \!\df v_R\,\cP^{\rm br}_R(t_R,v_R;\,t_*,t_\cut)
+ \sud_R(t_*,t_\cut) \biggr]
\nn \\
&= \cP_L^{\rm br}(t_L, v_L;\,t_M,t_\match) \times \sud_R(t_M,t_L)\, \sud_R(t_*,t_\match)
\nn \\
&\equiv \cP^{\rm br\,nb}_M(t_L,v_L;\, t_M, t_\match)
\,,\end{align}
where we used the definitions in \eqs{eq:doublebranch_brbr}{eq:doublebranch_rest}, and in the third step we used \eq{eq:singlebranchintegral} and again that $\cP_L^{\rm br}$ only depends on the shower cutoff through a $\theta$ function. The crucial feature of \geneva's shower algorithm that makes this calculation possible is that the double-branch probability can be integrated analytically over the smaller virtuality $t_R$ without knowing the value of the larger virtuality $t_L$. An analogous calculation holds for $t_L < t_\match < t_R$.

\begin{figure}
\includegraphics[scale=0.58]{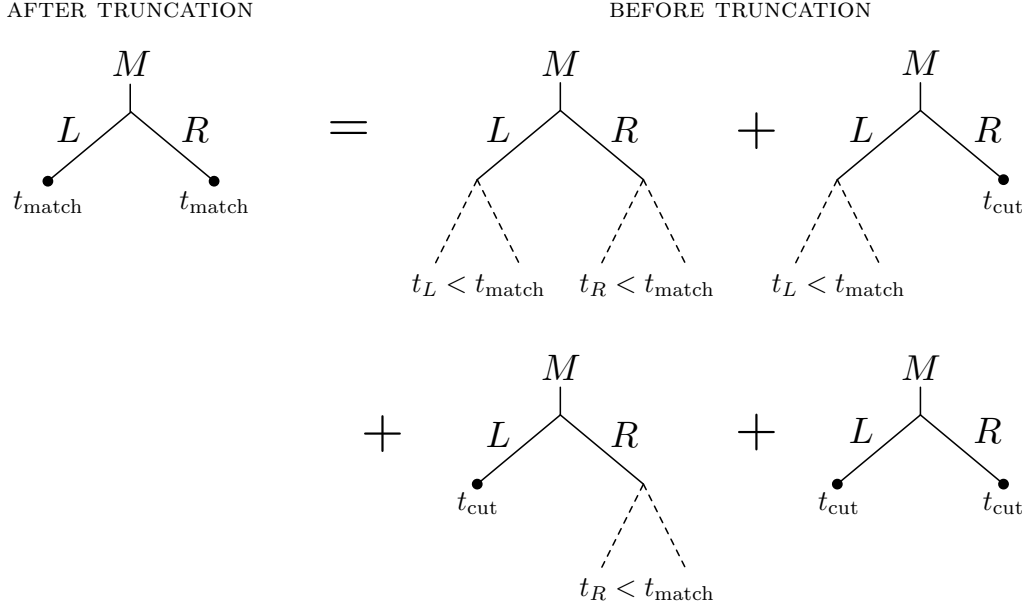}
\caption{To get a truncated event where neither daughter branches, the original event must have $t_{L,R} < t_\match$. Since, formally, an unbranched particle has $t < t_\cut$, there are now four different ways for this to happen. For the truncated event, the shower cutoff is effectively raised from $t_\cut$ to $t_\match$.}
\label{fig:truncationexamplec}
\end{figure}

Finally, take the case where after truncation, both daughters are unbranched. This occurs if both daughters either did not branch or branched below $t_\match$, as shown in \fig{fig:truncationexamplec}. In this case, the truncation probability is
\begin{align}
\cP^{\rm trunc \, nb\,nb}_M\,&(t_M,t_\cut;\, t_\match)
\nn\\
&= \int_{t_\cut}^{t_\match} \! \df t_L \int\! \df v_L \int_{t_\cut}^{t_\match}\! \df t_R \int\! \df v_R\,
\cP^{\rm br\,br}_M(t_L,v_L;\,t_R,v_R;\,t_M,t_\cut)
\nn\\ &\quad
+ \int_{t_\cut}^{t_\match} \! \df t_L \int \!\df v_L\, \cP^{\rm br\,nb}_M(t_L,v_L;\, t_M,t_\cut)
\nn\\ & \quad
+ \int_{t_\cut}^{t_\match} \! \df t_R \int\! \df v_R\, \cP^{\rm nb\,br}_M(t_R,v_R;\, t_M,t_\cut)
+ \cP^{\rm nb\,nb}_M(t_M,t_\cut)
\nn\\
&= \cP^{\rm nb\,nb}_M(t_M,t_\match)
\,,\end{align}
where we used \eq{eq:doubleunity} and the fact that the truncation algorithm preserves probability. One can also do the $t_L$ and $t_R$ integrals explicitly as in \eq{eq:truncation_integration} to verify this result. The  above calculations show that \eq{eq:truncgoal} indeed holds: The probability to get a truncated event is exactly the same as the probability for it to be produced by the shower with $t_\cut = t_\match$, consistent with the notion of a global evolution.

There  is one subtlety we have completely ignored so far. In step 2 of the shower algorithm outlined at the beginning of \sec{subsec:doublebranchprob}, we need to choose values for $E_{L,R}^\start$ to be used in the daughters' splitting functions. To not violate \eq{eq:truncgoal}, we need a method to determine values for $E_{L,R}^\start$ which gives the same values before and after truncation. In other words, we need an expression for $E_{L,R}^\start$ that is invariant under truncation.

Certainly  the easiest choice would be to take $E_{L,R}^\start \equiv E_\mathrm{CM}$. However, as noted in \sec{subsec:ChoiceofKinematicVariables}, we would like $E_{L,R}^\start$ to be of the order of the daughters' actual energies. In principle, there are many possible choices. In analogy with $t_\start = t_M$, we might attempt to take $E_{L,R}^\start = E_M$, where $E_M$ is the mother's energy. However, this does not work, because $E_M$ is not invariant under truncation.%
\footnote{As shown by \eq{eq:costheta_z}, the energies of two sisters depend on each others virtualities as well as their mother's energy and $\cos\theta$. Thus, truncating the aunt of a particle changes its mother's energy.}
Instead, we take $E_{L,R}^\start = E_M^\max$, where $E_M^\max$ is the maximum possible energy of the mother and is determined recursively as follows: Given the $E^\max_M$ used to branch the daughters, we define the daughters' $E^\max_{L,R}$, to be used in the subsequent branching of the granddaughters, by
\begin{equation}
\label{eq:Emax}
E^\max_{L,R}
= \frac{1}{2} E^\max_M \biggl[1 + \frac{t_{L,R}}{t_M}
+ \beta(E^\max_M) \cos\theta_M \Bigl(1 - \frac{t_{L,R}}{t_M} \Bigr) \biggr]
\,.\end{equation}
This corresponds to taking the minimum and maximum values of $z_M$ in \eq{eq:costheta_z}. For the initial hard scattering we simply take $E^\max_M = E_\mathrm{CM}$. This definition of $E^\max$ is indeed invariant under truncation, since it only depends on $E_\CM$ and the values of $\cos\theta$ and $t$ of a particle's direct ancestors, \ie\ only on those branches that cannot get truncated unless the particle itself is truncated first.

\begin{figure}
\includegraphics[scale=0.58]{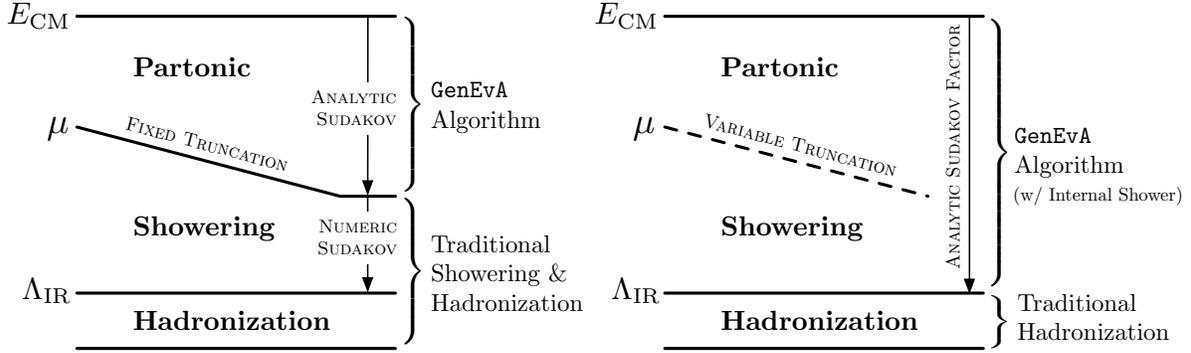}
\caption{Two different ways of using the \GenEvA\ algorithm in a complete event generation framework. In the left panel, the \GenEvA\ algorithm is only used above a scale $\mu = \sqrt{t_\cut} \gg \Lambda_\lab{IR}$, and traditional showering and hadronization is applied starting at $\mu$. In the right panel, \GenEvA\ is used to distribute events all the way down to $\sqrt{t_\cut} = \Lambda_\lab{IR}$ to interface directly with a hadronization scheme. The first case allows the use of already existing parton showers that have been tuned to data, while the second case allows the scales associated with the partonic calculations to vary even after hadronization and detector simulation. In the companion paper \cite{genevaphysics}, we discuss how both methods can give the same leading-logarithmic Sudakov improvements.}
\label{fig:regimes}
\end{figure}

With truncation, we now have two different ways of using the \GenEvA\ algorithm in the context of a complete event generation framework, as illustrated in \fig{fig:regimes}. One method is to run the \GenEvA\ algorithm to $t_\cut \gg \Lambda_\lab{IR}^2$ and truncate it to $\{t_\cut, n_\max\}$, which means that $(n \leq n_\max)$-parton events are unaffected, while $(n > n_\max)$-parton events get truncated to $n_\max$-parton phase space with a variable, event-specific $t_\match$ scale. The events can then passed to an external showering/hadronization program to start a new shower at $t_\cut$ for $n < n_\max$ and the variable $t_\match$ scale for $n = n_\max$. This uses \GenEvA\ more-or-less as a traditional phase space generator. Alternatively, we can run the \GenEvA\ algorithm all the way down to $t_\cut = \Lambda_\lab{IR}^2$, interface with a hadronization routine, and then after the fact truncate the shower history back up to $\{t_\match \gg \Lambda_\lab{IR}^2, n_\max\}$ to obtain a point in $n$-body phase space which now already has a complete shower attached to it.

Because truncation preserves probability, both of these methods will give statistically identical results, however, the second method allows multiple different truncation schemes on the same already hadronized (and even detector-simulated) event. Of course, in practice an external showering program can be tuned to data, whereas the \GenEvA\ shower only guarantees leading-logarithmic accuracy, but the possibility to improve perturbative theoretical distributions after detector simulation is an intriguing possibility, and for this reason it might be useful to develop an analytically calculable but realistic parton shower.

\section{The Jacobian Factor}
\label{sec:jac}

In the previous section, we saw that the analytic parton shower covers all of phase space. However, the parton shower distributes events in terms of the shower variables $\{t_, \cos \theta, \phi \}$ as in \eq{eq:showerspace}, and there is a Jacobian factor from converting these to the variables describing Lorentz-invariant phase space.

The  complete parton shower space $\df\vS_n$ for a shower history with $n$ final state particles has two pieces. First, there are the variables $\df\Phi^{(0)}_2$ of the initial hard scattering that yields two on-shell (and for us massless) final states $A$ and $B$. Second, there is the product of the individual $\df\vS_2$ splitting variables that describe the subsequent $1\to 2$ parton shower branches. This gives
\begin{equation}
\df\vS_n = \df\Phi^{(0)}_2 (p_\CM; p_A^{(0)}, p_B^{(0)}) \prod_{i=1}^{n-2} \df\vS_2(M_i\to L_i R_i)
\,,\end{equation}
where $p_{A,B}^{(0)}$ are the four-momenta of the particles $A$, $B$ \emph{before} running the parton shower, and $i$ runs over all parton shower branches. We define that the way the parton shower takes the momenta of the $A$ and $B$ particles off shell is by leaving the angles $\df \Omega_2$ unchanged. Therefore, the phase space for the initial on-shell final state particles is related to that of the ultimate off-shell, showered particles via
\begin{equation}
\df\Phi^{(0)}_2(p_\CM; p_A^{(0)}, p_B^{(0)})
= \frac{1}{\lambda(t_\CM; t_A, t_B)}\, \df\Phi_2(p_\CM; p_A, p_B)
\,,\end{equation}
where the ordinary phase space factor $\lambda$ is given in \eq{eq:betalambda}.

To convert the natural shower variables $\df\vS_n$ to Lorentz-invariant $n$-body phase space $\df\Phi_n$, we can use the well-known method to recursively decompose $n$-body phase space in terms of multiple copies of $2$-body phase space:
\begin{align}
\label{eq:phasespace_recursion}
\df\Phi_n(p_\CM; p_1,\ldots, p_n)
&= (2\pi)^4 \delta^{4} \biggl(p_\CM - \sum_{i=1}^n p_i \biggr) \prod_{i=1}^n \frac{\df^4 p_i}{(2\pi)^3} \, \delta(p_i^2 - m_i^2) \, \theta(p_i^0)
\nn\\
&= \df\Phi_{n-1}(p_\CM; p_1,\ldots, p_{n-2}, Q)\, \frac{\df Q^2}{2\pi} \, \df\Phi_2(Q; p_{n-1}, p_{n})
\nn\\
&= \df\Phi_{n-1}(p_\CM; p_1,\ldots, p_{n-2}, Q)\, \frac{1}{J(t_Q;t_{n-1},t_n)} \,
\df t_Q \, \df\!\cos\theta_Q \, \df \phi_Q
\,,\end{align}
where the 2-body Jacobian factor is given by
\begin{equation}
\label{eq:Jindividual}
J(t_M;t_L,t_R) = \frac{64 \pi^3}{\lambda(t_M;t_L,t_R)}
\,.\end{equation}
The variables $t_Q \equiv Q^2$, $\cos\theta_Q$, and $\phi_Q$ in the last line of \eq{eq:phasespace_recursion} are exactly the same variables used in $\df\Sigma_2$ in \eq{eq:showerspace}. Their integration ranges are such that for any $\df\Phi_2(p_M; p_L, p_R)$ we have
\begin{equation}
\label{eq:trestriction}
-1 \leq \cos\theta_M \leq 1
\,,\qquad
0 \leq \phi_M \leq 2\pi
\,,\qquad
0 \leq \sqrt{t_L} + \sqrt{t_R} \leq \sqrt{t_M}
\,,\end{equation}
identical to \eq{eq:showerspace_limits}. Lorentz-invariant phase space can thus be written suggestively as
\begin{equation}
\df\Phi_n(p_\CM; p_1,\ldots, p_n) = \df\Phi_2(p_\CM; p_{A}, p_{B}) \prod_{i=1}^{n-2} \frac{1}{J(t_{M_i},t_{L_i},t_{R_i})}\, \df\vS_2(M_i\to L_i R_i)
\,,
\end{equation}
where $A$ and $B$ can be thought of as the particles originating from the hard interaction and $t_{M_i}$, $t_{L_i}$ and $t_{R_i}$ denote the virtualities of the mother and daughter particles in the $i$-th phase space recursion.

Though both the parton shower and Lorentz-invariant phase space can be built recursively, there is one important difference between the parton shower $\df\vS_n$ and Lorentz-invariant phase space $\df \Phi_n$ regarding identical particles. A parton shower distinguishes all final state particles by a series of ``$L$'' and ``$R$'' labels, regardless of whether those particles have the same quantum numbers. In contrast, Lorentz-invariant phase space does not distinguish between identical particles. Thus, an additional factor has to be included in the Jacobian to account for this difference. For a parton shower history with $n$ final state particles that contains $n_1, n_2, \ldots, n_k$ sets of identical particles, with $n_1 + n_2 + \dotsb + n_k = n$, we have to divide the Jacobian by an extra factor of $n_1!\, n_2! \dotsb n_k!$.

Putting all this information together, the total Jacobian relating $n$-body parton shower space to Lorentz-invariant $n$-body phase space is
\begin{equation}
\label{Jtotal}
J(\vS_n) \equiv \biggl\lvert\frac{\partial (\vS_n)}{\partial (\Phi_n)} \biggr\rvert
= \frac{1}{\lambda(t_\CM; t_A, t_B)}\,\frac{1}{n_1!\, n_2! \dotsb n_k!}\, \prod_{i=1}^{n-2} J(t_{M_i},t_{L_i},t_{R_i})
\,.\end{equation}
This factor is a crucial piece of the master formula \eq{eq:finalweight}.

\section{The Overcounting Factor}
\label{sec:oc}

\subsection{Overview}
\label{subsec:ocoverview}

Different parton shower histories can give rise to the same set of final state four-momenta, so the parton shower covers phase space multiple times. Therefore, if we wish to reweight a parton shower event to a matrix element, which only depends on the phase space $\Phi$ of the external particles but knows nothing about the underlying parton shower history $\vS$, we have to account for the fact that the mapping $\Phi(\vS)$ from parton shower space $\vS$ to phase space $\Phi$ is not one-to-one. This is achieved by the overcounting factor $\ha(\vS)$ in \eq{eq:finalweight}.

Since there is only a fixed number of parton shower trees that can yield a given set of external particles, there is also only a fixed number $n(\Phi)$ of points $\vS_i(\Phi)$ that map to a given point $\Phi$. Certainly the easiest choice for $\ha(\vS)$ is thus
\begin{equation}
\label{eq:alpha_trivial}
\ha(\vS) = \frac{1}{n[\Phi(\vS)]}
\,,\end{equation}
which only requires the knowledge of $n(\Phi)$ for $\Phi = \Phi(\vS)$, but not the precise form of each $\vS_i(\Phi)$. Furthermore, we have defined our parton shower without any dead zones, such that for a given set of $n$ final state particles, each compatible parton shower tree fully covers the corresponding $n$-body phase space. Hence, $n(\Phi)$ only depends on the particle types in $\Phi$ and thus equals the number of compatible parton shower trees, a quantity that could be calculated in advance. On the other hand, the reweighting efficiency of this choice will usually be poor, because if two points $\vS_{1,2}(\Phi)$ have $\cP(\vS_1) \gg \cP(\vS_2)$ then \eq{eq:finalweight} yields weights $w(\vS_1) \ll w(\vS_2)$ that are very different.

The optimal choice in terms of reweighting efficiency is to have the same weight $w[\vS_i(\Phi)]$ for all $i$ and a given $\Phi$. This is achieved by
\begin{equation}
\label{eq:alpha_ideal}
\ha(\vS) = \frac{\cP(\vS)J(\vS)}{\sum_i \cP(\vS_i)J(\vS_i)}
\,,\end{equation}
which just reduces \eq{eq:finalweight} back to \eq{eq:optimalweight}. Unfortunately, calculating the sum in \eq{eq:alpha_ideal} requires the construction of all possible parton shower histories $\vS_i[\Phi(\vS)]$ and the computation of their probabilities $\cP(\vS_i)$. Since the number of parton shower histories $\vS_i$ for $n$ final state particles grows like $n!$, this quickly becomes computationally expensive.

In general, we can assign
\begin{equation}
\label{eq:alpha_general}
\ha(\vS) = \frac{\alpha(\vS)}{\sum_i \alpha(\vS_i)}
\,,\end{equation}
such that $\sum_i \ha(\vS_i) = 1$ automatically, and choose $\alpha(\vS)$ such that $\sum_i \alpha(\vS_i)$ is more readily calculable. The choices $\alpha(\vS) = 1$ and $\alpha(\vS) =\cP(\vS)J(\vS)$ recover \eq{eq:alpha_trivial} and \eq{eq:alpha_ideal}, respectively. The better $\alpha(\vS)$ approximates $\cP(\vS)J(\vS)$ the more uniform the total weights will be, resulting in a better reweighting efficiency. At the same time, it is desirable to have a fast way to compute $\sum_i \alpha(\vS_i)$, such that the computational time used to correct the overcounting issue is less than (or at least comparable with) the computational time needed to evaluate the matrix element $\sigma(\Phi)$. In the next subsection we show that, with a suitable choice of $\alpha(\vS)$, the \texttt{ALPHA} algorithm of Refs.~\cite{Caravaglios:1995cd,Caravaglios:1998yr} can be applied to calculate the sum $\sum_i \alpha(\vS_i)$, reducing the $n!$ problem to a $2^n$ problem, which is a significant gain.

Before going on, we want to point out that the overcounting factor $\ha(\vS)$ has appeared in other Monte Carlo programs, most notably \madevent. Indeed, our solution to the phase space overcounting by the parton shower was inspired by the technique used in Ref.~\cite{Maltoni:2002qb} to slice phase space into complementary integration ranges. Given a set of $n$ Feynman diagrams
\begin{equation}
\lvert\mathcal{M}\rvert^2 = \biggl\lvert\sum_{i=1}^n \mathcal{M}_i\biggr\rvert^2
\,,\end{equation}
\madevent\ performs independent phase space integrations over $n$ different functions
\begin{equation}
f_i = w_i \lvert\mathcal{M}\rvert^2
\qquad \text{with} \qquad w_i
= \frac{\lvert\mathcal{M}_i\rvert^2}{\sum_i \lvert\mathcal{M}_i\rvert^2}
\,,\end{equation}
such that the sum over the $n$ integrations $\sum_i w_i |\mathcal{M}|^2 = |\mathcal{M}|^2$ gives the desired integrated squared-amplitude. The ``overcounting factor'' $w_i$ is not gauge invariant, but it does have the singularity structure of the $i$-th Feynman diagram, allowing a phase space integration that is aware of the relevant poles. \GenEvA\ does essentially the same, but now $\ha(\vS)$ will capture the singularity and color factor structure of a given parton shower history.

\subsection{The ALPHA Algorithm}
\label{subsec:alpha}

\begin{figure}
\includegraphics[scale=0.7]{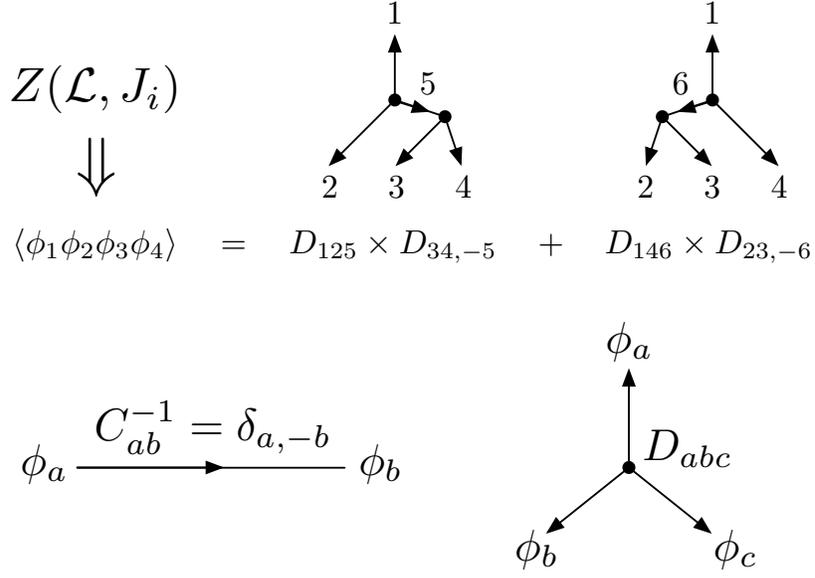}
\caption{Schematic picture of the \texttt{ALPHA} algorithm. Given a quantity that can be described as the sum over all diagrams, one can create a Lagrangian $\mathcal{L}$ whose generating functional $Z$ yields the appropriate sum over ``Feynman diagrams''. The \texttt{ALPHA} algorithm gives an efficient way of computing these sums as long as the fields $\phi_i$ only appear linearly in $\mathcal{L}$, hence the need for the off-diagonal propagator $C_{ab}^{-1} = \delta_{a,-b}$. In the above example, particle 1 could correspond to a photon, 2 to a quark, 3 to a gluon, and 4 to an anti-quark. The $125$ and $146$ vertices could correspond to the electromagnetic couplings, the $34,-5$ vertex to the $\bar{q} \to \bar{q}g$ splitting function, and the $23,-6$ vertex to the $q \to qg$ splitting function.
\label{fig:alpha}}
\end{figure}

We want to calculate a quantity that corresponds to a sum over all possible graphs. This same issue arises in the computation of quantum amplitudes which are sums over all possible Feynman diagrams. Therefore, we can use known techniques to efficiently calculate the sum $\sum_i \alpha(\vS_i)$ by constructing a toy ``Lagrangian'' whose ``generating functional'' provides the desired ``amplitude'' as in \fig{fig:alpha}.

One simple algorithm for doing this is the \texttt{ALPHA} algorithm~\cite{Caravaglios:1995cd,Caravaglios:1998yr}, which is the core algorithm used in the \texttt{ALPGEN} event generator~\cite{Mangano:2002ea}. The \texttt{ALPHA} algorithm essentially works by reducing common subexpressions in the sum over all graphs. In a more familiar language, it exploits the fact that the calculation of a tree-level amplitude (which for us is the exact result) can be reduced to recursively solving equations of motion in the presence of external sources.

Consider the following Lagrangian:
\begin{equation}
L(\{\phi_a\}) = \frac{1}{2} C_{ab} \phi_a \phi_b + \frac{\lambda }{6}\, D_{abc} \phi_a \phi_b \phi_c
\,,\end{equation}
where $\lambda$ is a dummy coupling constant and $\phi_a$ are simple fields with no space-time dependence (\ie\ field theory in $0+0$ dimensions). The generating functional $Z(\{J_a \})$ of this theory is\footnote{The factor of $C_{ab}$ is included in the source term such that $J_a$ is a source for $\phi_a$, not a source for $C^{-1}_{ab} \phi_b$.}
\begin{equation}
\label{eq:Z_def}
Z(\{J_a \}) = - L(\{\phi_a\}) + \phi_a C_{ab} J_b
\,,\end{equation}
where the fields $\phi_a$ satisfy the variational condition (the equations of motion)
\begin{equation}
\label{eq:phi_eom}
\frac{\delta L}{\delta \phi_a}
= C_{ab} J_b
\,.\end{equation}
The $n$-body correlation function is equal to
\begin{equation}
\label{eq:correlator}
\langle \phi_1 \phi_2 \cdots \phi_n \rangle
= \lim_{J_i \to 0} \frac{\partial^n Z(\{J_a \})}{\partial J_1\partial J_2 \cdots \partial J_n}
\,,\end{equation}
and is identical to the sum over all Feynman diagrams (or in our case, the sum over all parton shower histories) with propagators $C_{ab}^{-1}$, vertices $D_{abc}$, and $n$ external legs.

To calculate $Z$ at tree level, we can solve \eq{eq:phi_eom} for the fields $\phi_a$. We start by expanding $\phi_a$ in a series in $\lambda$
\begin{equation}
\label{eq:phi_expanded}
\phi_a = \phi_a^0 + \lambda\, \phi_a^1 + \dotsb + \lambda^{n-2}\, \phi_a^{n-2}
\,,\end{equation}
where $\phi_a^i$ is the coefficient of $\phi_a$ at order $\lambda^i$, and the series need only go to order $\lambda^{n-2}$, because an $n$-point correlator needs only $n-2$ three-point interactions. The equation of motion following from \eq{eq:phi_eom} is
\begin{equation}
\phi_a = J_a - \frac{\lambda}{2}\, C^{-1}_{ab} D_{bcd}\, \phi_c \phi_d
\,,\end{equation}
which we can easily solve recursively order by order in $\lambda$ using \eq{eq:phi_expanded}. The result is
\begin{align}
\label{eq:phi_gensol}
\phi_a^0 &= J_a
\,,\nn\\
\phi_a^k
&= -\frac{1}{2} C_{ab}^{-1} D_{bcd}\,
 \bigl(\phi_c^0 \phi_d^{k-1} + \phi_c^1 \phi_d^{k-2} + \dotsb + \phi_c^{k-1} \phi_d^0 \bigr)
= -\frac{1}{2} C_{ab}^{-1} D_{bcd} \sum_{i + j = k - 1} \phi_c^i \phi_d^j
\,,\end{align}
which can then be plugged back into \eq{eq:Z_def}.

In the special case that $\mathcal{L}$ is only linear in the $\phi_a$ fields, the expression for the correlator in \eq{eq:correlator} simplifies. For $n$ external fields $\phi_1, \phi_2, \cdots, \phi_n$:
\begin{equation}
\label{eq:correlatorsimple}
\langle \phi_1 \cdots \phi_n \rangle
= Z(J_1 = 1, J_2 = 1, \ldots, J_n = 1, J_\lab{other} = 0)
\,.\end{equation}
We will see below that the ``Lagrangian'' relevant for \GenEvA\ is linear in the $\phi_a$, so \eq{eq:correlatorsimple} does indeed hold.

\subsection{A Shower ``Lagrangian''}

To build a Lagrangian useful for the \GenEvA\ algorithm, we introduce a separate field $\phi$ for every particle in a parton shower history (both for internal and external fields) and label each field by the specific four momentum $p_a$ and additional quantum numbers $f_a$ (flavor, spin, color, etc.) of the particle it represents,
\begin{equation}
\phi = \phi_{p_a, f_a} \equiv \phi_a
\,.\end{equation}
We stress that $p_a$ and $f_a$ are just labels; the field itself does not carry any momentum or quantum numbers. Note also that as far as the parton shower is concerned, all particles are distinguishable by their ``$L$'' and ``$R$'' labels. In addition, we define all momenta as outgoing and introduce two separate fields for the same particle going into a vertex and coming out of a vertex, with the convention that if $\phi_a$ represents an outgoing particle with outgoing momentum $p_a$ then $\phi_{-a}$ represents the same particle incoming with outgoing momentum $-p_a$. The reason to have two fields for every particle is in order to use the result of \eq{eq:correlatorsimple} which requires $\mathcal{L}$ to be linear in $\phi_a$.

Assume we have a parton shower history with $n$ external (incoming or outgoing) particles with momenta $P_i$, where $i$ runs from $1$ to $n$. We again define all $P_i$ as outgoing, such that overall momentum conservation reads $\sum_i P_i = 0$. For example, for a final state parton shower with $n - 1$ outgoing partons we include an ``outgoing'' center-of-mass particle with momentum
\begin{equation}
P_n = - \sum_{i=1}^{n-1} P_i
\,.\end{equation}
The crucial point is that the momenta $P_i$ span all possible momenta $p_a$ that can occur in any parton shower history with the same set of external particles. Explicitly, we introduce $2^n - 2$ different fields $\phi_a$ corresponding to the momenta
\begin{equation}
\label{eq:pa_def}
p_a = \sum_{i = 1}^n c^i_a P_i \qquad \text{with} \qquad c^i_a = 0 \text{ or } 1
\,,\end{equation}
excluding the case where all $c^i_a$ are $0$ or $1$, which would give $p_a = 0$. The index $a$ runs over $1 \leq \lvert a \rvert \leq 2^{n - 1} - 1$, and by definition we take $c^i_{-a} = 1 - c^i_a$ such that
\begin{equation}
p_{-a} = \sum_{i = 1}^n (1 - c^i_a) P_i = 0 - \sum_{i = 1}^n c^i_a P_i = - p_a
\,,\end{equation}
as defined above. For simplicity, we also define $c^i_a = \delta^i_a$ for $1 \leq a \leq n$ such that the first $n$ fields $\phi_a$ represent the external particles of the history.%
\footnote{Except for the trivial case $n = 2$ where $p_1 = P_1$ and $p_{-1} = P_2$.}
Hence, while the number of parton shower histories grows like $n!$, the number of $\phi_a$ fields needed to calculate the sum over parton shower histories only grows like $2^n$, and is therefore computationally tractable for reasonable numbers of external particles.

To include the additional quantum numbers $f_a$ in the above discussion, we can replace $p_a \to (p, f)_a$ and $P_i \to (P, f)_i$ everywhere, supplementing \eq{eq:pa_def} with appropriate quantum number information (\eg\ to include flavors). We only need to include a field $\phi_a$ if there is at least one way to perform the specific sum over $f$. For example, if we have two outgoing quarks with momenta $P_1$ and $P_2$ then we do not need a field with momentum label $p_{12} = P_1 + P_2$ because there is no particle that can decay into two quarks. On the other hand, we can add two quarks and an antiquark of the same flavor by first adding the quark and the antiquark to give a gluon which can be added to the remaining quark.%
\footnote{In practice, the required fields $\phi_a$ are constructed recursively by always combining two previously constructed fields, so adding flavors is trivial and requires no additional computational cost.}

The addition of momenta is always unique. However, including additional quantum numbers, the final result of the sum over $f$ can be ambiguous. For instance, including QED radiation and ignoring color, adding a quark and an antiquark can give a gluon or a photon. In this case, we either have to introduce a separate field $\phi$ for each final result, or alternatively, include the additional quantum number, color in this case, that breaks the ambiguity.

With the above conventions, momentum conservation in the parton shower requires
\begin{equation}
\label{eq:CD_conditions}
C_{ab} = C(a) \delta_{a,-b}
\,,\quad
C^{-1}_{ab} = \frac{1}{C(a)} \delta_{a,-b}
\,,\qquad
D_{abc} = 0 \quad \text{for} \quad p_a + p_b + p_c \neq 0
\,.\end{equation}
The first condition on the propagator means that an outgoing particle with $p_a$ can only connect to its incoming version with outgoing momentum $p_{-a} = -p_a$, while the second condition enforces momentum conservation at each vertex in the parton shower history. Overall momentum conservation now implies that $Z$ can only be linear in the sources for the external fields so \eq{eq:correlatorsimple} holds.%
\footnote{To see this, assume that $Z$ is quadratic or higher in one of the sources for the external fields with momentum $P_i$. This implies that there exists a correlator which has two or more copies of one of the original external fields. By construction, this cannot be the case, because it would require at least one field with momentum label containing $2P_i$ for some $i$.}
Note that this relies crucially on the fact that we introduced separate fields for incoming and outgoing particles, such that $C_{ab}$ would be off-diagonal.

To obtain the desired sum over all parton shower histories we can now simply use \eq{eq:correlatorsimple} to plug the expansion of the fields in \eq{eq:phi_expanded} back into \eq{eq:Z_def} and extract the order $\lambda^{n - 2}$ piece:
\begin{equation}
\begin{split}
\label{eq:Z_finalsol}
\sum_i \alpha(\vS_i)
&\equiv \langle \phi_1 \cdots \phi_n \rangle
\\
&= -\frac{1}{2} \sum_{i+j= n-2} C(a) \phi_a^i \phi_{-a}^j
 -\frac{1}{6} \sum_{i+j+k=n-3} D_{abc} \phi_a^i \phi_b^j \phi_c^k
 + C(a) \phi_a^0 \phi_{-a}^{n-2}
\,,\end{split}
\end{equation}
where a sum over $a,b,c$ is understood and we also used \eq{eq:CD_conditions} for $C_{ab}$. The expressions for the fields $\phi_a^k$ in \eq{eq:Z_finalsol} are obtained from \eqs{eq:phi_gensol}{eq:CD_conditions} and setting $J_{1,\ldots,n} = 1$ and all other sources to zero,
\begin{align}
\label{eq:phi_finalsol}
\phi_a^0 &=
\left\{\begin{aligned}
1 & \qquad \text{for} \quad 1 \leq a \leq n \,,\\
0 & \qquad \text{otherwise}
\,,\end{aligned}\right.
\nn\\
\phi_a^k
&= -\frac{1}{2C(a)} D_{-a,bc} \sum_{i + j = k - 1} \phi_b^i \phi_c^j
\,.\end{align}
Once $C(a)$ and $D_{abc}$ are specified, then \eq{eq:Z_finalsol} together with \eq{eq:phi_finalsol} is our final result for $\sum_i \alpha(\vS_i)$. To obtain $\alpha(\vS)$ for a particular parton shower history (\ie\ the numerator in \eq{eq:alpha_general}) we can simply multiply the appropriate factors of $1/C(a)$ and $D_{abc}$ for the given parton shower history. We will discuss the precise from of $C_a$ and the nonzero $D_{abc}$ next.

\subsection{Choice of Overcounting Function}
\label{subsec:ocfunc}

As discussed in \sec{subsec:ocoverview}, ideally $\alpha(\vS)$ should equal the probability for generating a given history, but if we want to use the \texttt{ALPHA} algorithm outlined above, we have to be in a situation where that probability can be cast into a propagator $C(a)^{-1}$ and a local vertex $D_{abc}$, which can only depend on the momenta and quantum numbers associated with the indices $a, b, c$. As we have seen in \sec{sec:ps}, the splitting of a mother into two daughters requires information from the grandmother, which cannot be encoded in the local vertex $D_{abc}$. Also, from the computational point of view, calculating the probability for a history requires numerically expensive Sudakov factors. Therefore, the compromise we will take is to choose
\begin{equation}
C(a) = 1
\end{equation}
and have all nonzero $D_{abc}$ equal to a Jacobian-improved splitting function.

We stress again that we are free to make any choice for $C(a)$ and the nonzero $D_{abc}$. For example, choosing $C(a) = 1$ and $D_{abc} = 1$ the algorithm simply counts the number of parton shower histories, which is equivalent to using \eq{eq:alpha_trivial} and provides a cross-check on the implementation. The choice of overcounting ``Lagrangian'' only determines the efficiency of the program and not the resulting distributions.

By definition, if $D_{abc}$ is nonzero then the flavors $f_a, f_b, f_c$ form a valid splitting with $p_a + p_b + p_c = 0$. In a parton shower, the mother particle is treated differently from the daughters, so to determine which of the three particles acts as the mother of the splitting, we calculate the invariant masses $t_{a,b,c} = p_{a,b,c}^2$. Since the parton shower is ordered in $t$, we know that the largest of $t_{a,b,c}$ belongs to the mother. Assuming for simplicity that this is $t_a$, we use
\begin{equation}
\label{eq:Dabc_choice}
D_{abc} = f_{f_a \to f_b f_c}(t_a, \cos\theta_a,\phi_a)\, J(t_a; t_b, t_c)
\,.\end{equation}
where $J(t_a;t_b,t_c)$ is the Jacobian defined in \eq{eq:Jindividual} and $f_{f_a \to f_b f_c}$ is the splitting function defined in Section \ref{subsec:ChoiceofKinematicVariables}.\footnote{There is a technical issue that the splitting functions are functions of $E^\start_M$, so strictly speaking there is not enough local information to even determine the correct splitting functions, because in the shower algorithm we set $E_M^\start$ to the maximum energy of the grandmother. However, regardless of which $E^\start_M$ we choose to define $D_{abc}$, the singularities will still be correct. In practice, we calculate $E^\start_M$ assuming that the mother's sister is unbranched and that the grandmother's rest frame is the CM frame.} Since this choice captures all the singularities of QCD, we can expect the corresponding $\alpha(\vS)$ to be reasonably close to $\cP(\vS) J(\vS)$.

If $t_a = E_\lab{CM}^2$, then the splitting corresponds to the initial hard interaction, in which case we use
\begin{equation}
D_{abc}^\lab{CM} = J(t_a; t_b, t_c)
\,.\end{equation}
We can also add extra matrix element information to $D_{abc}^\lab{CM}$ such as angular information or charges. In practice this is only important if there are two or more different hard-interaction vertices that can contribute to the same final state, such as $e^+ e^- \to u \bar{u} d\bar{d}$, which can have both a $e^+ e^- \to u \bar{u}$ and $e^+ e^- \to d \bar{d}$ core.

\section{The Matrix Element}
\label{sec:me}

The last missing ingredient in calculating the weight for a given parton shower history is the desired distribution $\sigma[\Phi(\vS)]$. The appropriate choice for this distribution is entirely determined by physics considerations, and one should implement the best possible theoretical calculations available. Since this work focuses on the generation of the phase space, we will only consider two simple theoretical distributions as examples and discuss more elaborate choices in the companion paper~\cite{genevaphysics}.

We briefly discuss phase space issues for tree-level (LO) matrix elements $\sigma(\Phi)$. We then consider tree-level matrix elements improved to leading-logarithmic accuracy (LO/LL). This corresponds to the type of calculation one does in parton shower/matrix element merging~\cite{Catani:2001cc, Lonnblad:2001iq, Krauss:2002up, MLM, Mrenna:2003if, Schalicke:2005nv, Lavesson:2005xu, Hoche:2006ph, Alwall:2007fs, Giele:2007di, Lavesson:2007uu, Nagy:2007ty, Nagy:2008ns}. To arrive at a simple expression for the LO/LL cross section, we will use the notion of $\sigma_i(\Phi)$ and the alternative form of the event weight in \eq{eq:finalweight_i}.

\subsection{Tree-Level Matrix Elements}
\label{subsec:lo}

The simplest theoretical distributions one could use in an event generator are given by fixed-order tree-level matrix elements. These results depend only on the scalar products of the four-momenta of the external particles and are defined entirely by Lorentz-invariant phase space variables $\Phi$. While such tree-level calculations give at best a rudimentary description of the true QCD distributions, we will use them to validate our phase space generator in \sec{subsec:madcompare} and test its efficiency in \sec{subsec:efficiency}.

We want to touch on two additional subtleties having to do with phase space vetoes. Recall that at tree level, there are singularities in QCD matrix elements when two partons get close in $t_{ij} = (p_i + p_j)^2$. In full QCD, these singularities are regulated by Sudakov factors, but a tree-level event generator has to cut off these $t_{ij}$ singularities by hand. In the context of the \GenEvA\ algorithm, the natural phase space cut variable is virtuality. After all, if we run the shower down to $t_\cut$, then the shower never encounters the singular region of QCD.

\begin{figure}[t]
\includegraphics[width=0.8\columnwidth]{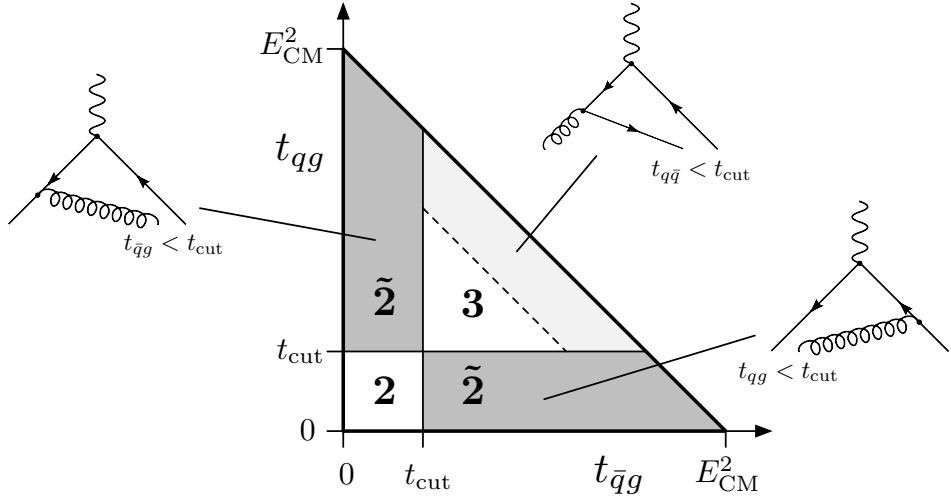}
\caption{
3-body phase space for $e^+e^-\to q\bar{q}g$. Region $\mathbf{2}$ corresponds to the 2-jet region because the parton shower will not populate regions with both $t_{qg}$ and $t_{\bar{q}g}$ less than $t_\cut$. However, the parton shower can ``accidentally'' distribute events in the $\mathbf{\tilde{2}}$ regions where either $t_{qg}$ or $t_{\bar{q}g}$ is less than $t_\cut$. These regions must be vetoed because they occupy a QCD singularity away from a shower singularity. While there is no QCD singularity associated with $t_{q\bar{q}} < t_\cut$, it is convenient to also veto that region for simplicity (\eg\ when comparing to \texttt{MadEvent} in \sec{subsec:madcompare}). The remaining part of phase space defines the 3-jet region $\mathbf{3}$.}
\label{fig:supertruncate}
\end{figure}

However, when we run the shower, it is possible for two particles to ``accidentally'' have virtualities smaller than $t_\cut$, and the matrix element will return a singular answer. This sickness also appears in the overcounting factor, because the \aalpha\ algorithm will encounter a value of $D_{abc}$ that cannot be generated by a shower with a $t_\cut$ ending scale. Consider $3$-body phase space for $e^+e^-\to q\bar{q}g$ as in \fig{fig:supertruncate}. There are regions of phase space for which a large angle emission from $q \to q g$ can put the gluon closer to the anti-quark than $t_\cut$. The $\bar{q} \to \bar{q} g$ splitting cannot generate those same four-momenta because it is below the cutoff scale, so there is no ``partner'' shower history for the event. Moreover, the shower generated those four-momenta far away from any QCD singularity, while the four-momenta are indeed in the singular region, so the event weight will be very large.

The solution, of course, is to veto events from the shower for which the minimum $t_{ij}$ between singularity-producing partons is less than $t_\cut$. Note that we need not veto events for which the $q$ and $\bar{q}$ are closer than $t_\cut$, because in the context of $e^+ e^- \to q \bar{q} g$, there is no $q\bar{q}$ singularity.\footnote{There is a singularity in $e^+ e^- \to q \bar{q} q' \bar{q}'$, of course.} If for simplicity we do not wish to impose a flavor-dependent $t_\cut$ scale, then we can always veto events where any pair of particles (singular or not) are closer than $t_\cut$. Denoting  the fixed-order tree-level matrix elements by $\sigma^\lab{tree}(\Phi)$, this is equivalent to taking
\begin{equation}
\label{eq:losigma}
\sigma^{\rm LO}(\Phi) = \sigma^{\rm tree}(\Phi)\, \prod_{ij} \theta(t_{ij} > t_\cut)
\,.\end{equation}
Using \eq{eq:finalweight}, the corresponding event weight is
\begin{equation}
\label{eq:loweight}
w^{\rm LO}(\vS) = \frac{\sigma^{\rm tree}[\Phi(\vS)]\, \hat{\alpha}(\vS)}{\cP(\vS)J(\vS)}\,
\prod_{ij} \theta(t_{ij} > t_\cut)
\,,\end{equation}
which is what we will use in \sec{subsec:madcompare}.

The  second subtlety has to do with the fact that the scale $t_\match$ is ambiguous for $n_\max$-body phase space. Recall from \sec{subsec:truncation} that to create fully exclusive events with no dead zones, we want to start a parton shower approximation at $t_\match$. When $n_\max$-body phase space is generated, the scale $t_\match$ will vary on an event-by-event basis to ensure that the attached parton shower will populate all of $(n > n_\max)$-body phase space. However, since the same $n_\max$ four-momenta can be created by different shower histories during the phase space generation, they can be associated with several different $t_\match$ scales.

As far as phase space vetoes are concerned, the variable $t_\match$ can simply be ignored, because even after truncating to $\{t_\cut,n_\max\}$, the shower does populate all of $n_\max$-body phase space down to $t_\cut$. However, if we start a parton shower approximation at $t_\match$, we do have to be mindful that events having the same $n_\max$ partonic four-momenta will start the shower at different scales. Far from being a bug, this $t_\match$ ambiguity will be exploited in the next subsection to create the leading-logarithmically improved event sample. As we discuss in the companion paper \cite{genevaphysics}, this ambiguity is a physical ambiguity that has to be resolved with calculations.

\subsection{Logarithmically-Improved Matrix Elements}
\label{subsec:logimprove}

It is well known that fixed-order calculations do not give an adequate description of true QCD distributions in singular regions of phase space. This is because fixed-order calculations only reproduce the power singularities that occur when intermediate propagators go on shell. However, in addition to these power singularities there are logarithmically enhanced terms, which start to show up one higher order in perturbation theory. These logarithmically enhanced terms have the form $[\alpha_s \log^2(t/E_\CM^2)]^n$. Thus, the effective expansion parameter in the singular regions of phase space is not $\alpha_s$, but rather $\alpha_s \log^2(t/E_\CM^2)$. These double-logarithmic terms can be resummed to all orders in perturbation theory, giving an effective expansion parameter $\alpha_s \log(t/E_\CM^2)$, which is only enhanced by a single logarithm. However, such resummed calculations are more difficult to implement, since they not only depend on the four-momenta of the external particles, but also on a choice of scales which determine the exact form of the resummation.

Resummed calculations are crucial if one wants to merge partonic calculations with parton shower algorithms to produce fully exclusive events, as required for meaningful comparisons with experimental collider data. The parton shower automatically includes the resummation of the double-logarithmic terms via the Sudakov factor discussed in \eq{eq:sudakov}, and therefore has double-logarithmic sensitivity to the scale at which it is started. Thus, the absence of resummed partonic calculations leads to a double-logarithmic dependence on the unphysical matching scale $t_\match$ between the partonic calculation and the parton shower. This issue is discussed in much more detail in the companion paper~\cite{genevaphysics}.

Most implementations of resummed calculations in event generators define the relevant scales in terms of the external four-momenta. An example is the so-called CKKW prescription~\cite{Catani:2001cc} to merge fixed order calculations with parton showers. In this prescription, one first generates an event based on tree-level matrix elements and then uses a flavor-aware $k_T$ algorithm on the external particles to find the most likely parton shower history that could have generated this event. The resummed partonic result is then obtained by reweighting the fixed-order event by the appropriate no-branching probabilities associated with this specific parton shower history.\footnote{CKKW also corrects for the running of $\alpha_s$.}

There are two main issues with this approach and other similar ones. First, this kind of algorithm only allows partonic calculations which are the simple product of a tree-level matrix element and Sudakov factors. In general, the resummed expression can contain various terms with differing logarithmic structures. A recent study~\cite{Bauer:2006qp,Bauer:2006mk} of the production of highly energetic partons in soft-collinear effective theory (SCET)~\cite{Bauer:2000ew,Bauer:2000yr,Bauer:2001ct,Bauer:2001yt} has shown that the logarithmic structure is obtained by a sequence of matching and running calculations, which leads to a sum over terms, each with a different resummation of logarithms. These different terms can be identified with the different possible shower histories, which is due to the fact that the effective theory and the parton shower are both derived from the soft-collinear limit of QCD.

Second, reweighting tree-level calculations by Sudakov factors may result in inefficient event generators. This is because the product of many Sudakov factors can give numerically very different results for different events depending on the precise scales found by the $k_T$ algorithm or any other jet identification procedure. Thus, if the initial set of events was distributed only with the correct power-singularity structure, the resulting events after including logarithmic information can have widely differing event weights, leading to a poor statistical efficiency. Of course, one can easily avoid this issue by accounting for both the power and logarithmic singularities at the time of initial event generation.

In  \geneva\, both of these issues are solved automatically, because the underlying phase space generator is itself a parton shower. First, we can assign different weights for different parton shower histories, allowing an easy implementation of the sum over different logarithmically resummed terms. Second, the events are directly generated based on the singularity structure of QCD and already include the leading-logarithmic effects. Thus, reweighting the parton shower to resummed partonic expressions will actually give much more uniform weights than reweighting to fixed-order matrix elements. We will come back to this issue in \sec{subsec:efficiency} when discussing the efficiency of \geneva\, and focus for now on how to implement resummed partonic calculations in \geneva.

A particularly useful expression for resummed partonic distributions is given in the companion paper~\cite{genevaphysics}. Starting with \eq{eq:finalweight_i}, note that the parton shower by itself is creating events with $w(\vS) = 1$. Therefore, the internal \GenEvA\ parton shower distributes events with an effective $\vS$-dependent ``matrix element''
\begin{equation}
\sigma_i^{\rm shower}[\Phi(\vS)] = \cP(\vS) J(\vS) \equiv Q(\vS) \Delta(\vS)
\,,\end{equation}
where $Q(\vS)$ is the product of  a hard scattering matrix element, splitting functions, and Jacobians, and $\Delta(\vS)$ is the product of Sudakov factors. Starting  from the fixed-order tree-level matrix elements $\sigma^{\rm tree}(\Phi)$, an LO/LL sample can be defined by (see Eq.~$(111)$ in Ref.~\cite{genevaphysics})
\begin{equation}
\label{eq:lollsigma}
\sigma_i^{\rm LO/LL}[\Phi(\vS)]
= \sigma^{\rm tree}[\Phi(\vS)]\, \frac{Q(\vS)}{\sum_i Q[\vS_i(\Phi)]}\, \Delta(\vS)
\,,\end{equation}
where the sum runs over all shower histories $\vS_i(\Phi)$ that yield the same phase space point $\Phi$. Therefore, the formula for the LO/LL event weight is quite  simple%
\footnote{Since we use the \aalpha\ algorithm as described in \sec{sec:oc} to compute the sum over parton shower histories, the precise splitting functions $Q(\vS)$ in \eq{eq:lollsigma} are those used in \eq{eq:Dabc_choice}. The actual splitting functions $\tilde{Q}(\vS)$ used in the shower slightly differ from these by the choice of $E^\max$. Therefore, in practice, \eq{eq:lollweight} contains an additional factor $Q(\vS)/\tilde{Q}(\vS)$.}
\begin{equation}
\label{eq:lollweight}
w^{\rm LO/LL}(\vS)
= \frac{\sigma^{\rm LO/LL}_i[\Phi(\vS)]}{\cP(\vS) J(\vS)}
= \frac{\sigma^{\rm tree}[\Phi(\vS)]}{\sum_i Q[\vS_i(\Phi)]}
\,.\end{equation}

Although  \eq{eq:lollsigma} includes a Sudakov factor, the LO/LL event weight no longer has any Sudakov information in it, because the internal \GenEvA\ shower already distributes events with the correct double-logarithmic information. Since the Sudakov factors $\Delta(\vS)$ contained in $\cP(\vS)$ in the LO weight in \eq{eq:loweight} come at some computational cost, creating this LO/LL merged sample is computationally even more efficient than the LO result alone.

Note the difference between this result and CKKW. Here, we are using every shower history to define an appropriate Sudakov factor. The relative contribution of each Sudakov factor is given by
\begin{equation}
\frac{Q(\vS)}{\sum_i Q[\vS_i(\Phi)]}
\,.\end{equation}
In those regions of phase space that are dominated by one shower history, one $Q(\vS) \to \infty$ and therefore one Sudakov dominates. This corresponds to the approach of Ref.~\cite{Lonnblad:2001iq}. In CKKW~\cite{Catani:2001cc}, the chosen Sudakov factor always corresponds to the dominant history, whereas in \eq{eq:lollsigma}, away from the singular regions, an average Sudakov is used, more closely mimicking the SCET calculation \cite{Bauer:2006qp,Bauer:2006mk}.

It would be interesting to study which approach better corresponds to experimental data, though we emphasize that the \GenEvA\ algorithm can of course distribute events that are morally equivalent to the CKKW procedure, by simply choosing the partonic cross section
\begin{equation}
\sigma^{\rm CKKW}(\Phi) = \sigma^{\rm tree} (\Phi)\, \Delta[\vS_{\rm dom}(\Phi)]
\,,\end{equation}
where $\vS_{\rm dom}(\Phi)$ is the dominant shower history for the phase space point $\Phi$.

\subsection{Further Logarithmic Possibilities}

\GenEvA\ also allows a numerical treatment of an interesting single-logarithmic issue. In the LO matrix element \eq{eq:losigma}, we vetoed events for which two partons were ``accidentally'' closer than $t_\cut$. This made sense, because we were not using the Sudakov information in the LO calculation. However, imagine the case that the tree-level matrix element were identically equal to the product of splitting functions
\begin{equation}
\sigma^{\rm tree}[\Phi(\vS)] \to \sum_i Q[\vS_i(\Phi)]
\,,\end{equation}
such that the event weight in \eq{eq:lollweight} is always one. Because the parton shower preserves probability, the $t_{ij} < t_\cut$ veto will now decrease the total cross section for $e^+ e^- \to n \text{ jets}$ from the Born value. Putting in the correct expression for $\sigma^{\rm tree}(\Phi)$, this will result in a single-logarithmic decrease in the cross section. While this effect is formally beyond LO/LL order, it is numerically important for high-multiplicity matrix elements.

As  we argued in \fig{fig:supertruncate}, events that violate the $t_\cut$ constraint should be thought of as actually contributing to lower-dimensional phase space. Imagine we take an event sample and truncate to $\{t_\cut,n_\max\}$. As explained earlier, this gives $(n < n_\max)$-parton events with $t_\match = t_\cut$, and $n_\max$-parton events for which $t_\match$ changes from event to event. Even after truncation, there will be some $n$-parton events with $3 \leq n \leq n_\max$ left that violate the $t_\cut$ constraint. Instead of vetoing these ``rogue'' events, we can further truncate them to some scale $t' > t_\match$ until the $t_\cut$ violation disappears. This will take all rogue $n$-parton events and make them into $(<n)$-parton events, recovering the cross section that would otherwise be lost to the $t_\cut$ veto. We will refer this additional truncation up to $t'$ as ``truncation-prime''.

Remember that these events were originally truncated to $\{t_\cut,n_\max\}$, so we have to restart the shower at the originally determined $t_\match$ scale (and not $t'$), otherwise we would double-count existing regions of phase space from the ordinary $n$-parton events. If the probability of getting a given $n$-body shower history $\vS_n$ after truncation is $\cP^\lab{trunc}(\vS_n)$, then the probability $\cP^{\rm trunc'}(\vS_n)$ to get the same shower history after truncation-prime is given schematically by
\begin{equation}
\label{eq:truncprimeprob}
\cP^{\rm trunc'}(\vS_n)
= \cP^\lab{trunc}(\vS_n) + \sum_ {i = 1}^{n_\max-n} \int_{\rm \mathbf{\tilde{n}}} \!\df(\vS_{n+i}/\vS_n)\, \cP^\lab{trunc}(\vS_{n+i})
\,,\end{equation}
where $\df(\vS_{n+i}/\vS_n)$ represents an integral over those $(n+i)$-body shower variables that are not shared by the $n$-body shower, and $\mathbf{\tilde{n}}$ represents those regions of $(n+i)$-body phase space that really look like $n$-jet events (\eg\ the $\mathbf{\tilde{2}}$ regions in \fig{fig:supertruncate}). Note that for $n = n_\max$, truncation-prime does not affect the event probability.

If we now use the same event weight as in \eq{eq:lollweight} for the resulting event sample after truncation-prime, then this is equivalent to choosing
\begin{equation}
\label{eq:supertruncsigma}
\sigma_i^{\rm LO/LL'}[\Phi(\vS)]
= \sigma^{\rm tree}[\Phi(\vS)]\, \frac{Q(\vS)}{\sum_i Q[\vS_i(\Phi)]}\, \Delta(\vS)\, \frac{\cP^{\rm trunc'}(\vS)}{\cP^\lab{trunc}(\vS)}
\,.\end{equation}
In other words, we can do the integrations in \eq{eq:truncprimeprob} numerically by truncating the shower to $t'$. This $\cP^{\rm trunc'}/\cP^\lab{trunc}$ enhancement is formally beyond leading-logarithmic order, so it is safe to include it. Numerically, it helps yield more stable total cross sections as $t_\cut$ is varied. The main downside to using the numerical $\cP^{\rm trunc'}/\cP^\lab{trunc}$ method is that it breaks the reversibility of further truncation. That is, once truncation-prime is applied, the probability for obtaining a set of four-momenta is proportional to $\cP^{\lab{trunc}'}$ and thus extremely convoluted, so one can no longer use the argument at the end of \sec{subsec:truncation} that reweighting can occur after hadronization.

\section{Results}
\label{sec:results}

In  this section, we show results for the process $e^+e^- \to n \text{ jets}$ using the partonic cross sections defined in the previous section. As input for the fixed-order tree-level result $\sigma^\lab{tree}(\Phi)$, we use \madgraph~\cite{Stelzer:1994ta} to obtain numerical \fortran\ \texttt{HELAS}~\cite{Murayama:1992gi} routines for $e^+ e^- \to n \text{ partons}$%
\footnote{For simplicity, we are only including the diagrams with $e^+ e^- \to \gamma^* \to \text{partons}$. An intermediate $Z$-boson could easily be included.},
where $2 \leq n \leq 6$, and a ``parton'' is either a gluon or a $u$, $d$, $s$, or $c$ quark (neglecting quark masses). In this section, we always use the center-of-mass energy and phase space cut
\begin{equation}
E_\CM = 1000 \GeV
\,, \qquad
\sqrt{t_\cut} = 100 \GeV
\,,\end{equation}
unless otherwise noted.

We  first validate the \GenEvA\ algorithm by using the simple LO cross section from \eq{eq:losigma} and comparing the output of \geneva\ with that of \madevent~\cite{Maltoni:2002qb}. We then show some rudimentary results using the LO/LL cross sections in \eqs{eq:lollsigma}{eq:supertruncsigma}, with a more thorough discussion given in the companion paper \cite{genevaphysics}. Finally, \madevent\ offers a useful benchmark to compare the efficiency of the \GenEvA\ algorithm. We will see that \GenEvA's efficiency is at or above \madevent\ levels, and that \GenEvA\ itself is more efficient at distributing  logarithmically-improved LO/LL results than pure tree-level LO results.

\subsection{Comparison with MadEvent}
\label{subsec:madcompare}

Since  \geneva\ uses the same \madgraph/\texttt{HELAS} matrix-element engine as \madevent, both programs should and do give identical results. For the purpose of this comparison we are not attaching any additional parton shower emissions, and simply compare the partonic cross sections for a fixed number of final state partons.

\begin{table}[t]
 \begin{tabular}{c | c | c || c | c | c}
 process & \madevent & \geneva &process & \madevent & \geneva \\
 \toprule
 LO 3 (fb) & $216.71\phantom{0} \pm 0.21\phantom{0}$ & $216.77\phantom{0} \pm 0.22\phantom{0}$ & LO 5 (ab) & $2542\phantom{.000} \pm 3\phantom{.000^{-}}$ & $2543\phantom{.000} \pm 3\phantom{.000}$ \\
 \hline
 $u \bar u g$ & $\phantom{0}86.62\phantom{0} \pm 0.13\phantom{0}$ & $\phantom{0}86.60\phantom{0}\pm 0.18\phantom{0}$ & $u \bar u ggg$ & $\phantom{0}912\phantom{.000} \pm 2\phantom{.000^{-}}$ & $\phantom{0}912\phantom{.000} \pm 2\phantom{.000}$ \\
 $d \bar d g$ & $\phantom{0}21.75\phantom{0} \pm 0.07\phantom{0}$ & $\phantom{0}21.55\phantom{0}\pm 0.10\phantom{0}$ & $d \bar d ggg$ & $\phantom{0}227.5\phantom{00} \pm 0.9\phantom{00^{-}}$ & $\phantom{0}228.3\phantom{00}\pm 0.8\phantom{00}$ \\
 $s \bar s g$ & $\phantom{0}21.63\phantom{0} \pm 0.06\phantom{0}$ & $\phantom{0}21.73\phantom{0}\pm 0.10\phantom{0}$ & $u \bar u d \bar d g$ & $\phantom{00}33.8\phantom{00} \pm0.2\phantom{00^{-}}$ & $\phantom{00}34.3\phantom{00} \pm 0.4\phantom{00}$ \\
 $c \bar c g$ & $\phantom{0}86.71\phantom{0} \pm 0.13\phantom{0}$ & $\phantom{0}86.70\phantom{0}\pm 0.18\phantom{0}$ & $u \bar u u \bar u g$ & $\phantom{00}25.6\phantom{00} \pm 0.2\phantom{00^{-}}$ & $\phantom{00}25.7\phantom{00} \pm 0.3\phantom{00}$ \\
 \hline\hline
 LO 4 (fb)& $\phantom{0}36.44\phantom{0} \pm 0.04\phantom{0}$ & $\phantom{0}36.49\phantom{0} \pm 0.04\phantom{0}$ & LO 6 (ab) & $\phantom{00}67.9\phantom{00} \pm 0.3\phantom{00^{-}} $ & $\phantom{00}68.0\phantom{00} \pm 0.2\phantom{00}$ \\
 \hline
 $u \bar u g g$ & $ \phantom{0}14.00\phantom{0} \pm 0.03\phantom{0}$ & $ \phantom{0}14.00\phantom{0} \pm 0.02\phantom{0}$ & $u \bar u gggg$ & $\phantom{00}22.41\phantom{0}\pm0.09\phantom{0^{-}}$ & $\phantom{00}22.29\phantom{0}\pm 0.12\phantom{0} $ \\
 $d \bar d g g$ & $ \phantom{00}3.504 \pm 0.013$ & $ \phantom{00}3.511 \pm 0.011$ & $u \bar u u \bar u gg$ & $\phantom{000}1.117\pm0.006\phantom{^{-}}$ & $\phantom{000}1.14\phantom{0}\pm 0.03\phantom{0}$ \\
 $u \bar u d \bar d$ & $\phantom{00} 0.175 \pm 0.001$ & $\phantom{00}0.180 \pm 0.003$ & $u \bar u u \bar u u \bar u$ & $\phantom{000}0.005 \pm 0.001^{-}$ & $\phantom{000}0.005\pm 0.001$ \\
 $u \bar u u \bar u$ & $\phantom{00}0.132 \pm0.001$ & $\phantom{00}0.132 \pm 0.002$ & $u \bar u d \bar d s \bar s$ & $\phantom{000}0.019\pm 0.001^{-}$ & $\phantom{000}0.020\pm 0.005$ \\
 \hline
 \end{tabular}
\caption{Comparison of total cross sections between \geneva\ and \madevent. Both programs agree within statistical uncertainties ($0.001^{-}$ indicates an error too small to be reported by \madevent). In \madevent, the relative uncertainty on each sample is roughly constant, whereas in \geneva, less populated channels are allowed to have higher relative uncertainty. This happens because the ratio of various subprocesses is given in \GenEvA\ by the QCD symmetry structures, and no extra integration time is spent determining the ratio between different channels.}
\label{tab:crosssection}
\end{table}

We first consider the results for the total cross sections, which are shown in Table~\ref{tab:crosssection} for the $n$-parton samples combining all flavor subprocesses as well as for various individual flavor subprocesses. One can clearly see that the two programs agree within statistical uncertainties. Notice that while in \madevent\ the relative uncertainty for each subprocess is roughly the same, in \geneva\ the relative uncertainty scales as the square root of the cross section for the given subprocess. This is because \geneva\ automatically produces a sample containing the sum of all subprocesses in a single run, with the ratio of the various subprocesses roughly given by the QCD symmetry structures. This means that the most important channels automatically get the most statistics and thus the smallest relative uncertainties, and no extra integration time is spent on less important channels. In \madevent\ the individual subprocesses are generated separately and combined afterwards.

\begin{figure}[t]
\includegraphics[width=0.5\textwidth]{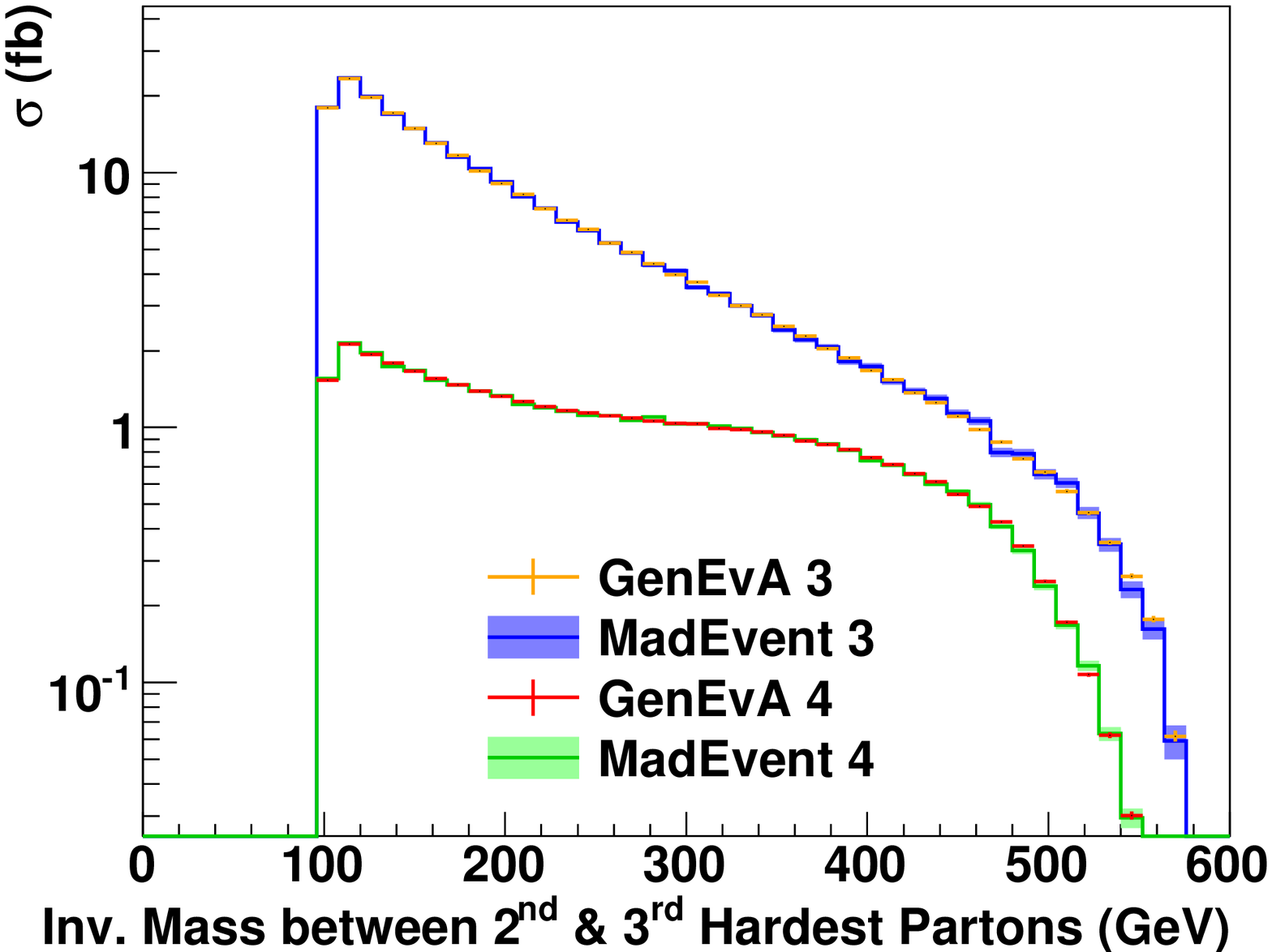}%
\hfill%
\includegraphics[width=0.5\textwidth]{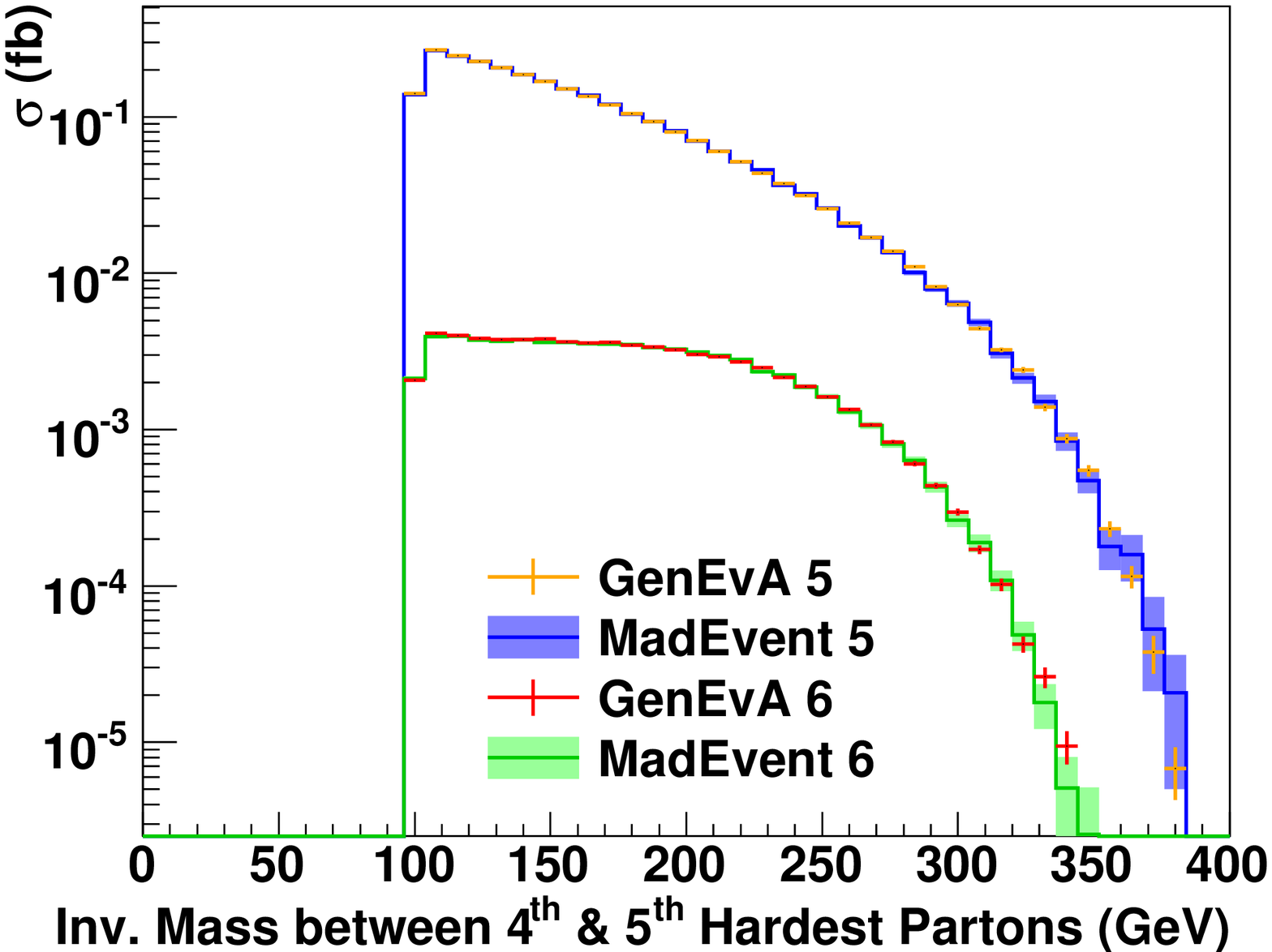}%
\caption{Comparison between \madevent\ and \GenEvA\ for 3-, 4-, 5-, and 6-parton tree-level matrix elements. Left panel: The pairwise invariant mass of the 2nd and 3rd hardest partons for 3- and 4-parton matrix elements. Right panel: The pairwise invariant mass of the 4th and 5th hardest partons for 5- and 6-parton matrix elements. The two programs agree within the statistical uncertainties, validating the \GenEvA\ algorithm since both programs use the same \texttt{HELAS}/\texttt{MadGraph} matrix elements. Note the hard cutoff at $\sqrt{t_\cut} = 100 \GeV$.}
\label{fig:comparison}
\end{figure}

Next, we compare the differential distributions generated by both programs for a few simple partonic observables. We have checked a large range of partonic observables and in all cases the distributions obtained from \geneva\ and \madevent\ agree perfectly within the statistical uncertainties, verifying that the \geneva\ phase space generator is working properly. As an illustration, we show in \fig{fig:comparison} the distribution of the pairwise invariant mass between the 2nd and 3rd hardest partons for 3- and 4-parton matrix elements (left panel) and between the 4th and 5th hardest partons for 5- and 6-parton matrix elements (right panel). Here, hardest is defined by the parton with the largest energy. We will discuss the different efficiencies of the two programs later on in \sec{subsec:efficiency}.

\subsection{Logarithmically-Improved Results}

The main usefulness of an LO/LL partonic calculation is for creating fully merged LO/LL samples interfaced with a parton shower. We do this in the companion paper \cite{genevaphysics} and here only give partonic results, which by themselves have only limited usefulness, but show the effect of Sudakov suppression.

From \eq{eq:lollsigma}, we see that the LO/LL partonic calculation depends on a Sudakov factor, which in turn depends on the $t_\match$ scale of the event after truncation. For an $n$-body matrix element with fixed $n$, there are two different cases, as discussed in \sec{subsec:truncation}: If $n = n_\max$, then $t_\match > t_\cut$ is an event-specific scale corresponding to the last emission and the result should be regarded as an inclusive sample. If $n < n_\max$, we always have $t_\match = t_\cut$ and the result should be regarded as exclusive. For $n < n_\max$, we also have the choice of whether to include the truncation-prime correction factor in \eq{eq:supertruncsigma}. These three possibilities are labeled as%
\footnote{In the companion paper \cite{genevaphysics}, we only show LO/LL inc.\ and LO/LL$'$ exc.\ results and will therefore remove the prime symbol from the LO/LL notation.}
\begin{align}
\text{LO/LL inc.\ $n$} &\equiv \text{\eq{eq:lollsigma} with $n = n_\max$}
\,,\nn\\
\text{LO/LL exc.\ $n$} &\equiv \text{\eq{eq:lollsigma} with $n < n_\max$}
\,,\nn\\
\text{LO/LL$'$ exc.\ $n$} &\equiv \text{\eq{eq:supertruncsigma} with $n < n_\max$}
\,.\end{align}

\begin{figure}[t]
\includegraphics[width=0.5\textwidth]{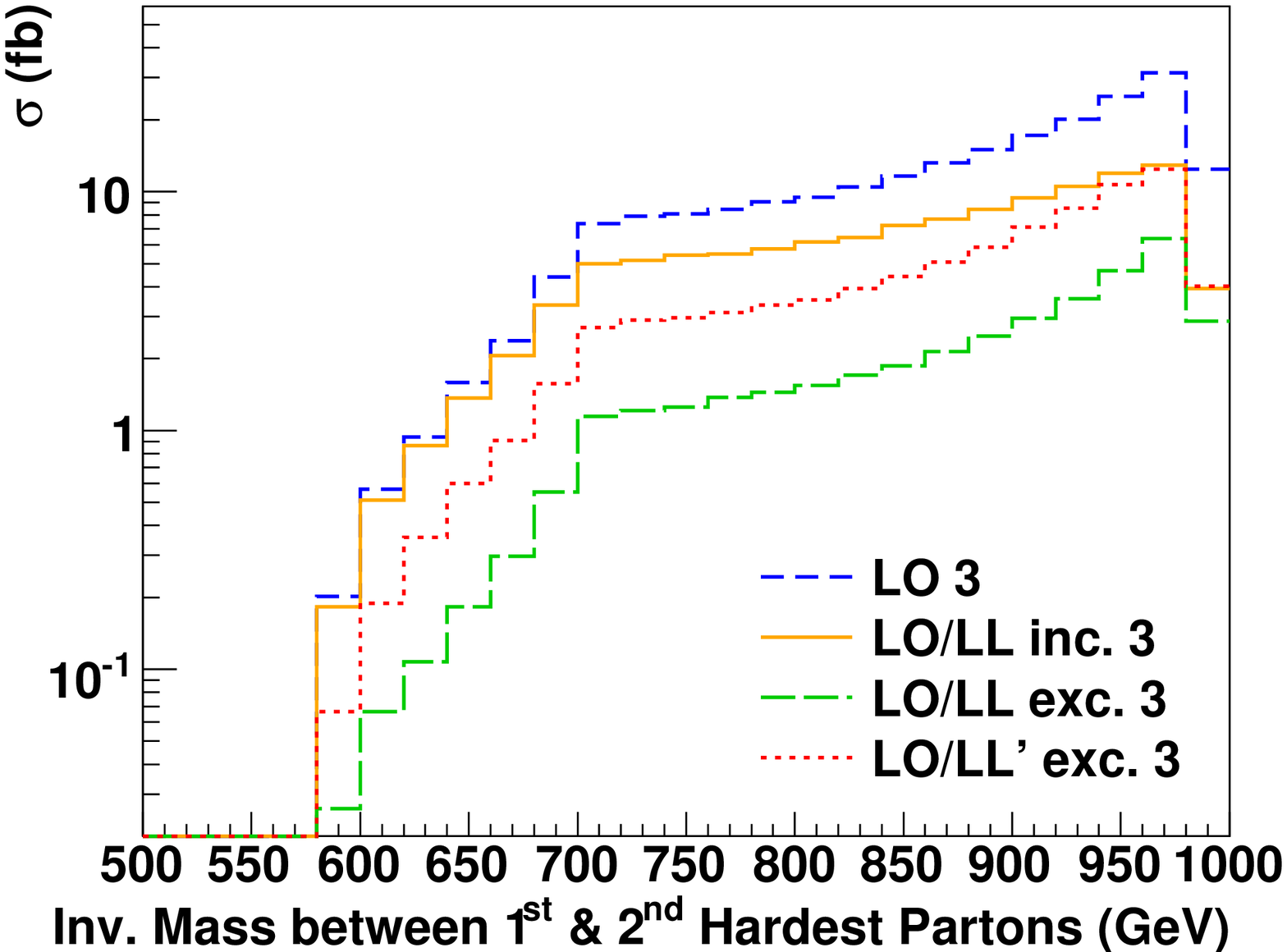}%
\hfill%
\includegraphics[width=0.5\textwidth]{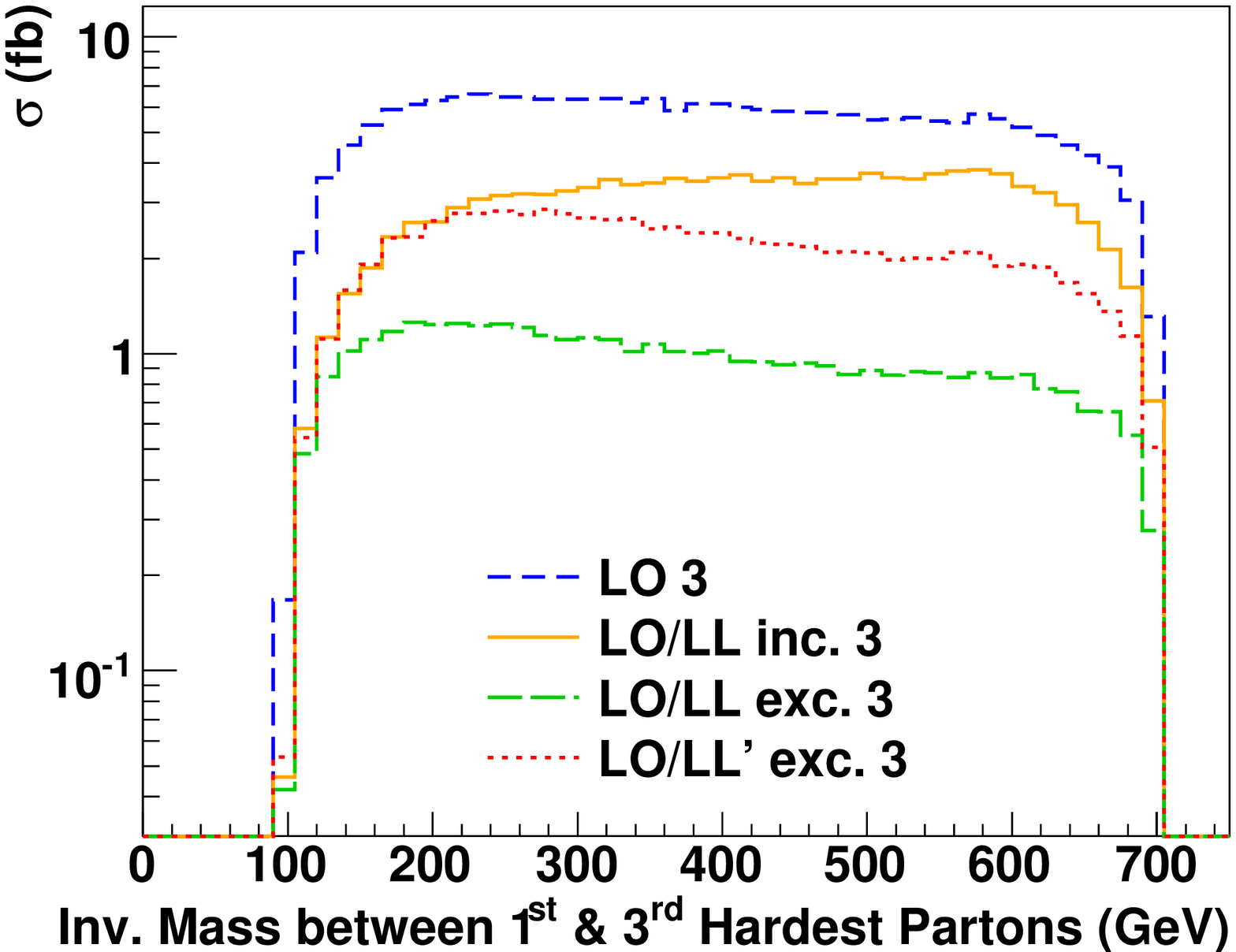}%
\caption{The effect of Sudakov factors on 3-parton matrix elements $e^+ e^- \to q \bar{q} g$. In the LO/LL inc.\ sample, the Sudakov suppression only captures the logarithms between $E_\CM$ and the gluon emission scale. In the LO/LL exc.\ sample, there are additional no-branching probabilities from the gluon emission scale down to the fixed $t_\match = t_\cut$ scale, giving effectively an ``exclusive'' 3-jet result with an overall Sudakov suppression. The LO/LL$'$ exc.\ sample includes the $\mathbf{\tilde{n}}$ correction factor from \eq{eq:supertruncsigma}, while the LO/LL exc.\ sample does not. Left panel: Invariant mass between 1st and 2nd hardest partons. The LO and LO/LL inc.\ samples agree near the lower endpoint, because this is the 3-jet region where Sudakov suppression is not active. The upper endpoint is the 2-jet region, and the $\mathbf{\tilde{n}}$ integrations in the LO/LL$'$ exc.\ sample restore lost cross section compared to the LO/LL exc.\ sample, asymptoting to the LO/LL inc.\ result. Right panel: Invariant mass between 1st and 3rd hardest partons, which shows similar effects. In the companion paper \cite{genevaphysics}, we will combine the LO/LL partonic results to create a merged LO/LL sample.}
\label{fig:logcompare}
\end{figure}

Differential  distributions for these three possibilities are shown in \fig{fig:logcompare} for $n = 3$, \ie\ $e^+ e^- \to q \bar{q} g$. Again, the hardest parton is defined as the parton with the largest energy. When the invariant mass between the 1st and 2nd hardest partons is at the lower threshold, the three partons are well separated, with a large emission scale for the gluon. Since the LO/LL inc.\ sample only captures the logarithms between $E_\CM$ and the gluon emission scale, it contains no Sudakov suppression in this part of phase space, and hence asymptotes to the LO result. The LO/LL exc.\ sample has an additional overall Sudakov suppression from leading-logarithmic running from the gluon emission scale down to the fixed matching scale $t_\match = t_\cut$. The upper endpoint is the 2-jet region with a low gluon-emission scale, so all three LO/LL samples are Sudakov suppressed compared to the LO sample. The LO/LL$'$ exc.\ sample has roughly the same shape as the LO/LL exc.\ sample but the overall scale is increased, because the $\mathbf{\tilde{n}}$ integrations it contains restore some of the lost cross section from phase space vetoes. The reason it asymptotes to the LO/LL inc.\ sample at the upper endpoint is because in this region it is most likely for two partons to be ``accidentally'' closer than $t_\cut$. The identical effect can be seen in the right panel when the invariant mass between the 1st and 3rd hardest partons is small. Note that the Sudakov suppression from the LO result to the LO/LL results is formally double-logarithmic, while the increase from the LO/LL exc.\ to the LO/LL$'$ exc.\ result due to the inclusion of the $\mathbf{\tilde{n}}$ integrations is formally single-logarithmic.

\subsection{The Efficiency of GenEvA}
\label{subsec:efficiency}

We argued in \sec{subsec:benefits} that \geneva\ has several advantages over more traditional algorithms, so it is interesting to see how its efficiency compares to other event generators. We begin by addressing the speed of \GenEvA\ for fixed-order calculations by comparing it to \madevent~\cite{Maltoni:2002qb}, and then show how \geneva\ is optimized to generate resummed partonic results.

Comparing these two programs is not straightforward, since the different methods used for phase space generation make it difficult to design a fair comparison. First, \geneva\ is written in C++, whereas \madevent\ is written in \fortran. Thus different levels of optimization in the compilers can affect the results. However, given that both programs use the same \fortran\ \texttt{MadGraph}/\texttt{HELAS} matrix elements and because evaluating the matrix element is the slowest part of \GenEvA, the comparison is reasonable though not optimal. Second,  in a single run \geneva\ directly produces a fully inclusive sample, containing a sum over all subprocesses and multiplicities, whereas \madevent\ generates events for each subprocess and multiplicity separately and combines them at the end of the run. Thus, \madevent\ is producing many events that are thrown out at the end of the day. This is a source of inefficiency that could easily be ameliorated by first calculating the approximate cross sections for the various subprocesses and then generating only the events required. In our speed comparisons, we therefore give timing information for $n$-parton samples with fixed $n$ and summed over all subprocesses for both \geneva%
\footnote{To obtain an $n$-parton sample with fixed $n$ in \geneva\ we simply set $n = n_\max$ and discard events with fewer than $n$ final state partons.}
and \madevent, as well as for \madevent\ samples with only one flavor subprocess.

\begin{table}[t]
 \begin{tabular}{l || c | c | c }
 & $\eta_{\rm eff}$ & $T_{\rm eff}$ (msec) & $T_{0.9}$ (msec) \\
 \hline\hline
 \geneva\ LO 3 & 0.789 & \phantom{000}0.57 &\phantom{000}0.62 \\
\geneva\ LO/LL inc.\ 3 & 0.965 & \phantom{000}0.47 & \,\,$<$ 0.47\\
 \hline
\madevent\ 3 & 0.982 & \phantom{000}2.6\phantom{0} & \,\,$<$ 2.6\phantom{0}\\
\madevent\ $u\bar{u}g$ & 0.994 & \phantom{000}3.0\phantom{0} & \,\,$<$ 3.0\phantom{0}\\
 \hline\hline
 \geneva\ LO 4 & 0.525 & \phantom{000}1.7\phantom{0} &\phantom{000}2.2\phantom{0} \\
\geneva\ LO/LL inc.\ 4 & 0.713 & \phantom{000}1.3\phantom{0} &\phantom{000}1.5\phantom{0}\\
 \hline
\madevent\ 4 & 0.809 & \phantom{00}11.1\phantom{0} &\phantom{00}11.4\phantom{0}\\
\madevent\ $u\bar{u}gg$ & 0.752 & \phantom{000}5.4\phantom{0} & \phantom{000}5.7\phantom{0}\\
 \hline\hline
 \geneva\ LO 5 & 0.390 & \phantom{00}10.0\phantom{0} &\phantom{00}15\phantom{.00} \\
\geneva\ LO/LL inc.\ 5 & 0.557 & \phantom{000}8.6\phantom{0} &\phantom{00}10.8\phantom{0}\\
 \hline
\madevent\ 5 & 0.843 & \phantom{00}62\phantom{.00} & \phantom{00}64\phantom{.00}\\
\madevent\ $u\bar{u}ggg$ & 0.833 & \phantom{00}27\phantom{.00} & \phantom{00}27\phantom{.00}\\
\hline \hline
 \geneva\ LO 6 & 0.298 & \phantom{0}160\phantom{.00} & \phantom{0}250\phantom{.00} \\
\geneva\ LO/LL inc.\ 6 & 0.396 & \phantom{0}150\phantom{.00} & \phantom{0}230\phantom{.00}\\
 \hline
\madevent\ 6 & 0.809 & 1900\phantom{.00} & 2300\phantom{.00}\\
\madevent\ $u\bar{u}gggg$ & 0.784 & \phantom{0}330\phantom{.00} & \phantom{0}350\phantom{.00}\\
\hline
 \end{tabular}
\caption{Comparison of the speed and efficiency between \geneva\ and \madevent. $\eta_{\rm eff}$ is the statistical efficiency defined in \eq{eq:neff}, and $T_{\rm eff}$ is the time required to create one statistical event. $T_{0.9}$ is the time to create one statistical event in a partially unweighted sample with $\eta_\lab{eff}(w_0) = 0.9$, as defined in \eq{eq:etaeff_w0}. Note that \GenEvA\ performs better on Sudakov-improved (LO/LL) results than on tree-level (LO) matrix elements. \GenEvA\ is in all cases faster than \madevent, in most channels by a factor of a few, despite the fact that \GenEvA's efficiencies are lower since it is a fixed-grid algorithm.
 \label{tab:speed}}
 \end{table}

The  efficiency results are shown in Table~\ref{tab:speed}, where 3-, 4-, 5-, and 6-parton matrix elements are compared. Shown are the statistical efficiencies $\eta_{\rm eff}$ defined in \eq{eq:neff}, as well as the time $T_{\rm eff}$ needed to create one statistically equivalent event. Because \GenEvA\ is a fixed-grid algorithm, its statistical efficiency $\eta_{\rm eff}$ is lower than that of \madevent, where grid optimization is used. However, the time needed to create a statistically equivalent weighted event sample is smaller for \GenEvA\ than \madevent, presumably because of the speed of the parton-shower-based phase space algorithm. To assess this efficiency/speed tradeoff, we also report $T_{0.9}$, the time needed to create one statistically equivalent event in a partially unweighted event sample with $\eta_\lab{eff}(w_0) = 0.9$ as defined in \eq{eq:etaeff_w0}. That is, we unweight just those events with weights below a threshhold weight $w_0$, with $w_0$ chosen such that the final event sample has a 90\% statistical efficiency. By this $T_{0.9}$ measure, \GenEvA\ is still faster than \madevent\ in all channels, often by a factor of a few.

We  do not show comparisons of the unweighting efficiencies between \GenEvA\ and \madevent, because there is no reasonable way to do so. As argued in \sec{subsec:eff}, unweighting efficiencies are sensitive to the exponential tails of the weight distribution. Depending on how long one runs the two programs, one may or may not sample these tails, giving wildly varying unweighting efficiencies during one run and between different runs. Furthermore, it is completely sufficient for all practical purposes to have \emph{partially} unweighted samples with a reasonably high statistical efficiency.

As we discussed above, the fact that \geneva\ is using a parton shower as a phase space generator makes it ideal to distribute events according to logarithmically-improved partonic calculations. This is because the parton shower already contains the correct double-logarithmic behavior, such that the resulting weights are expected to be uniform, giving rise to a high statistical efficiency. To see this, consider the event weight given in \eq{eq:lollweight}. Since the splitting functions reproduce the tree-level matrix elements of QCD in the collinear limit, the denominator has exactly the same singular behavior as the numerator, with no logarithmic differences.

\begin{figure}[t]
\includegraphics[width=0.5\textwidth]{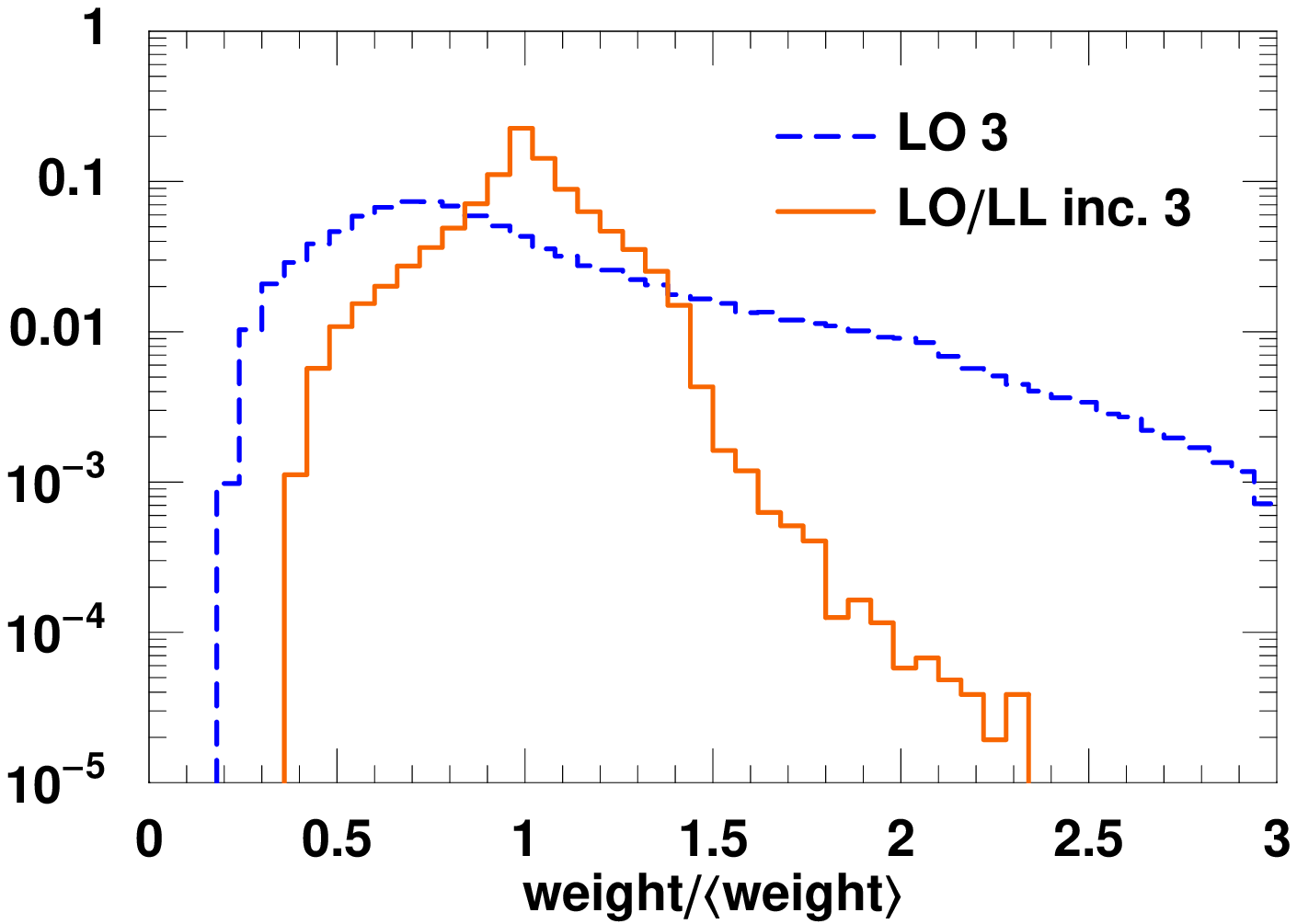}%
\hfill%
\includegraphics[width=0.5\textwidth]{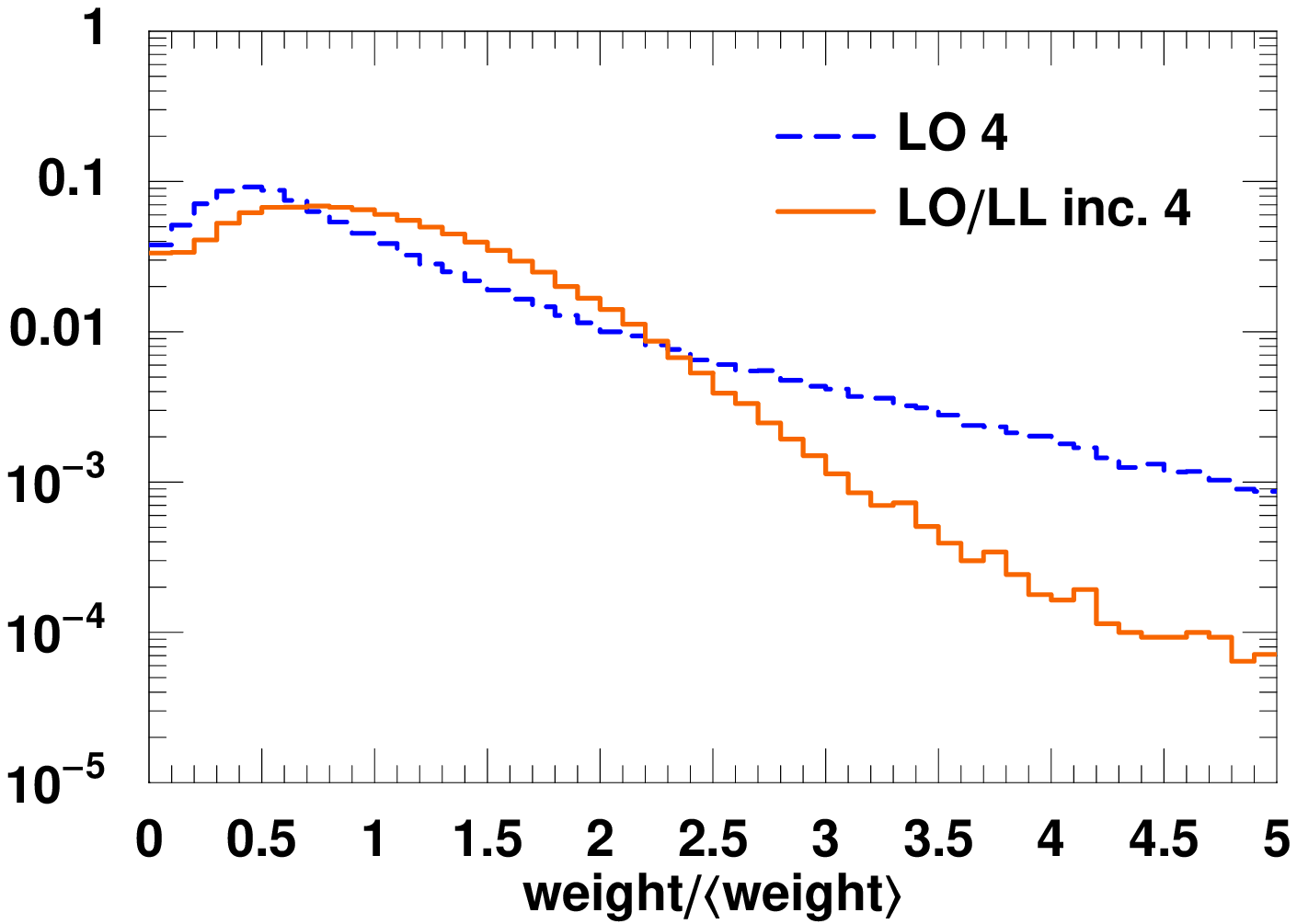}%
\\
\includegraphics[width=0.5\textwidth]{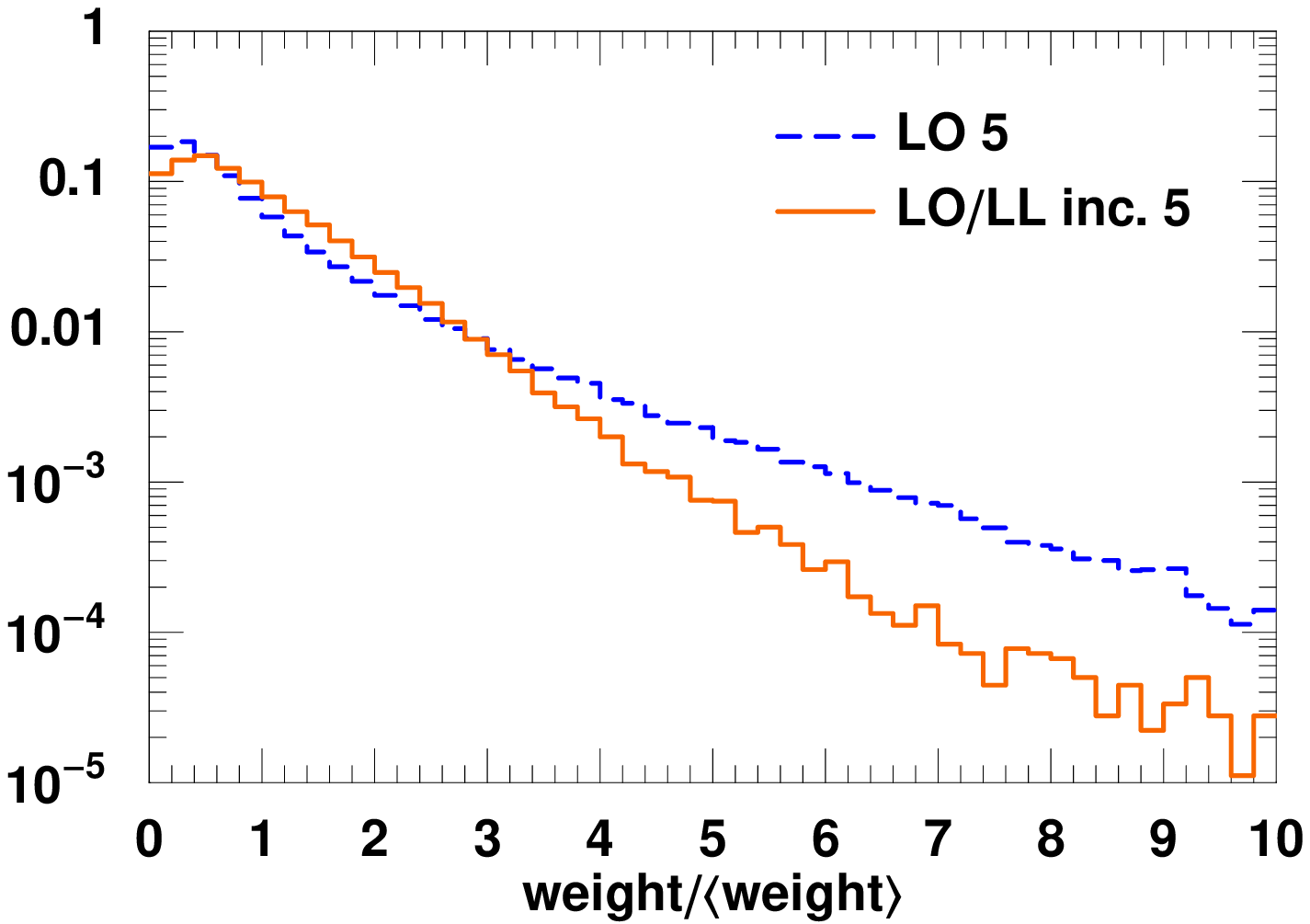}%
\hfill%
\includegraphics[width=0.5\textwidth]{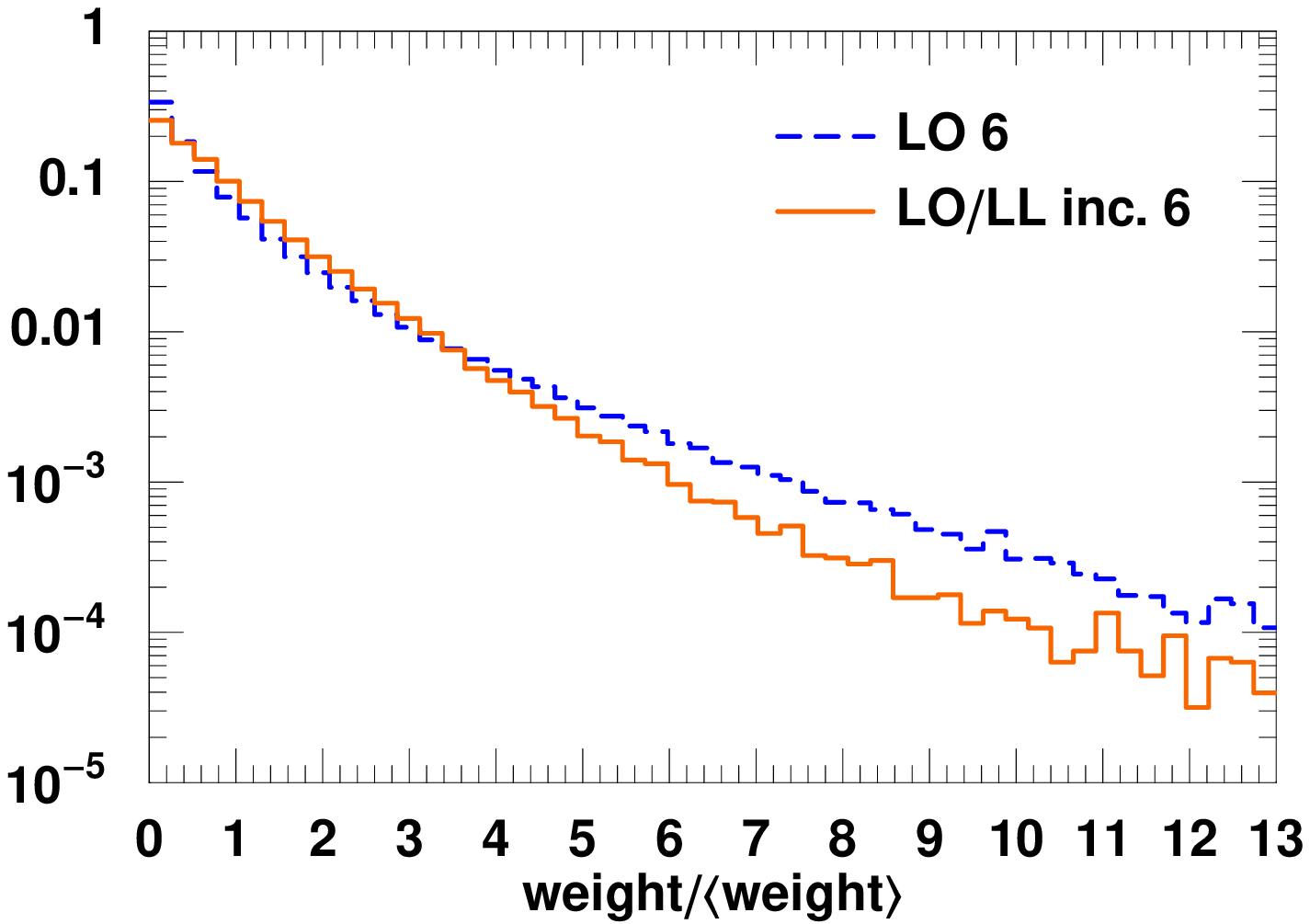}%
\caption{Comparison of the weight distributions between \GenEvA\ LO and LO/LL samples for 3- to 6-parton matrix elements. Because the \GenEvA\ phase space generator already includes leading-logarithmic information, the LO/LL weight distributions are more strongly peaked at their average weights. More importantly, there is a considerable suppression of the high-weight tail when going from the LO to LO/LL samples. As expected, the effect of leading logarithms is less relevant for the higher-multiplicity samples, as phase space suppression keeps the double-logarithms from growing too large.}
\label{fig:weightcompare}
\end{figure}

\begin{figure}[t]
\includegraphics[width=0.5\textwidth]{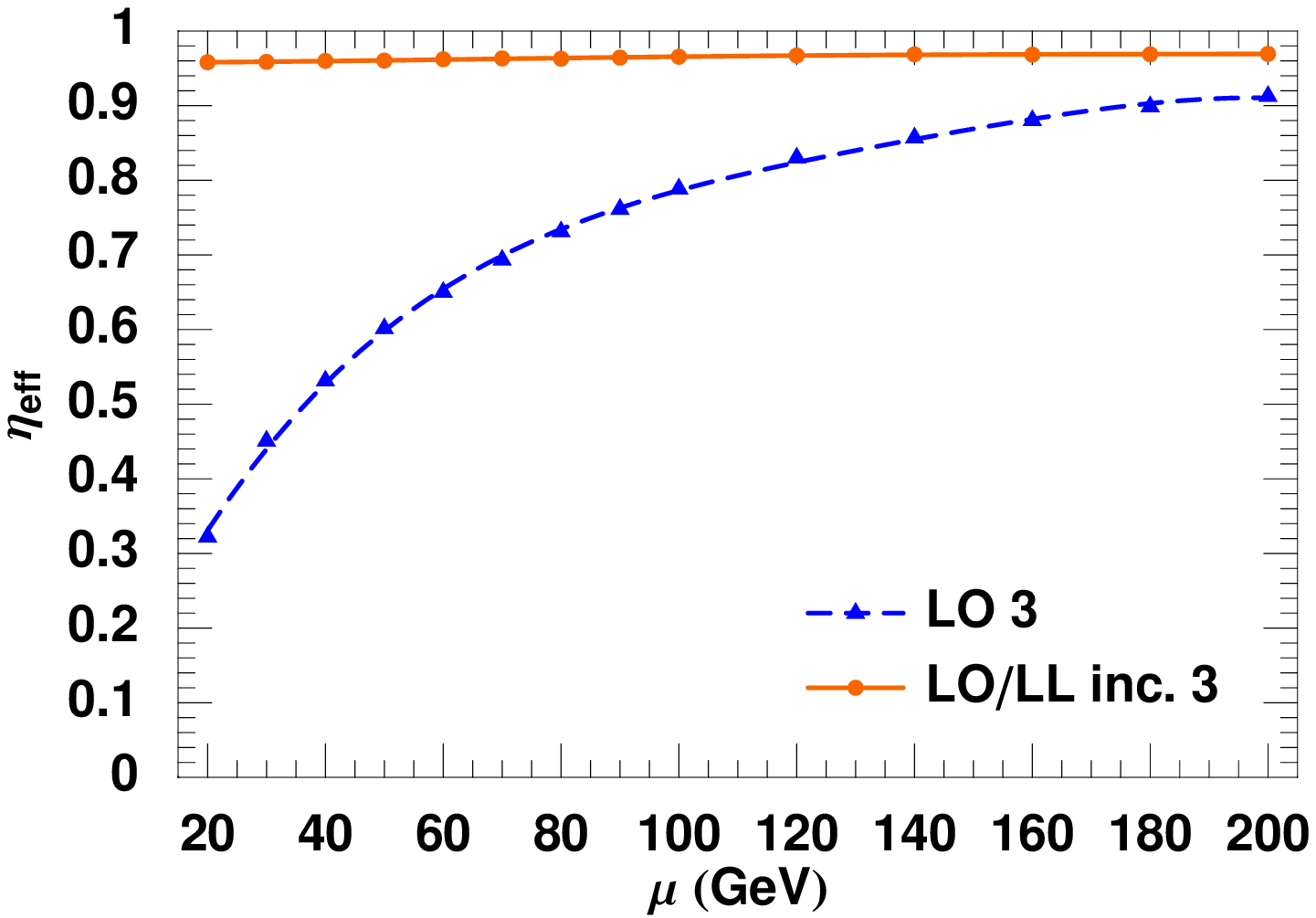}%
\hfill%
\includegraphics[width=0.5\textwidth]{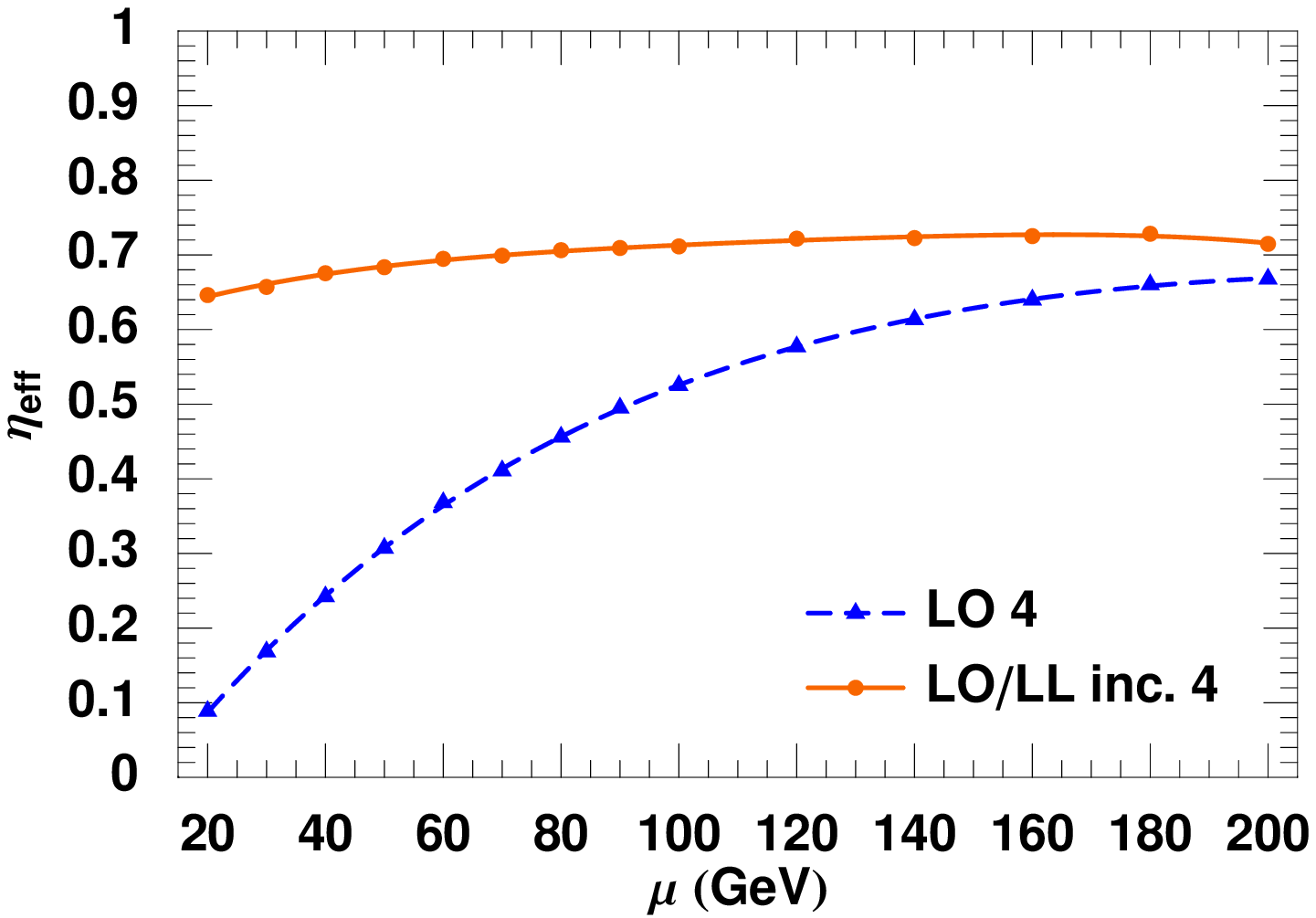}%
\\
\includegraphics[width=0.5\textwidth]{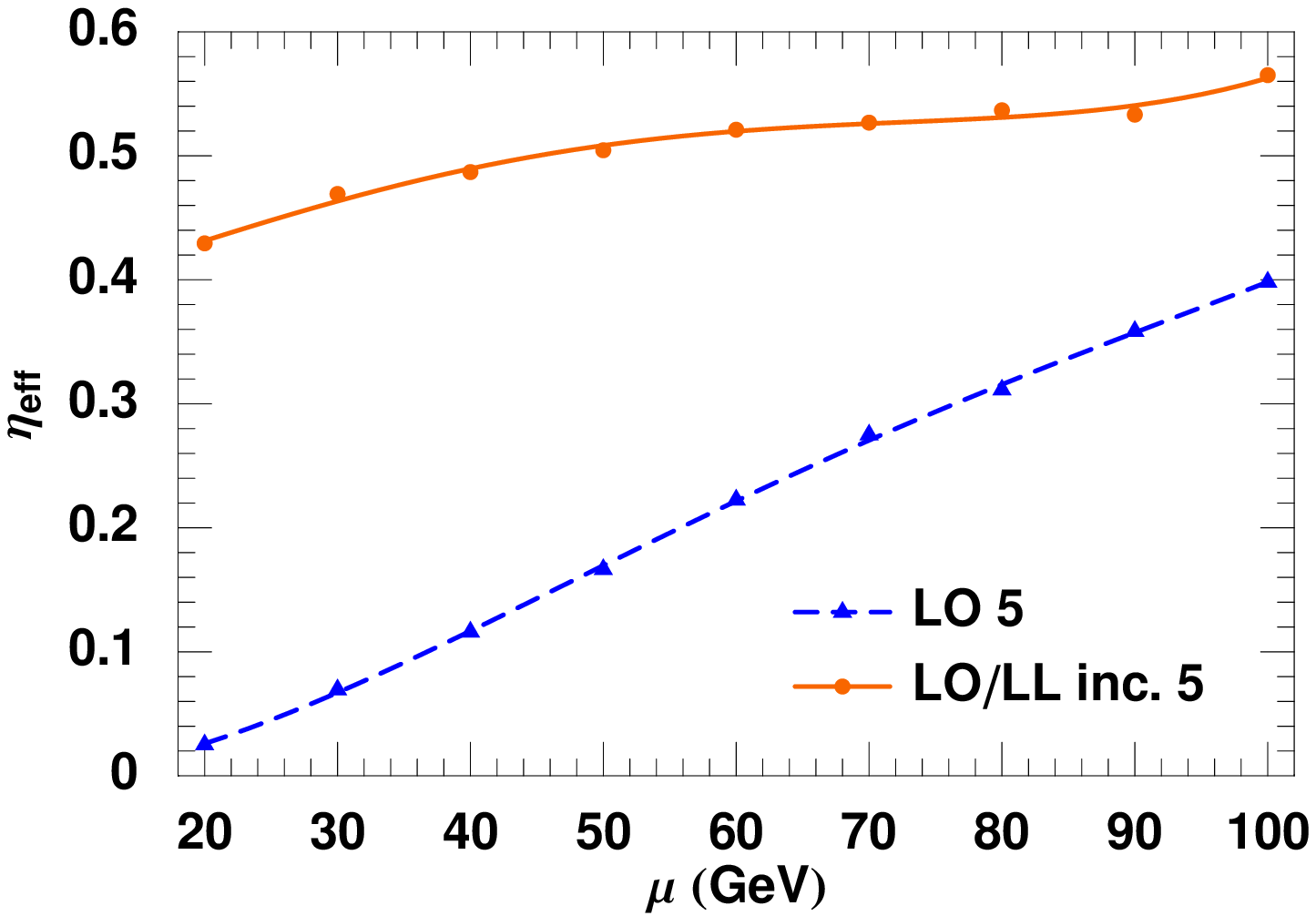}%
\hfill%
\includegraphics[width=0.5\textwidth]{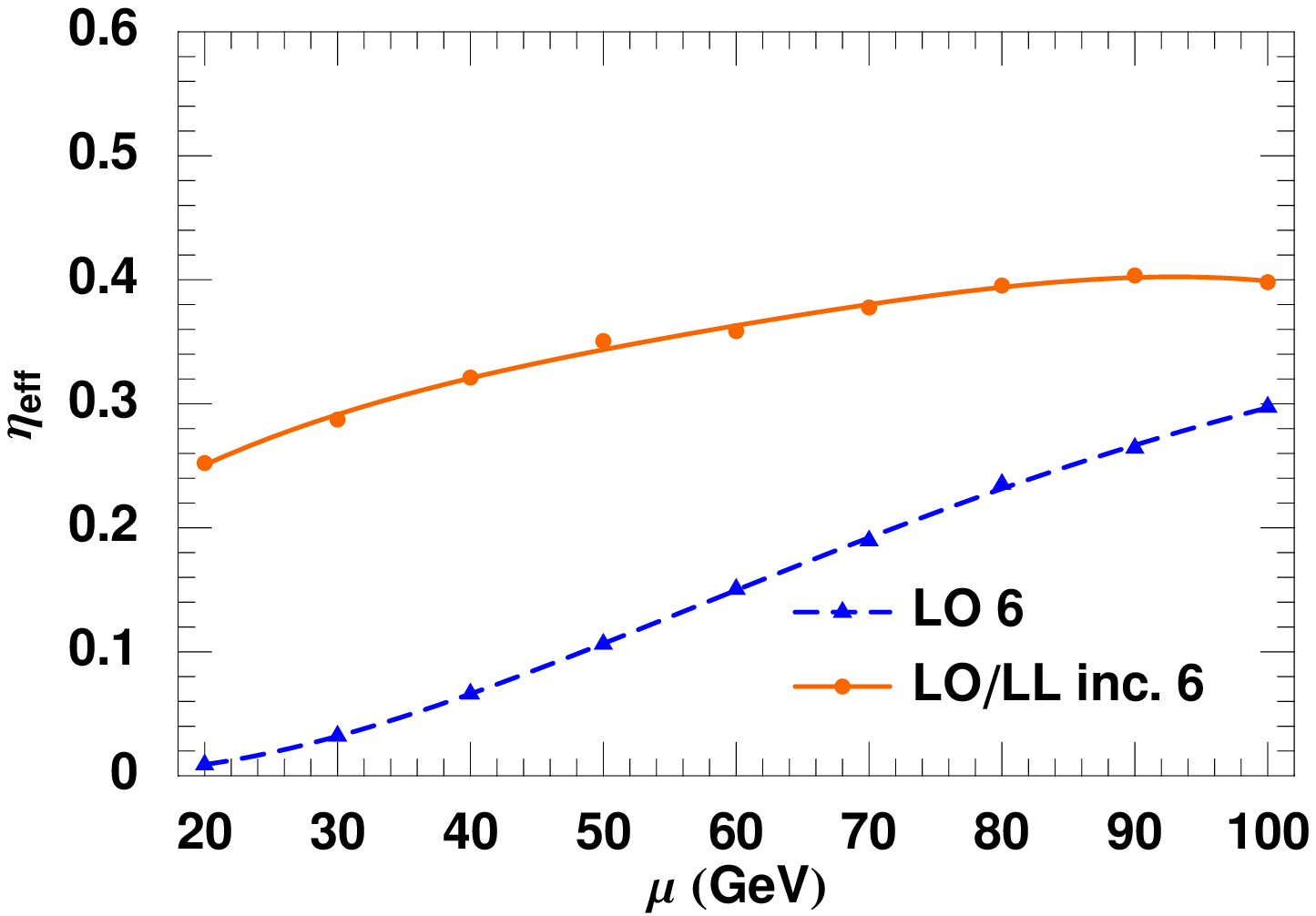}
\caption{Comparison of the statistical efficiencies $\eta_\mathrm{eff}$ between \GenEvA\ LO and LO/LL samples for 3- to 6-parton matrix elements. As the matching scale $\mu = \sqrt{t_{\rm match}}$ is lowered, the efficiency of the LO/LL sample stays roughly constant, while the efficiency of the LO sample drops dramatically.}
\label{fig:effcompare}
\end{figure}

This increase in efficiency for LO/LL event generation can be seen in a few different ways, the simplest being that in Table~\ref{tab:speed}, the three different efficiency/speed measures are all better for the \GenEvA\ LO/LL samples than for the \GenEvA\ LO samples. In \fig{fig:weightcompare}, we show the distribution of event weights between the LO/LL and LO samples. For $n=3$ especially, the weight distribution is far more peaked near the average weight in the LO/LL sample compared to the LO sample, showing that the inclusion of leading-logarithmic information softens the high-weight tails. Even for $n=6$, where Sudakov logarithms play a much smaller role in determining the differential distributions, there is still a factor of 5 decrease in the number of high-weight events. The change in efficiency is shown even more dramatically in \fig{fig:effcompare}, which compares the statistical efficiency between the LO/LL and LO samples as the matching scale $\mu$ is changed. For the LO/LL samples, the efficiency is nearly constant as $\mu$ varies, while the LO samples exhibit a sharp decrease in efficiency for $\mu \to 0$.

Thus, \geneva\ is faster at distributing resummed results than tree-level results. We note that the efficiency of \GenEvA\ could improve further with the use of a more realistic internal parton shower. For example, tree-level matrix elements include the effect of color coherence which is absent in the virtuality-ordered shower currently used in \GenEvA. If a $p_{\perp}$-ordered shower could be implemented as discussed in \sec{subsec:evolvar}, then the leading color coherence effects would already be captured by the phase space generator, which should lead to more uniform weights.

\section{Conclusions}
\label{sec:conclusions}

We have shown how to construct a phase space generator from a parton shower. The resulting \geneva\ algorithm uses a very different strategy compared to more conventional approaches. Most available generators use a non-recursive, multi-channel, adaptive-grid algorithm to distribute events in phase space directly according to a desired distribution, which are often tree-level matrix elements. In contrast, \geneva\ uses a parton shower to generate events, and in a second step reweights the events to the desired distribution.

Though the \geneva\ algorithm uses a fixed grid with no adaptation, the recursive nature of the parton shower allows for an incredibly fast generation of phase space points. Since the parton shower has the same singularity structure as QCD, the effectively fixed grid used in the \geneva\ algorithm is presumably very close to the final adapted grid in traditional phase space generators. As a result, the \geneva\ algorithm is able to compete with and likely even outperform traditional generators.

The main advantage of the \geneva\ algorithm arises when one goes beyond fixed-order calculations. It is well known that leading-logarithmic terms have to be taken into account in order to obtain realistic QCD predictions. In particular, when merging matrix element calculations with subsequent parton shower evolution, leading-logarithmic information is essential to avoid strong dependence on the unphysical matching scale. Unfortunately, analytic expressions for logarithmically improved, fully differential distributions are prohibitively difficult to obtain for a large number of external particles, since the required scales in the logarithms depend in a complicated way on all kinematic invariants in the event.

For this reason, traditional generators often implement leading-logarithmic improvements after the initial phase space generation is complete, and use the known kinematics of the event in order to calculate the required scales in the logarithms. In contrast, the parton shower used to generate phase space points in \geneva\ automatically contains the leading logarithmic behavior of QCD by construction, and for this reason \geneva\ is in fact more efficient in distributing logarithmically improved distributions than fixed-order calculations and thus has the potential to outperform traditional techniques.

To conclude, we discuss some potential extensions to the \GenEvA\ algorithm to make it useful for LHC physics.

\subsection{Towards Hadronic Collisions}
\label{subsec:towardsLHC}

Currently, the \GenEvA\ algorithm is only designed to work for $e^+ e^-$ collisions with no initial state photon radiation. The technical reason for building \GenEvA\ for $e^+ e^-$ first is that the final state parton shower with fixed center-of-mass energy is easier to understand and verify. However, the name \GenEvA\ is obviously inspired by the site of the upcoming LHC experiment at CERN. We anticipate extending the \GenEvA\ algorithm to hadronic collisions, and one might wonder whether there are any additional technical hurdles to overcome.

To build \GenEvA\ for $e^+ e^-$ collisions, there were three technical problems that needed to be solved. First, in order to generate events that conserve energy-momentum, parton showers need to adjust the kinematics of the parton splittings after they are generated, which typically makes it impossible to calculate the probability analytically. Second, different parton shower histories can give rise to the same points in Lorentz-invariant phase space, such that a parton shower covers phase space multiple times. Third, one needs a way to ``truncate'' a high-multiplicity event generated by the parton shower to a lower-multiplicity event for which the probability is still known analytically.

In this paper we explained in detail how these technical issues are solved in the \geneva\ algorithm. The first issue is solved by employing the analytic parton shower developed in Ref.~\cite{Bauer:2007ad}. The second issue is solved by introducing the overcounting factor which multiplies the generated event weight. Finally, the double-branch probability used in this paper enforces a global evolution, which allows for a simple truncation algorithm.

These technical solutions should have straightforward generalizations to hadronic collisions. While we know of no analytically calculable initial state shower, building one should not be too difficult as it is well known that phase space can be recursively described both in terms of an $s$-channel ``final state'' tree and a $t$-channel ``initial state'' tree. The trick to creating the analytic final state shower was to use $\cos \theta$ as the primary variable instead of $z$, and an analogous trick should be possible for the initial state shower. The issue of double-counting will require figuring out how to deal with interference between initial state and final state shower configurations, but as a last resort, one can always do a brute force sum over parton shower histories to solve the double-counting issue. Truncation should be more-or-less identical to the final state shower, because many initial state showers are based on backwards evolution, and therefore can have a notion of well-ordering.

Hence, the main new issue in hadronic collisions is the proper treatment of parton distribution functions (PDFs). However, apart from issues of efficiency, this is really not an issue, because parton distribution functions simply introduce two new variables $x_1$ and $x_2$ and define a new choice of partonic calculation
\begin{equation}
\df\sigma = f_1(x_1,Q_1^2) f_2(x_2,Q_2^2)\, \df x_1\, \df x_2 \, \df \hat{\sigma}_{12 \to X}
\equiv \sigma'(\Phi')\, \df \Phi'
\,,\end{equation}
where $\Phi'$ includes $x_{1,2}$ in addition to the ordinary phase space variables, and $\sigma'(\Phi')$ is the fully differential cross section including PDF information. The point of reweighting is that no matter how events are selected in $\Phi'$, events can be distributed according to $\sigma'(\Phi')$, and it is the user's responsibility to, for example, use NLO PDFs with NLO calculations, use the proper scales for $Q_{1,2}^2$, and so on.

\subsection{Heavy Resonances}

At the LHC, the most interesting Standard Model backgrounds to Beyond the Standard Model signals come not from QCD itself but from heavy resonances like $W$, $Z$, and $t$ created in association with QCD radiation. Because \GenEvA\ is based on a parton shower, one might wonder whether it is possible to include heavy resonances in the \GenEvA\ algorithm.

For this purpose, it is useful to think of \GenEvA\ not as a parton shower but as simply a recursive phase space generator, and resonances corresponds to just a funny choice of ``splitting function''. Because events will be reweighted at the end of the day to the desired distribution, all what is important is that the ``splitting functions'' have the right singularity structure, which for the purposes of heavy resonances is the Breit-Wigner line shape.

What will be particularly bizarre about the \GenEvA\ algorithm with heavy resonances is that there will be ``Sudakov factors'' associated with resonances, but as long as the resonances are narrow enough, this Sudakov factor will either be unity if $t_\cut > m^2$ or zero if $t_\cut < m^2$, meaning that a resonance is either on shell or it is decayed. This also means that even if $t_\cut > m^2$ and a resonance is treated as on shell for the matrix element calculation, if the \GenEvA\ shower is restarted at $t_\cut$, the Breit-Wigner distribution will be recovered for the resonance with the rest of the shower history compensating for the change in energy-momentum.

Once the logistics for handling mass thresholds with heavy resonance is established, the treatment of not-so-heavy resonances like the bottom quark will be straightforward, and one would use the ordinary massive Altarelli-Parisi splitting functions for $b$ and $c$ quarks. The top quark is an interesting case, because in \GenEvA, the top would both decay via $t \to Wb$ but also radiate through $t \to tg$. Both of these processes need to coexist in \GenEvA\ in order to cover all of multiplicity, flavor, and phase space, and this dual nature of the top may be interesting for understanding Monte Carlo effects in top-quark mass measurements.

\subsection{Choice of Evolution Variable}
\label{subsec:evolvar}

For the \GenEvA\ algorithm to be relevant in the LHC era, it will have to have a smooth interface with a hadronization/underlying event model. Regardless of the phase space strategy, every parton-based Monte Carlo program eventually has to interface to some external program that is responsible for hadronizing the initial and final state partons into the hadrons observed in the experiment and also to deal with secondary interactions, beam remnants, pile up, and other nonperturbative effects present in any hadronic collision.

In \fig{fig:regimes} we argued that because \GenEvA\ is based on a parton shower and because there already exists a smooth interface between the showering regime and the hadronization regime in existing programs, then \GenEvA\ should inherit the same smooth interface. However, as mentioned already, the internal \GenEvA\ shower is optimized for speed and is not really meant to provide the best possible modeling of small angle QCD emissions. For example, the \GenEvA\ algorithm currently evaluates $\alpha_s$ at a fixed scale and does not implement color coherence. Also, it uses virtuality as the evolution variable and not $p_\perp$ as modern showers do. While it is in principle possible to construct an analytic parton shower which includes all these effects, this might not be the most efficient strategy.

As discussed in \sec{subsec:truncation}, instead of using the internal shower in the showering regime, \GenEvA\ can rely on an external parton shower as long as the external parton shower starts evolving from the event-defined matching scale $t_\match$. From a physics point of view, it is important that the double-logarithmic dependence on $t_\match$ cancel between the partonic and showering regimes, but this is difficult if the evolution variables differ in the two regimes. In this paper, we have been using virtuality as the matching variable, so the current version of the \GenEvA\ algorithm can therefore only be smoothly interfaced with a virtuality-ordered parton shower. How then, can \GenEvA\ interface with more modern $p_\perp$-ordered showers?

This turns out to be an algorithmic as opposed to a conceptual problem, though. All \GenEvA\ requires to map out phase space is a parton shower with no dead zones for which an analytic formula exists for calculating the probability for generating a given point in phase space. As a phase space generator, a $p_\perp$-ordered shower is no better or worse than a virtuality-ordered shower, because at the level of phase space, $p_\perp$-ordering is equivalent to virtuality-ordering with an additional veto to eliminate certain unwanted shower histories.

The main problem with a $p_\perp$-ordered shower is that there is no ``local'' way to determine whether a parton shower history is $p_\perp$-ordered or not. With a virtuality-ordered shower, every way of combining four-momenta that is consistent with flavor gives a valid shower history, allowing \GenEvA\ to use the numerically efficient \texttt{ALPHA} algorithm~\cite{Caravaglios:1995cd,Caravaglios:1998yr} to ``count'' the number of parton shower histories that give identical final sate four-momenta. It is impossible to use the \texttt{ALPHA} algorithm to count $p_\perp$-ordered shower histories (or even virtuality-ordered shower histories with an angular-ordering veto), because whether or not a history is $p_\perp$-ordered (or angular-ordered) requires information about previous branches.

It would be straightforward to rewrite \GenEvA\ to use a $p_\perp$-ordered shower to map out phase space as long as one could find a way to efficiently count $p_\perp$-ordered shower histories. One idea is to generate the full $p_\perp$-ordered expression for $\sum_i \hat{\alpha}(\vS_i)$, and then use some variant of the \texttt{O'Mega} algorithm \cite{Moretti:2001zz} to generate numerically efficient code to calculate this sum. In other words, we can treat $\sum_i \hat{\alpha}(\vS_i)$ as a ``matrix element'' by itself, and construct an expression for it offline.

One additional subtlety comes from the new interleaved evolution introduced in \texttt{Pythia} \cite{Sjostrand:2006za}. In order for reweighting to make sense, \GenEvA\ assumes that any external showering program will not affect the existing \GenEvA\ shower history. In interleaved evolution, the primary interaction is not independent of secondary ones and the underlying event competes with the hard scattering for the momentum fraction of the incoming partons. Again, this is an algorithmic and not a conceptual problem. One could always write a more complicated version of \GenEvA\ around an interleaved shower, and as long as the probability to obtain a given (interleaved) shower history is known, the \GenEvA\ algorithm will be more or less unchanged. Practically speaking, though, the matching scale can probably be taken sufficiently large such that \texttt{Pythia} interleaved evolution starting from the \GenEvA\ cutoff will give a reasonable description of the hadronic regime.

\subsection{Outlook}

Once the technical issues involving hadronic collisions, heavy resonances, and evolution variables are solved, there is a final technical question about how the \GenEvA\ algorithm will implement more advanced matrix elements, such as those that involve complicated loop calculations~\cite{Bern:2007dw}, or those based on effective theories \cite{Bauer:2006mk,Bauer:2006qp,Schwartz:2007ib}. As currently written, the \GenEvA\ algorithm assumes that there is a unique answer for the value of $\sigma(\Phi)$, but depending on the analytic form of $\sigma(\Phi)$, it may be more convenient to use Monte Carlo methods not only to sample phase space but also to determine $\sigma(\Phi)$. While the specific implementation will affect the performance of the \GenEvA\ algorithm, the idea of additional ``hidden'' Monte Carlo variables being used to determine $\sigma(\Phi)$ is still compatible with the \GenEvA\ algorithm's use of reweighting. \GenEvA\ can in principle implement any desired differential cross section as long as it is well-defined on paper.

Combined with the \GenEvA\ framework \cite{genevaphysics}, the \GenEvA\ algorithm has the potential to significantly improve our understanding of Standard Model backgrounds. The ability to conceive of an event simultaneously as a shower history and as a fixed-body phase space point allows one to answer questions that are difficult to address in either language individually. The efficiency of the \GenEvA\ algorithm makes it a worthy competitor to existing methods, and the automatic inclusion of leading-logarithmic effects raises the bar for what the experimental collaborations can expect from the Monte Carlo community in the LHC era.

\begin{acknowledgments}
We would like to thank Johan Alwall, Lance Dixon, Walter Giele, Beate Heinemann, Zoltan Ligeti, Michelangelo Mangano, Michael Peskin, Matthew Schwartz, Torbj\"{o}rn Sj\"{o}strand, Peter Skands, Iain Stewart, and Dieter Zeppenfeld for many useful discussions. We would also like to thank Johan Alwall, Fabio Maltoni, and Tim Stelzer for patient assistance with \texttt{MadGraph}, and Jeffrey Anderson for help with general computing questions.
This work was supported in part by the Director, Office of
Science, Office of High Energy Physics of the U.S.\ Department of Energy under
the Contract DE-AC02-05CH11231.
CWB acknowledges support from a DOE OJI award and an LDRD grant from LBNL.
JT is supported by a fellowship from the Miller Institute for Basic Research in Science.
 \end{acknowledgments}

\bibliographystyle{../physrev4}
\bibliography{../geneva}

\end{document}